\newcommand{\code}[1]{\texttt{#1}}

\documentclass[a4paper,12pt,times,numbered,print,index]{Classes/PhDThesisPSnPDF}

\usepackage[dvipsnames]{xcolor}

\usepackage{fancyvrb}

\RecustomVerbatimCommand{\VerbatimInput}{VerbatimInput}%
{fontsize=\footnotesize,
 frame=lines,  
 framesep=2em, 
 rulecolor=\color{Gray},
}

\input{Preamble/preamble}

\title{Numerical analysis of the Primordial Power Spectrum for inflationary potentials}

\author{Ira Wolfson}

\dept{Department of Physics}

\university{Ben Gurion University of the Negev}
\crest{\includegraphics[width=0.2\textwidth]{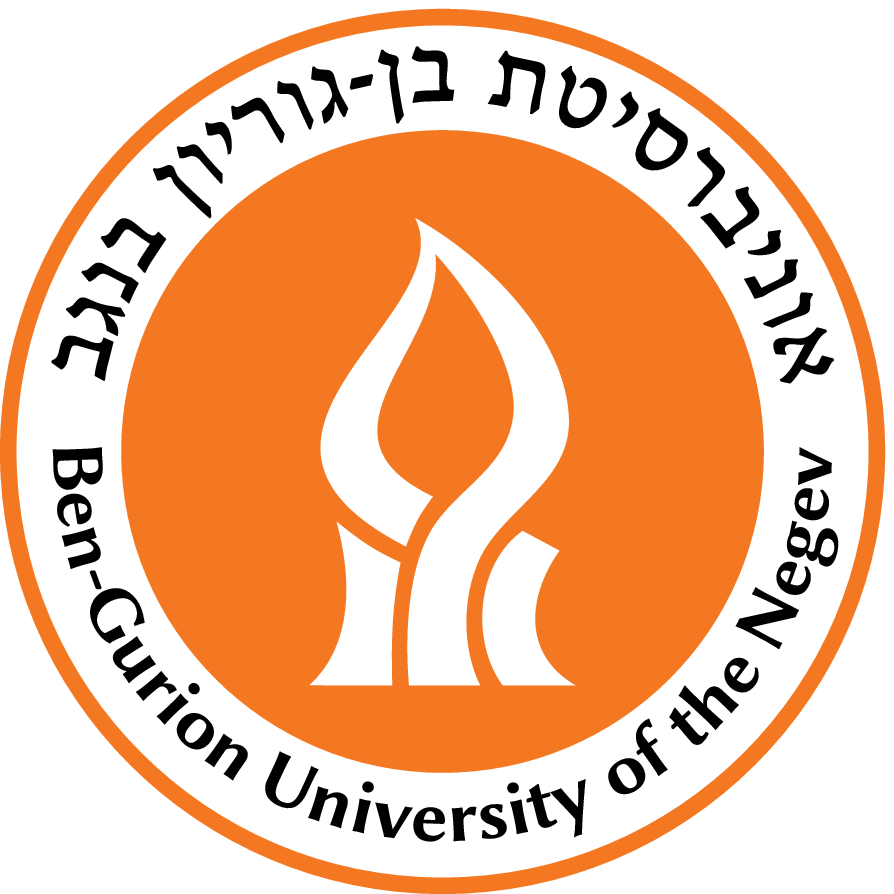}}

\degreetitle{Doctor of Philosophy}

\subject{LaTeX} \keywords{{LaTeX} {PhD Thesis} {Engineering} {University of
Cambridge}}

\ifdefineAbstract
\fi

\ifdefineChapter
\fi

\begin{document}
\frontmatter

\maketitle


\begin{dedication} 

For Netta, Tamar, Meir and Daya. \\

\end{dedication}


\begin{declaration}

I hereby declare that except where specific reference is made to the work of 
others, the contents of this dissertation are original and have not been 
submitted in whole or in part for consideration for any other degree or 
qualification in this, or any other university. This dissertation is my own 
work and contains nothing which is the outcome of work done in collaboration 
with others, except as specified in the text and Acknowledgements. This 
dissertation contains fewer than 65,000 words including appendices, 
bibliography, footnotes, tables and equations and has fewer than 150 figures.


\end{declaration}


\begin{acknowledgements}      

I want to acknowledge, first and foremost my advisor, Prof. Ramy Brustein, for taking a chance on a young student coming hat-in-hand to start working in the mostly barren field of numerical cosmology in Israel. As is the case oftentimes, professional disagreements were not hard to come by, but the personal and professional guidance was always available. Instrumental in my scietific coding `upbringing' were Dr. Rahul "write your own" Kumar, and Dr. Daniel "you're wrong" Ariad. These people were there when I was but a seedling of the computational physicist I am today, and I am forever in their debt. \\
In the later stages of my doctoral adventure, several people were instrumental in guiding me or, at the very least, providing a sounding board for my craziness. Dr. Ido Ben-Dayan contributed time and effort to bring me up to speed in all matters of observational and experimental cosmology. Snir "Erdelyi" Cohen, Tomer "Deadpool" Ygael , and Yossi "Pigal" Naor were the usual victims of my (temporary, physics induced) insanity.\\
When PhD has grown terminal, two main people were instrumental in `bringing it all together'. Prof. Eiichiro Komatsu, who, with a 30-minute talk crystallized many insights, and sharpened them into a point-like focus of what it meant to be a physicist. And Dr. Yuval Ben-Abu, who never, not even for a minute, let me rest. He always pushed me towards the next great challenge. \bigskip\\
There is still a long list of thanks due, which is in no way exhaustive:\\
Prof. Antonio Riotto, for his support and invaluable advice.\\
Members of my advisory committee, professors Eduardo Guendelman and Uri Keshet.\\
Sifu and Shermu, Donald and Cheryl-Lynn Rubbo.\\
And my family and friends, chief among them are my parents - Joshua and Pnina Wolfson.\bigskip \\
In this as in everything I owe my life to my wife, Netta Schramm.\\

\end{acknowledgements}

\begin{abstract}
In this work we study small field models of inflation, which, against previous expectations, yield significant Gravitational Wave (GW) signal, while reproducing other measured observable quantities in the Cosmic Microwave Background (CMB). We numerically study these, using previously published analytic works as general guidelines. We first discuss the framework necessary to understand the model building procedure and some of its motivations. We review the slow-roll paradigm, derive the slow-roll parameters and discuss different formulations thereof. We further present the Lyth bound and its theoretical descendants and finally, we outline the small/large field taxonomy and their characterization in the current nomenclature.\\
We proceed to present our models and the methods used in their building and examination. We employ MCMC simulations to evaluate model likelihoods and by process of marginalization extract the most probable coefficients for these inflationary potentials. An additional method applied is a multinomial fit, where we create a functional correspondence between coefficients and observables. This allows us to use the observable values directly to yield the most likely coefficients. We compare the results of the two methods and evaluate the level of tuning required for these models.\\
We discuss an apparent discrepancy between analytical approaches of evaluating Primordial Power Spectrum (PPS) observables and the precise numerical results in our models. We identify some of the sources of this discrepancy and remark on their meaning in the age of precision cosmology.\\
Finally, we present the results of our study, for the most likely inflationary models with polynomial potentials of degree 5, and 6. We demonstrate our ability to produce potentials that yield GW with a \textit{tensor-to-scalar} ratio $r=0.03$. This is a realistic expectation of GW detection sensitivity in the near future. Detecting GW of a primordial source will provide a direct indication of the energy scale of inflation, and therefore an interesting probe into physics beyond the standard model.\\
\end{abstract}


\tableofcontents

\listoffigures

\listoftables


\printnomenclature

\mainmatter


    \graphicspath{{Chapter2/}}
\chapter{Scientific Introduction - State of the art}\label{science_intro}
Even from the infancy stages of humanity, the question of origins was an ever-present one. Every religion in existence and in fact, every culture, incorporates an origin story for the universe.\\\\
Up until the new era of scientific renaissance, this field was dominated predominantly by theologists, mystics and story-tellers. Some may attribute this to proper education being exclusive to members of the upper religious castes.\\
\section[The Big Bang]{The evidence for a hot Big Bang}
It is not until Einstein's theory of relativity that we can construct an origin story of sufficient precision, to exclude divine intervention. At least a manifest one.\\\\
In the early 20th century,  Alexander Friedmann, using Einstein's own theory of relativity, was able to derive an equation of motion for the evolution of the universe\cite{FriedmannEqs,1999GReGr..31.1991F} . It took about seven years, Friedmann's death, and Georges Lema\^{i}tre's independent work \cite{1927ASSB...47...49L,1931MNRAS..91..483L}, for Einstein to accept the idea that his new physics proposed a mechanism for the evolution of the cosmos itself. Coupled with Edwin Hubble's observations and the Hubble law \cite{HubbleLaw}, this initiated a flood of scientific theories regarding the origin of our universe, with three main competing scenarios:
\begin{itemize}
	\item A static universe - Most famously connected to Einstein's ``biggest blunder", the cosmological constant \cite{Einstein:1917:KBA}.
	\item An ever expanding universe where matter is constantly created, to fill in the newly created space.
	\item An evolving universe which is currently expanding, originated by a primordial explosion-like event.
\end{itemize}
The last of which was ridiculed by the pre-eminent cosmologist Fred Hoyle, who coined the then derogatory phrase ``Big Bang theory" - in response to the notion of matter and energy created ex-nihilo at the onset of our universe, to give rise to rapid expansion.\\\\
This last ``Big Bang theory", became the lead contender in 1965 when Arno Penzias and Robert Wilson, of Bell labs, stumbled upon the Cosmic Microwave Background Radiation (CMBR or CMB) \cite{Penzias_Wilson1965}. Dicke, Peebles and Wilkinson, themselves working on a microwave band antenna at the time, were able to provide the theoretical framework that would explain this as a remnant of the ``Big Bang" \cite{DickeWilkinsonPeebles1965}. This, together with different mounting evidence of light-element abundance in the universe was sufficient confirmation of the ``Big Bang" as the most probable physical scenario. 
\section[Problems in heaven]{Problems with the old big bang model}
Several measurements of the CMB temperature spectrum were made starting with Penzias and Wilson \cite{Penzias_Wilson1965}, through Thaddeus \cite{Thaddeus:1972an} and others, but the first overwhelming evidence of the black-body spectrum of the CMB was found by the Far-Infrared Absolute Spectrophotometer (FIRAS) instrument on the Cosmic Background Explorer (COBE) satellite \cite{Mather:1998gm}. From these measurements, with systematic dominated error margins of no more than 0.3\% at peak brightness, and an rms value of 0.01\%. This extraordinary precision in measuring the black-body nature of the CMB put the "Big Bang theory" on as sound a footing as anyone could hope for. However, as with all scientific discoveries, one answer gives rise to a multitude of other problems.
\subsection[Horizon]{The horizon problem}
\begin{figure}[!h]
\begin{center}
	\includegraphics[width=0.9\textwidth]{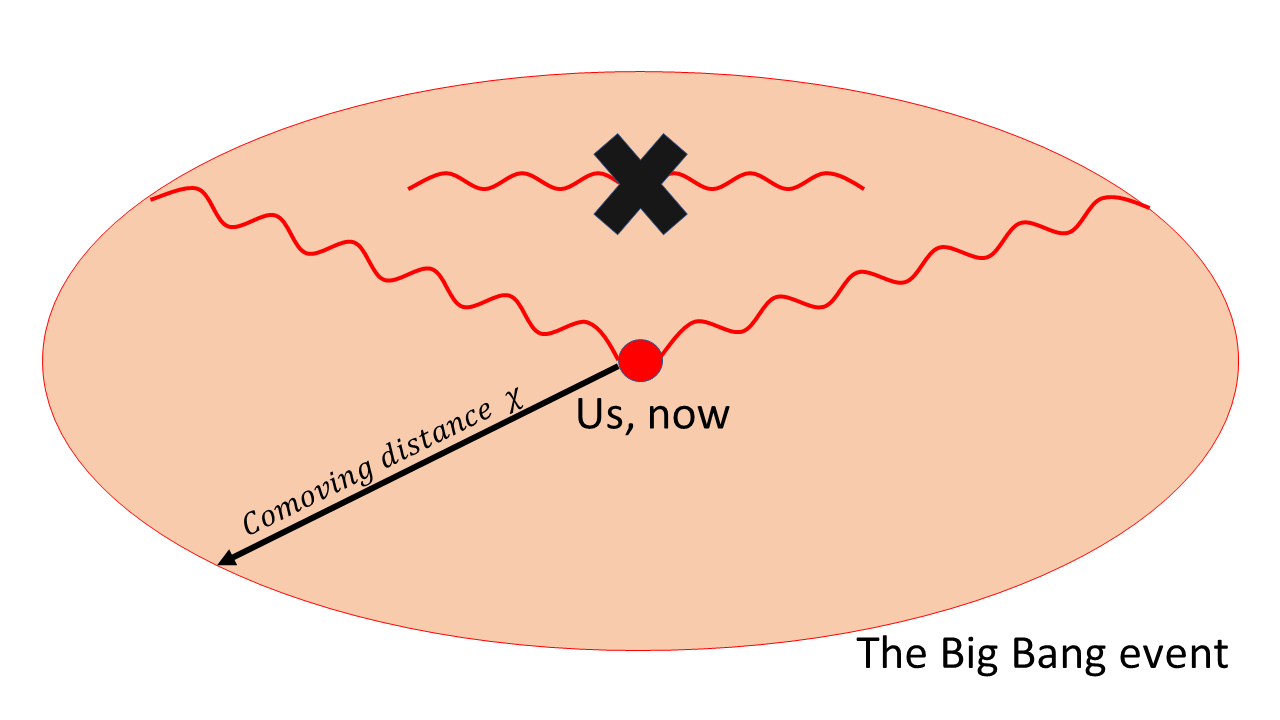}
	\caption[{\bf The horizon problem}]{ Light emitted from the "Big Bang" event is just now reaching our eyes, after travelling a distance equal to the comoving distance $\chi(t)=c\int_{t_{BB}}^{t_{now}}\frac{dt'}{a(t')}$. There are regions in the CMB surface that are thus causally disconnected but are nonetheless homogeneous to one part in $10^5$. This is suggestive of information exchange, where no exchange is possible.} \label{fig:Horizon_problem}
\end{center}
\end{figure}

The horizon problem can roughly be put as the following: if we are now at the light horizon of the ancient "Big Bang" event, the information that is propagated to us from one end of the CMB, cannot propagate to the other end of the CMB. Thus these two patches of CMB cannot possibly have exchanged information during the expansion period of the universe (See Fig. \ref{fig:Horizon_problem}). However, they are thermalized and homogeneous to one part in $10^{5}$. If they can't exchange information how is that possible?
Formalising this requires to consider the Friedmann-Robertson-Walker (FRW) metric:
\begin{align}
	&ds^2=-c^2dt^2 +a(t)^2 d{\bf x}^2.
\end{align} 
When one considers the distance travelled by light, one sets the interval to zero, thus for light travelling along the $x$ axis yielding:
\begin{align}
	&x=\int dx = \int c\frac{dt'}{a(t')}.
\end{align}
So the comoving distance is defined as:
\begin{align}
	&\chi(t)=\int_{t_0}^{t_{now}}  c\frac{dt'}{a(t')} \Rightarrow \chi(a)=\int_a^1 \frac{da'}{a'^2 H(a')}, \label{eq:comoving_distance}
\end{align}
where the speed of light $c$ is set to $1$, $a(t)$ is the scale factor of the FRW metric, which is set such that the current scale is $a(t_{now})=1$ and $H$ is the Hubble parameter, which varies for different historical stages of the universe. One of the useful quantities to define is \textit{conformal time}, in which we rewrite the FRW metric as follows:
\begin{align}
	&ds^2=dt^2-a^2d{\bf x}^2 \equiv a^2 (d\tau^2 -d{\bf x}^2 ),
\end{align}
so $\int cd\tau$ is the comoving distance.  
So, roughly speaking, unless we are in a closed universe which is enclosed within a fully connected sphere of radius no more than the observed $\chi$, there are patches in the CMB that are causally disconnected. Since these patches are unable to exchange information since the Big Bang event, how can they be in thermal equilibrium with each other? Yet we observe them to be in thermal equilibrium up to one part in $10^5$.
\subsection[Flatness]{The flatness problem}
Suppose we want to measure distances between different CMB patches. Since we are limited to a two-dimensional spherical view of the CMB, it is intuitive to measure distance by the connection between angle and arc length:
\begin{align}
	&l=d_A \theta,
\end{align}
where $l$ is the arc length, $d_A$ is called the angular diameter distance, and $\theta$ is the angle subtended. In a completely flat universe Eq. \eqref{eq:comoving_distance} is connected to $d_A$ by:
\begin{align}
	&d_A^{flat}=a\chi .
\end{align}
However, if we live in an open or closed universe in which the curvature is non-zero, the proper generalization is given by:
\begin{align}
	&d_A= \left\{\begin{array}{lcr}
	\frac{a}{H_0}\sqrt{|\Omega_k|}\sinh\left(\sqrt{\Omega_k}H_0 \chi\right)& & \Omega_K>0\\
	\frac{a}{H_0}\sqrt{|\Omega_k|}\sin\left(\sqrt{-\Omega_k}H_0 \chi\right)& &\Omega_K<0
	\end{array}\right. ,
\end{align}
where $\Omega_k$ is the curvature density, and $H_0$ is the current Hubble parameter.
While the energy density in radiation scales as $a^{-4}$, and matter density scales as $a^{-3}$, the energy density in curvature scales as $a^{-2}$. This means that if we can construct a scale ladder spanning several epochs, in which $a$ sufficiently changes, we can assess the energy density in curvature. Such a ladder is provided to us by Type Ia Supernovae \cite{Riess:1994nx,Dunkley:2008ie}, and so we are able to evaluate the luminosity distance $d_L=\frac{\chi}{a}$ to an object and compare it to the angular distance $d_A$. By observing a large sample of these pairs we can extract the overall behaviour of $a$ at different times, and which of the three angular distance functions are the best fit. Using this method the most recent Planck data sets the curvature density $\Omega_k$ at
\begin{align}
	&\Omega_k=0.001\pm 0.002.
\end{align}  
This value is very close to zero, which means a flat universe. Of all possible values the curvature density can take, which can be of the order $\pm 1$, corresponding to a universe dominated by the curvature component, why is it that we find our universe so close to a flat one? This is not the most finely-tuned quantity in our universe \cite{Carroll:2014uoa}, but it is fine-tuned nonetheless. 
\subsection[Relic]{The relic problem}
Every Grand-Unified-Theory (GUT) that includes electromagnetism inevitably produces super-heavy magnetic monopoles \cite{Zeldovich:1978wj,PhysRevLett.43.1365,PhysRevLett.44.631,mukhanov2005physical}, typically of magnetic charge $\propto \frac{1}{e}$, where $e$ is the fundamental electric charge and is of order $e\simeq 10^{-19}\; C$. Such a large magnetic charge would have observable effects on the universe and would be easily detectable. However so far we have not found any such effect and deduce the lack of any monopoles in the observed universe. During the GUT phase of the universe, the energy density in magnetic monopoles should have been such that today we should have measured $\Omega_{\mathrm{Mono}}\sim 10^{13}$. How is it possible that we detect no monopoles then? 
\section[Inflation]{Inflation saves the day}
It was Starobinsky \cite{Starobinsky:1979ty} closely followed by Guth \cite{GuthInflation}, that suggested the idea of inflation, an epoch of rapid and accelerated expansion of the universe. This mechanism at once fixes the above problems. Let us consider:
\begin{align}
	&c\tau=\int\frac{da}{a}\frac{c}{aH},
\end{align}
which is the logarithmic integral of the comoving Hubble radius. When the distance between two objects is larger than the comoving Hubble radius, they cannot currently be in causal connection with each other. If particles are however separated from each other by more than $c\tau$ they could have never exchanged information. If we can find a way in which two particles are currently separated more than the comoving Hubble radius, but separated by less than $c\tau$, we can solve the horizon problem. In other words, we want today:
\begin{align}
	&\frac{1}{aH(a)}<1,
\end{align}
and a scale $a'$ at some time in the evolution of the universe such that 
\begin{align}
   &\frac{1}{a'H(a')}>1.
\end{align}
Thus there is a phase in cosmic history where the term $aH$ is increasing:
\begin{align}
	&\frac{d(aH)}{dt}=\frac{d^2 a}{dt^2}>0,
\end{align} 
which means the scale factor of the universe was growing in an accelerated fashion. However, this could not have occurred during matter or radiation dominated eras, so there is something \textit{else} that gives rise to this phenomenon.
This solution to the horizon problem naturally solves the flatness problem as well as the relic problem. Rarefying the energy density stored in curvature, by accelerated growth of the scale of the universe, the currently observed curvature strongly tends towards zero. This is similar to the process of deriving a smooth function. By sufficiently 'zooming-in' ($dx\rightarrow 0$), the curvature vanishes, and we are left with the linear term alone. Similarly, a universe initially populated by relics could have been sufficiently diluted such that we do not observe any relics today. 
\begin{figure}[!h]
\begin{center}
	\includegraphics[width=0.9\textwidth]{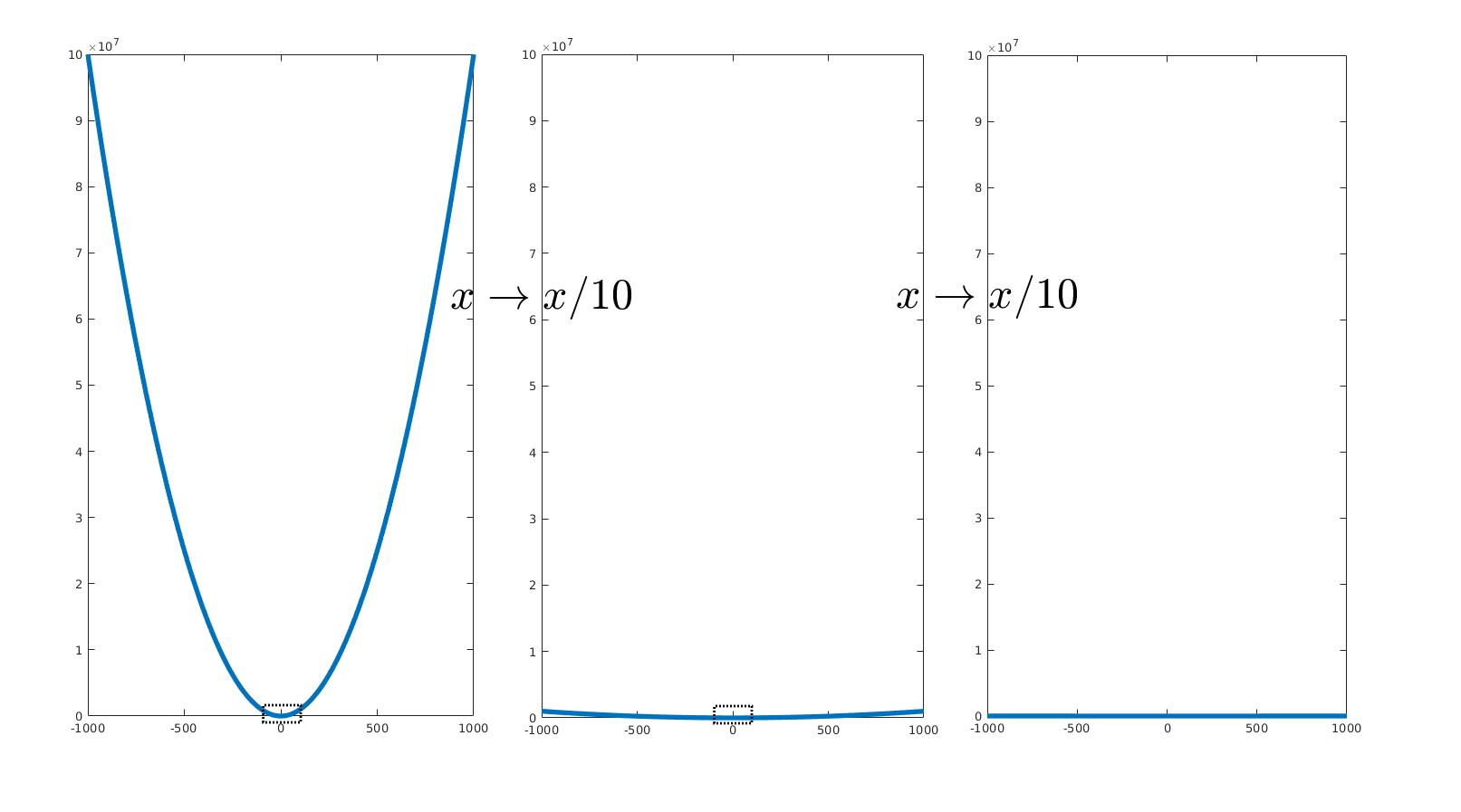}
	\caption[{\bf Solving flatness }]{By continually rescaling the universe, distances which might have been connected by a curved interval are now connected by a less curved interval, thus diluting the energy density in curvature. This is analogous to 'zooming-in' in order to examine the linear approximation of a function at a point.} \label{fig:Flatness_solution}
\end{center}
\end{figure}

\subsection{Features of inflation}
There are additional consequences for a period of accelerated growth of the scale factor, beyond the observations suggested above. Perhaps one of the most looked at today is the notion of primordial gravitational waves. In a later section we will quantitatively look at the production of gravitational waves from inflation. However, at this point we want to look at the general idea. Gravitational waves are generated whenever the metric at some point in space undergoes acceleration and is not completely uniform. To see that clearly we consider small perturbations over the flat Minkowski metric:
\begin{align}
	&g_{\mu\nu}=\eta_{\mu\nu}+h_{\mu\nu},
\end{align}
We then define the trace-reversed perturbation:
\begin{align}
	&\tilde{h}_{\mu\nu}=h_{\mu\nu}-h\eta_{\mu\nu}.
\end{align}
Moving to the transverse-traceless gauge we get:
\begin{align}
	&\tilde{h}^{TT}_{\mu\nu}=h^{TT}_{\mu\nu}.
\end{align}
By then using the Lorentz gauge we set:
\begin{align}
	&\partial_{\mu}\tilde{h}^{\mu\nu}=0
\end{align}
We thus get the following a linearized version of the Einstein equation, which is just a non-homogeneous wave equation:
\begin{align}
	&\square \tilde{h}_{\mu\nu}=-16\pi G T_{\mu\nu},
\end{align}
where $T_{\mu\nu}$ is the stress-energy tensor.
Solving this equation is possible by finding the Green's function for the wave equation, and incorporating a retarded quantity $t_r=t-|{\bf x-y}|$. Finally, the term for gravitational wave production is proportional to the second temporal derivative of the Quadrupole moment of the energy density:
\begin{align}
	&I_{ij}(t)=\int y^i y^j T^{00}(t,{\bf y})d^3y\; ; \hspace{50pt} \tilde{h}_{ij}(t,{\bf x})=\frac{2G}{r}\frac{d^2 I_{ij}}{dt^2}(t_r)
\end{align}
This is not the main focus of this manuscript, however we clearly see that when we have an accelerating quadrupole moment, like in inflation, we have gravitational wave production. The physics that source primordial GW is slight spatial perturbation in curvature, sourced by slight quantum perturbation of the inflaton field. The common notion is that these perturbations are small such that the overall scale factor is given by $a(t)$ (no spatial dependence). The way to derive these equations is to consider the FRW metric with slight spatial perturbation such that:
\begin{align}
	g_{00}=-1\; ; \; g_{ij}=a^2(t)\left(\begin{array}{ccc}
	1+h_{+} &h_{\times}& 0\\
	h_{\times} & 1+h_{+} & 0\\
	0 & 0& 1
	\end{array}\right) .
\end{align}
Going on calculate the Christoffel symbols, Ricci tensor and scalar and throwing out second amd higher order terms, we get a linearized EFE. The zero-order contribution are the leading order Friedmann equations.  Replacing the spatial derivative terms with $k^2$ by going into spatial Fourier-space and after some algebra we arrive at:
\begin{align}
	\partial_{tt} h_{(+,\times)} + 3H \partial_t h_{(+,\times)} +\frac{k^2 h_{(+,\times)}}{a^2(t)}=0.
\end{align} 
Rephrased in conformal time, this takes on the follwoing form:
\begin{align}
	h''_{\zeta}+2\frac{a'}{a}h'_{\zeta} +k^2 h_{\zeta}=0, \label{eq:GW_prod}
\end{align}
where $\zeta\in (+,\times)$ and the prime repesents $\partial_{\tau}$. It is beneficial to recast this equation into a harmonic oscillator form. To do this we redefine:
\begin{align}
	\tilde{h}=\frac{ah}{\sqrt{16\pi G}},
\end{align}
where the numerical coefficient comes from physical Action considerations. Disregarding this numerical coffeicient we have:
\begin{align}
	&h=\frac{\tilde{h}}{a},\\
	\nonumber \\
	&h'=\frac{\tilde{h}'}{a}-\frac{a'}{a^2}\tilde{h},\\
	\nonumber \\
	&h''=\frac{\tilde{h}''}{a}-2\frac{a'}{a^2}\tilde{h}' -\frac{a''}{a^2}\tilde{h}+2\frac{(a')^2}{a^3}\tilde{h}\;.
\end{align}
Inserting these into Eq.~\eqref{eq:GW_prod}, yields:
\begin{align}
	\tilde{h}'' +\left(k^2-\frac{a''}{a}\right)\tilde{h}=0 .
\end{align}
This equation is similar to the Mukhanov-Sasaki (MS) equation, to be discussed later, that governs the production of scalar perturbation. However the pump field in this equation is the scale factor $a$, whereas the pump field in the MS equation is $z=a\dot{\phi}/H$.
\section{Solving inflation}
Let us first take a look at the simplest form of inflation: A constant energy density term over an FRW background metric. This is also known as the de-Sitter case. In this case the metric is given by $ds^2=dt^2 -a^2(t)d\bf{x}^2$ where the metric itself is time dependent. Since the system is explicitly time dependent, using a Hamiltonian formalism is disfavoured. Thus we use the Einstein Field Equations:
\begin{align}
	&G_{\mu\nu}=8\pi G T_{\mu\nu},
\end{align}
along with the Lagrangian formulation for the equations of motion of the associated fields.
When we 'solve' inflation we intend to say, we calculate the evolution of all inflationary quantities, which in this case are $t,H(t),\phi(t)$ from which we construct all associated quantities: $a(t),\dot{a}(t),N(t),\dot{\phi}(t)$, as well as the slow-roll parameters as defined ($\varepsilon_H=\frac{-\dot{H}}{H^2},\delta_H=\frac{\ddot{\phi}}{H\dot{\phi}}$, etc.) versus as potential and derivative terms ($\epsilon_V=\frac{1}{2}\left(\frac{V'}{V}\right)^2,\delta_V=\frac{V''}{V}$ etc.) We also construct the pump field $z$ that is later used for the quantum perturbations calculation:
\begin{align}
	&z=\frac{a\dot{\phi}}{H}. \label{eq:Pump_field}
\end{align}
\subsection[Friedmann equations]{General Relativity, Friedmann equations in conformal and cosmic time}
General Relativity, has shifted our understanding of gravitation from that of a force, to that of a characteristic of space-time, where the force of gravitation is nothing but Newton's first law as applied in curvilinear coordinates\cite[Chapter~1]{carroll2003spacetime}.\\\\
This is done by coupling a metric (i.e. a manifold that describes space-time) to the total energy present.\\\\
This is encoded mathematically by using (psuedo) Riemannian geometry, most commonly using the Einstein notation:\\
\begin{equation}\label{Einstein Field equations}
	R_{\mu\nu}-\frac{1}{2}g_{\mu\nu}\mathbf{R}=8\pi G T_{\mu\nu},
\end{equation}
where $R^{\mu\nu}$ is the Ricci Tensor (2nd order Tensor form), and $\mathbf{R}$ is the Ricci scalar which is given by:
\begin{equation}
	\mathbf{R}=g^{\mu\nu}R_{\mu\nu}.
\end{equation}
These might sometimes be simplified by using the Einstein tensor:
\begin{eqnarray}
G_{\mu\nu}\equiv R_{\mu\nu}-\frac{1}{2}g_{\mu\nu}\mathbf{R},
\end{eqnarray}
thus we arrive at the form:
\begin{eqnarray}
	G_{\mu\nu}=8\pi G T_{\mu\nu}.
\end{eqnarray}
These applied to the universe as a whole yield the Friedmann equations.
\subsubsection{FRW metric and Friedmann equations - cosmic time}
The Friedmann-Robertson-Walker metric (FRW metric) is given by allowing the spatial coordinates to change scale, as a function of time. In Cartesian coordinates, this takes on the form of:
\begin{eqnarray}
	g_{\mu\nu}=\left(\begin{array}{cccc}
	-1 &&&\\
	&a^2(t)&&\\
	&&a^2(t)&\\
	&&&a^2(t)
	\end{array}\right).
\end{eqnarray}
After going through the process of deriving the Ricci Tensor and Scalar, this metric gives two equations. Taking the temporal-temporal coordinate ($\mu\nu=00$), and any of the spatial-spatial coordinate ($\mu\nu=ii$) we have:
\begin{align}
	&\left(\frac{\dot{a}}{a}\right)^{2}=\frac{8\pi G}{3}\rho,\\
	&-2\left(\frac{\ddot{a}}{a}\right)-\left(\frac{\dot{a}}{a}\right)^{2}=8\pi G P,
\end{align}
where we use the diagonalized Stress-Energy tensor:
\begin{eqnarray}
	T_{\mu\nu}=g_{\mu\alpha}T^{\alpha}\,_{\nu}=\left(\begin{array}{cccc}
	\rho &&&\\
	&a^{2}P&&\\
	&&a^{2}P&\\
	&&&a^{2}P
	\end{array}\right),
\end{eqnarray}
with
\begin{eqnarray}
	T^{\mu}\;_{\nu}=\left(\begin{array}{cccc}
	-\rho &&&\\
	&P&&\\
	&&P&\\
	&&&P
	\end{array}\right).\label{Diagnonalized Stress-Energy tensor}
\end{eqnarray}
This is where one usually points out the possibility of an accelerated expansion/contraction solution to the Friedmann equation. This is given by an energy density term, which is constant, i.e.:
\begin{flalign}\label{exponential expansion}
	\left(\frac{\dot{a}}{a}\right)^{2}=\frac{8\pi G}{3}\rho\Rightarrow \frac{\dot{a}}{a}=\sqrt{\frac{8\pi G \rho}{3}}=Const,\\
	\nonumber \Rightarrow a(t)=\left(a_0 e^{\pm\sqrt{\frac{8\pi G \rho}{3}}t}\right).
\end{flalign}
Ultimately we are dealing with the positive exponential solution since this fits the observations, and solves the Horizon and Flatness problems as previously explained.\\\\
However a constant $\rho$ means permanent inflation, so $\rho$ has to change in time, albeit slowly enough to facilitate exponential inflation. This is called the slow-roll condition. It will be discussed at length in section \ref{sec:slowRollParadigm}.

\subsubsection{FRW metric and Friedmann equations - conformal time}\label{Conformal_time}
Moving to conformal time, puts the time coordinate $\tau$ on equal footing with the other spatial coordinates by defining:
\begin{eqnarray}
	dt=ad\tau .
\end{eqnarray}
Thus the interval, or line element, takes on the form of:
\begin{eqnarray}
	dS^2=-dt^2+a^2(t)\left(dx^2 +dy^2 + dz^2\right)=a^2(\tau)\left(-d\tau^2+dx^2 +dy^2 + dz^2\right),
\end{eqnarray}
so, in essence we are reducing the problem to Minkowski space over a time dependant overall scale factor. \\
In this context, the metric is given by:
\begin{eqnarray}
	g_{\mu\nu}=\left(\begin{array}{cccc}
	-a^2(\tau) &&&\\
	&a^2(\tau)&&\\
	&&a^2(\tau)&\\
	&&&a^2(\tau)
	\end{array}\right)=a^2(\tau)\left(\begin{array}{cccc}
	-1 &&&\\
	&1&&\\
	&&1&\\
	&&&1
	\end{array}\right),
\end{eqnarray}
and the derived Friedmann equations are given by:
\begin{eqnarray}
	\left(\frac{a^{\prime}}{a}\right)^2=\frac{8\pi G \rho}{3}a^2\\
	\frac{\left(- 2 aa^{\prime\prime} + {a^{\prime}}^{2}\right)}{a^{2}}  = 8 G P \pi a^{2}
\end{eqnarray}
It is interesting to see that a constant $\rho$ in conformal time produces the following solution:
\begin{flalign}
	a(\tau)=\frac{1}{1-\sqrt{\frac{8\pi G\rho}{3}}\tau}
\end{flalign}
which explodes to infinity at $\tau=\sqrt{\frac{3}{8\pi G\rho}}$, thus $a(\tau)$ experiences greater than exponential growth in conformal time. While analytically this stands on equal footing with a cosmic time solution, numerical integration considerations will later lead us to prefer cosmic over conformal time.
\subsection[Scalar field inflation]{The scalar field inflation}
Inflation is driven by an energy density term that permeates all of space and is slowly varying, even taking into account the rapid expansion of space.\\\\
There might be any number of physical drivers that manifest this behavior, but arguably the simplest form of mechanism which gives rise to inflation is the scalar field one. It has been shown by Albrecht and Steinhardt, that a Higgs-like field, can produce such a scalar effective field \cite{Albrecht:1982wi}.
The simple Lagrangian (density) form of a scalar field is given by:
\begin{equation}
	\mathcal{L}=\frac{1}{2}\left(\partial_{\mu}\phi\right)^2 -\frac{m^2}{2}\phi^2+ \mathcal{O}\left(\phi^{3}\right)
\end{equation} 
Where the linear term in the potential can be set to zero because we can always redefine $\tilde{\phi}$ such that the derivatives are unchanged and the quadratic term is changed accordingly. It is a similar procedure acting on the interplay between the second and fourth powers of the Lagrangian, that constitutes the Goldstone gauge Boson procedure \cite{NamboGoldstonBoson:1960,Goldston:1962}, and ultimately the Higgs Mechanism \cite{EnglertBrout:1964,Higgs:1964,theRestOfHiggs:1964}.\\\\
An almost straightforward generalization, using Legendre transform, and going to covariant form yields \cite[Page~152]{dodelson:2003}:
\begin{flalign}
	T^{\alpha}\,_{\beta}=g^{\alpha\nu}\frac{\partial\phi}{\partial x^{\mu}}\frac{\partial\phi}{\partial x^{\beta}}-g^{\alpha}\,_{\beta}\left[\frac{1}{2}g^{\mu\nu}\frac{\partial\phi}{\partial x^{\mu}}\frac{\partial\phi}{\partial x^{\nu}}+V\left(\phi\right)\right],
\end{flalign} 
where $V(\phi)$ is the interaction potential.
Considering this field to be mostly homogeneous, and disregarding the small perturbations to this quantity, the spatial derivatives vanish, and we are left with:
\begin{flalign}
	T^{(0)\alpha}\,_{\beta}=-g^{\alpha}\,_{0}g^{\beta}\,_{0}\left(\frac{d\phi^{0}}{dt}\right)^{2}+g^{\alpha}\,_{\beta}\left[\frac{1}{2}\left(\frac{d\phi^{0}}{dt}\right)^{2}-V\left(\phi^{0}\right)\right].
\end{flalign} 
Identifying this term with  Eq. (\ref{Diagnonalized Stress-Energy tensor}), we may now denote:
\begin{align}
	\rho=\frac{1}{2}\left(\frac{d\phi^{0}}{dt}\right)^{2}
	&+V\left(\phi^{0}\right),\\
	&P=\frac{1}{2}\left(\frac{d\phi^{0}}{dt}\right)^{2}-V\left(\phi^{0}\right),
\end{align}
which is in cosmic time FRW. The conformal time version is given in Eq. (\ref{rho in conformal},\ref{P in conformal})\\
from here on we will forego the superscript (0) for simplicity
\subsection{Equations of motion}
When one is tasked with simulating inflation, whether numerically or analytically, the first step is to evaluate the equations of motion for the scale factor ,$a$ , and the driving fields, in our case the inflaton $\phi$. This is done by following these steps:
\begin{itemize}
	\item Derive Friedmann equations from metric assumptions (FRW/FRW+conformal).
	\item Replace the quantities $\rho$ and $P$ with the driving fields equivalents.
	\item Identify the set of differential equations to work with.
\end{itemize}
We now proceed to do so, for a conformal background geometry, driven by a scalar field.\\\\
The metric is:
\begin{eqnarray}
	g_{\mu\nu}=\left(\begin{array}{cccc}
	-a(\tau)^2& & & \\
	&a(\tau)^2 & & \\
	& &a(\tau)^2 & \\
	& & & a(\tau)^2
	\end{array}\right)
\end{eqnarray}
and the Stress-Energy tensor for a scalar field is given by:
\begin{eqnarray}\label{Stress-Energy tensor for scalar field}
	T^{\alpha}\,_{\beta}=g^{\alpha\nu}\frac{\partial \phi}{\partial x^{\nu}}\frac{\partial \phi}{\partial x^{\beta}}-\delta^{\alpha}\;_{\beta}\left[\frac{1}{2}g^{\mu\nu}\frac{\partial \phi}{\partial x^{\mu}}\frac{\partial \phi}{\partial x^{\nu}}+V\left(\phi\right)\right].
\end{eqnarray}
These imply - in conformal time:
\begin{align}	
&T^{(0)0}\;_{0}\equiv\rho=\frac{1}{2a^2}\left(\frac{d\phi^{0}}{d\tau}\right)^2+V(\phi^{(0)}),\label{rho in conformal}\\
&T^{(0)i}\;_{i}\equiv P=\frac{1}{2a^2}\left(\frac{d\phi^{0}}{d\tau}\right)^2-V(\phi^{(0)}).\label{P in conformal}
\end{align}
With the two equations which we get from the Einstein field equations (EFE):
\begin{eqnarray}
	\left(\frac{1}{a}\frac{da}{d\tau}\right)^2=\frac{8\pi G \rho}{3}a^2,\\
	\frac{\left(- 2 a\cdot \frac{d^{2}a}{d\tau^{2}} + (\frac{da}{d\tau})^{2}\right)}{a^{2}}  = 8 G P \pi a^{2},
\end{eqnarray}
one can work through the algebra, to get:
\begin{eqnarray}\label{equation of motion for phi in conformal time}
	\left\{\begin{array}{l}
	\frac{d^2\phi}{d\tau^2}+2\left(\frac{1}{a}\frac{da}{d\tau}\right)\frac{d\phi}{d\tau}+a^2V'=0\\
	\\
	\frac{d^2 a}{d\tau^2}=\frac{4\pi G a}{3}\left[4a^2V-\left(\frac{d\phi}{d\tau}\right)^2\right]	\\
	\\
	\frac{d a}{d\tau}=\sqrt{\frac{8\pi G}{3}\left[\frac{a^2}{2}\left(\frac{d\phi}{d\tau}\right)^2+a^4V(\phi)\right]} .
	\end{array}\right. \label{Integration equations}
\end{eqnarray}
Since the first two are second order differential equations, and since $a$ is not observable, it might be useful to integrate over $A$ as defined by $A\equiv\frac{d\ln(a)}{d\tau}$. For this aim the second equation is given by:
\begin{eqnarray}\label{second friedman in H in conformal time}
	\frac{dA}{d\tau}=A^{2}\left(\frac{V-\left(\frac{d\phi}{d\tau}\right)^2}{V+\frac{1}{2}\left(\frac{d\phi}{d\tau}\right)^2}\right).
\end{eqnarray}
To conclude, in conformal time, a "good" integration scheme might use this set of equations:
\begin{align}
	\left\{\begin{array}{ccc}
	\phi''&=&-2A\phi'-\frac{dV}{d\phi}\\
	a'&=&A\cdot a\\
	A'&=&A^{2}\left(\frac{V-\phi'^{2}}{V+\tfrac{1}{2}\phi'^{2}}\right)
	\end{array}\right.
\end{align}
\\\\
Deriving the analogue in cosmic time gives:
\begin{large}
\begin{eqnarray}\label{Triad}
	\left\{\begin{array}{l}
	\ddot{\phi}+3H\dot{\phi}+V'=0\\
	\\
	\dot{a}=\sqrt{\frac{\left(\dot{\phi}^2 +2V\right)}{6M_{pl}^{2}}}a\\
	\\
	\ddot{a}=-\frac{a}{3M_{pl}^{2}}\left[\dot{\phi}^2-V\right]
	\end{array}\right. \; ,
\end{eqnarray}
\end{large}
with an equation for $\dot{H}$ given by:
\begin{eqnarray}
	\dot{H}=-\frac{\dot{\phi}^{2}}{2 M^{2}_{pl}}\; ,
\end{eqnarray}
where the quantity $\sqrt{\frac{1}{8\pi G}}$ is neatly wrapped in the term $\sqrt{\frac{1}{8\pi G}}=m_{pl}$, the reduced Planck mass.\\
Thus we are left to solve this set of coupled ordinary differential equations (ODE), to resolve the evolution of the background geometry:
\begin{large}
	\begin{align}
		&\left\{\begin{array}{l}
		\ddot{\phi}=-3H\dot{\phi}-V'\\
		\dot{H}=-\frac{\dot{\phi}^{2}}{2m^{2}_{pl}}
		\end{array}\right.
	\end{align}
\end{large}
\section[Slow-roll]{The slow-roll paradigm}\label{sec:slowRollParadigm}
As was alluded to before when we have a constant energy density term $\rho$ the solution for the temporal-temporal Friedmann equation over an FRW metric is given by:
\begin{align}
	&H^2=\frac{8\pi G}{3}\rho,\\
\end{align}
Since $\rho$ is just some constant we can now unpack $H$ and solve for $a(t)$:
\begin{align}
&H=\frac{\dot{a}}{a}=\sqrt{\frac{8\pi G}{3}\rho}\\
&\Rightarrow a=\sum_{j=1}^{2}A_j\exp\left((-1)^j\sqrt{\frac{8\pi G}{3}\rho}t\right).
\end{align}
It is customary to disregard the exponentially suppressed solution since they would be undetected in 2-3 efolds of inflation. Thus we arrive at the pure de-Sitter (dS) solution:
\begin{align}
	&a(t)=a_0 \exp\left(\sqrt{\frac{8\pi G}{3}\rho} \cdot t\right).
\end{align}
This gives rise to the idea of inflation as a phase of an exponent-like evolution of the universe, with an energy density that changes slow enough, to facilitate a perturbative solution, over the baseline of a pure de-Sitter evolution. 
\subsection{Quantifying slow-roll}
One way to quantify slow-roll would be to assume the basic solution for $a(t)$ is an exponent function, and then use perturbations over that baseline such that $a_0=\exp(C\cdot t)$, $a_1=\lambda f(t)$, and go on in a perturbative expansion. However, the usual formalism that is used to quantify slow-roll is based on the idea that the Hubble parameter $H$ should be very slowly changing with respect to cosmic time. This means that the perturbative degrees are applied to $H$ rather than $a$. It ,therefore, makes sense to write $H$ as some power series in $t$.
\subsubsection{Taylor expansion}
As it turns out since $H$ has units of $1/T$, i.e. $H$ is the reciprocal of cosmic time, it makes sense to develop the function $1/H$ in a Taylor series around $H_0$, where $H_0$ is a constant. Thus we have:
\begin{align}
&\frac{1}{H}\simeq \frac{1}{H_0} -\left.\frac{\dot{H}}{H^2}\right|_{H=H_0}t  -\left.\frac{\ddot{H}H^2 -2H\dot{H^2}}{H^4} \right|_{H=H_0}\frac{t^2}{2} +\hdots
\end{align}
The first slow-roll parameter can be read directly from the above expansion:
\begin{align}
	&\varepsilon_H\equiv -\frac{\dot{H}}{H^2}, \label{epsilon_H}
\end{align}
and it is usually, within the framework of slow-roll inflation, small and positive since it is understood that, naturally, the average energy density $\rho$ should decline as the universe inflates. This leads $H$ to slightly decrease, thus slightly increasing $1/H$. We are  left, then, with a positive first-order contribution to the Hubble horizon. Formally slow-roll takes place when
\begin{align}
	&0<\varepsilon_H \ll 1.
\end{align}
The second term in the Taylor expansion needs a bit of handling to find a more aesthetic term.
\subsubsection[$\varepsilon_H$ and $\delta_H$]{The slow-roll parameters $\varepsilon_H$ and $\delta_H$}
While the first slow-roll parameter $\varepsilon_H$ is a small, non-negative quantity, given simply by \eqref{epsilon_H}, in order to simplify the second slow-roll parameter we need to refer to the Klein-Gordon equation for a time-dependent scalar field (\ref{Triad}, top equation). Reason states that where the field $\phi$ is slowly rolling, the drag term that is given by $3H\dot{\phi}$, roughly counteracts the force term $\frac{dV}{d\phi}$, leading to a terminal velocity-like state. In that scenario the fields 'acceleration' is roughly zero, so we get:
\begin{align}
	&\ddot{\phi}=-3H\dot{\phi}-\frac{dV}{d\phi}\simeq 0,
\end{align}
differently stated as
\begin{align}
		&|3H\dot{\phi}|\simeq \left|\frac{dV}{d\phi}\right|.
\end{align}
It makes sense, then, to look at the dimensionless quantity 
\begin{align}
	&\delta_H\equiv \frac{\ddot{\phi}}{H\dot{\phi}},
\end{align}
and set an additional slow-roll condition as 
\begin{align}
	&\left|\delta_H\right|\ll 1.
\end{align}
The connection with the Taylor expansion is not immediately obvious, so we will explicitly derive it. The second order term in the Taylor expansion for $1/H$ is given by:
\begin{align}
	&\left.\frac{\ddot{H}H^2 -2H\dot{H^2}}{H^4} \right|_{H=H_0}\frac{t^2}{2}.\label{2ndTaylor_term}
\end{align}
Let us examine this term:
\begin{align}
	&\frac{\ddot{H}}{H^2} = \frac{1}{H^2}\frac{d}{dt}\left(-\frac{\dot{\phi}^2}{2}\right)\\ \nonumber
	&= \frac{-1}{H^2}\dot{\phi}\ddot{\phi}=-\frac{\dot{\phi}^2}{H}\frac{\ddot{\phi}}{H\dot{\phi}}\\ \nonumber
	&= 2H\frac{\dot{H}}{H^2}\delta_H=-2H\varepsilon_H\delta_H.
\end{align}
Thus the aforementioned term \eqref{2ndTaylor_term} is given by:
\begin{align}
	&\left.\frac{\ddot{H}H^2 -2H\dot{H^2}}{H^4} \right|_{H=H_0}\frac{t^2}{2}=-H\left.\varepsilon_H\left(\delta_H+\varepsilon_H\right)\right|_{H=H_0}t^2,
\end{align}
and the Taylor expansion up to second order is given by:
\begin{align}
	&\frac{1}{H_0}+\varepsilon_H t +\varepsilon_H\left(\delta_H+\varepsilon_H\right) H_0 t^2 ,
\end{align}
where the expansion is around $H=H_0$.
\subsubsection[$\epsilon_V$ and $\eta_V$]{Switching gears into $\epsilon_V$ and $\eta_V$}\label{sec:SR_potentials}
It is beneficial for theoretical physicists to be able to connect the slow-roll parameters directly to the inflationary potential. This is done because as we shall see the primordial power spectrum can be quantified up to several percent precision, directly by the slow-roll parameters. The first slow-roll parameter in potential and potential derivative language is usually given by
\begin{align}
	&\varepsilon_V\equiv \frac{1}{2}\left(\frac{V'}{V}\right)^2.
\end{align}
The two definitions, $\varepsilon_H$ and $\varepsilon_V$ coincide in both extreme limits, i.e. when slow-roll is extremely slow and $\ddot{\phi}$ strictly vanishes and $\dot{a}\propto\ddot{a}$, and at the other end of inflation, where $H$ is no longer slowly varying, and the accelerated evolution of the scale factor vanishes ($\ddot{a}=0$).
The derivation of the slow-roll parameter is as follows:
In slow-roll inflation, the potential overpowers the kinetic energy immensely thus:
\begin{align*}
	&V \gg \dot{\phi}^{2}
\end{align*}
This simplifies greatly the two major players of inflation, the first of which is:
\begin{align*}
	&\begin{array}{lcr}
	H^{2}=\frac{\rho}{3}&;(\rho=\frac{\phi^{2}}{2}+V);\Rightarrow& H^{2}\simeq \frac{V}{3}\; .\\
	\end{array}
\end{align*}
The second equation, the equation of motion for the scalar field is usually given by:
\begin{align*}
	&\ddot{\phi}+3H\dot{\phi}+V'=0,
\end{align*}
which is the result of derivation with respect to time of the more primitive form of:
\begin{align*}
	&\left(\frac{\dot{\phi}^{2}}{2}+V\right)-3H^{2}=0.
\end{align*}
So now, seeing as the first term on the left is dominated by $V$ we can set:
\begin{align*}
	&0=\left(\frac{\dot{\phi}^{2}}{2}+V\right)-3H^{2}\simeq V-3H^{2}
\end{align*}
And deriving with respect do time we have:
\begin{align*}
	&\dot{V}-6H\dot{H}\simeq 0
\end{align*}
Now we remind ourselves that $\dot{V}=V'\dot{\phi}$ and that $\dot{H}=-\frac{\dot{\phi}^{2}}{2}$ to get:
\begin{align*}
	&\left(V'\dot{\phi}+3H\dot{\phi}^{2}\right)\simeq0,
\end{align*}
or in its elegant form:
\begin{align*}
	&-V'\simeq 3H\dot{\phi}.
\end{align*}
So, to sum it up we have:
\begin{align*}
	&\left\{\begin{array}{c}
	H^{2}\simeq \frac{V}{3}\\
	3H\dot{\phi}\simeq -V'
\end{array}	 \right.
\end{align*}
From the second we get $\dot{\phi}=-\frac{V'}{3H}$, and since $\dot{H}=-\frac{\dot{\phi}^{2}}{2}$ we have:
\begin{align*}
	&\dot{H}=-\frac{V'^{2}}{18H^{2}},
\end{align*}
and so we have:
\begin{align*}
	&\varepsilon_H\equiv -\frac{\dot{H}}{H^{2}}=\frac{V'^{2}}{18H^{4}}=\frac{V'^{2}}{18}\cdot\frac{9}{V^{2}}=\frac{1}{2}\left(\frac{V'}{V}\right)^{2}\equiv \varepsilon_V,
\end{align*}
as long as we are in slow-roll. We want to pay attention though to the assumption in which $V \gg \dot{\phi}^{2}$, when this assumption is weaker, we need additional terms to mitigate between $\varepsilon_H$ and $\varepsilon_V$. In fact, later we will make the adjustments up to second order and supply a 'recipe' to extend this to higher orders as well.
~ By the same kind of process we can look at the term:
\begin{align}
	&\ddot{\phi}=\frac{d}{dt}\dot{\phi}=\frac{d\phi}{dt}\frac{d}{d\phi}\left(\dot{\phi}\right).
\end{align}
Since we know that $\dot{\phi}\simeq \frac{V'}{3H}\simeq \frac{V'}{\sqrt{3V}}$ we can now derive with respect to $\phi$ and get:
\begin{align}
	&\ddot{\phi}\simeq \frac{\dot{\phi}}{\sqrt{3}}\left(\frac{V''}{V^{1/2}}-\frac{1}{2}\frac{V'^2}{V^{3/2}}\right).
\end{align}
Thus the quantity
\begin{align}
	&\delta_H\equiv \frac{\ddot{\phi}}{H\dot{\phi}}\simeq\frac{\sqrt{3}}{\sqrt{V}}\frac{1}{\sqrt{3}}\left(\frac{V''}{V^{1/2}}-\frac{1}{2}\frac{V'^2}{V^{3/2}}\right)=\frac{V''}{V}-\frac{1}{2}\left(\frac{V'}{V}\right)^2,
\end{align}
which is to say, if we define $\eta_V$ as 
\begin{align}
	&\eta_V\equiv \frac{V''}{V},
\end{align}
we have 
\begin{align}
	&\delta_H\simeq \eta_V -\varepsilon_V\equiv \delta_V.
\end{align}
\subsection[Cosmic perturbations]{First-order scalar perturbation theory}\label{sec:scalar_perturbations}
In the previous sections, we have used the Einstein Field Equations, as well as a Stress-Energy tensor for a scalar field. The procedure for deriving the Mukhanov-Sasaki equation uses a first-order perturbation approach to the EFE's.
We shall go through the main parts, but only in general, whereas the detailed process is demoted to an appendix status. It should be noted that, in general, we follow Mukhanov's derivation as outlined in \cite{Mukhanov1992203}.\\\\
We use the conformal metric - $g_{\mu\nu}=a^{2}(\tau)\left(\eta_{\mu\nu}+h_{\mu\nu}\right)$, where $\eta_{\mu\nu}$ is the Minkowski metric, and $h_{\mu\nu}$ is a small perturbation of the form:
\begin{align}
	&\begin{array}{ccc}
	h_{00}&=&2\Phi\\
	h_{0i}&=&0\\
	h_{ij}&=&2\Psi\cdot \delta^{i}\;_{j}
	\end{array}.
\end{align}
The Action for the scalar field is given by:
\begin{align}
	&\begin{array}{ccc}
	I&=&\int d^{4}x \sqrt{-g}\left[\tfrac{1}{2}g^{\mu\nu}\partial_{\mu}\phi\partial_{\nu}\phi-V(\phi)\right]
	\end{array}.
\end{align}
Varying the Action with respect to $\phi$ yields the equations of motion for $\phi$ which are nothing but the Klein-Gordon equation for $\phi$ in conformal time:
\begin{align}
	&\left(\phi''-\nabla^{2}\phi\right)+2H\phi' +a^{2}\frac{dV}{d\phi}=0
\end{align}
Where, using a perturbed conformal metric, as well as a first-order perturbed $\phi=\phi^{(0)}(\tau)+\delta\phi(\tau,\vec{x})$ gives additional information:
\begin{align}
	&\begin{array}{ccc}
	2H\delta\phi' +\left(\delta\phi''-\nabla^{2}\delta\phi\right)-\left(\phi^{(0)\prime}(\Phi'+3\Psi')\right)&+a^{2}\frac{d^{2}V}{d\phi^{2}}+2a^{2}\Phi\frac{dV}{d\phi}&=0\\
	\end{array}.
\end{align}
The equations $G^{\mu}\;_{\nu}=8\pi G T^{\mu}\;_{\nu}$, can be broken into perturbative degrees:
\begin{align}
&\begin{array}{ccc}
		^{(0)}G^{\mu}\;_{\nu}&=&8\pi G ^{(0)}T^{\mu}\;_{\nu},\\
		\delta G^{\mu}\;_{\nu}&=& 8\pi G \delta T^{\mu}\;_{\nu},\\
		.&&\\
		\vdots&&\\
\end{array}
\end{align}
where the leading-order perturbations in the Stress-Energy tensor are:
\begin{flalign}
	\left\{\begin{array}{cc}
		\delta T^{0}\,_{0}=&\frac{1}{a^{2}}\left[\phi'^{0}\delta\phi'-\Phi\phi'^{(0)2}+a^{2}\frac{dV}{d\phi}\delta\phi\right]\\
		\delta T^{0}\,_{i}=&\frac{\phi'^{(0)}\delta\phi_{;i}}{a^{2}}\\
		\delta T^{i}\,_{j}=&-\frac{1}{a^{2}}\left[\phi'^{(0)}\delta\phi'-\Phi\phi'^{(0)2}-a^{2}\frac{dV}{d\phi}\delta\phi\right]\delta^{i}\,_{j} ,
	\end{array}\right.
\end{flalign}
and the corresponding perturbed Einstein tensor, after proving $\Psi=\Phi$ in our case, is given by:
\begin{align}
	&\begin{array}{lcl}
		\delta G^{0}\,_{0}&=&\frac{2}{a^{2}}\left[-3H\left(H\Phi+\Phi'\right)+\nabla^{2}\Phi\right]\\
		\\
		\delta G^{0}\,_{i}&=&\frac{2}{a^{2}}\partial_{i}\left[H\Phi+\Phi'\right]\\
		\\
		\delta G^{i}\,_{j}&=&\frac{-2}{a^{2}}\left[\left(2H'+H^{2}\right)\Phi+3H\Phi'+\Phi''\right]\delta^{i}\,_{j} .
	\end{array}
\end{align}
Using these to compile 3 equations, and using the equations of motion for $\phi$ we are left with:
\begin{align}\label{Mukhanov-Sasaki - raw1}
	&\Phi''-\nabla^{2}\Phi+2\left(H-\tfrac{2\phi''}{\phi'}\right)\Phi'+2\left(H\-\tfrac{2\phi''}{\phi'}H\right)\Phi=0 ,
\end{align}
Or in an equivalent form:
\begin{align}\label{Mukhanov-Sasaki - alternative1}
	&\Phi''-\nabla^{2}\Phi+2\left(\tfrac{a}{\phi'}\right)'\left(\tfrac{a}{\phi'}\right)^{-1}\Phi'+2\phi'\left(\tfrac{H}{\phi'}\right)'\Phi=0 .
\end{align}
Switching to a new variable $u\equiv\left(\tfrac{a}{\phi'}\right)\Phi$, with $\theta\equiv \frac{H}{a\phi'}$ this equation now takes the form:
\begin{align}\label{Mukhanov-Sasaki equation1}
	&u''-\nabla^{2}u-\left(\frac{\theta''}{\theta}\right)u=0 ,
\end{align}
This is the Mukhanov-Sasaki equation.\\\\
To simplify the solution of this equation we decompose $u$ unto Fourier modes $u_k$, going into Fourier space we get:
\begin{align}
	&u''_k +\left(k^2 -\frac{z''}{z}\right)u_k=0\;, \label{eq:MS_k-space}
\end{align}
and when we go to Fourier space the pump field in the original space, $\theta$, becomes the pump field in Fourier space 
\begin{align}
z=\frac{a\dot{\phi}}{H}. \label{Z}
\end{align} 
We can wrap up the quantity $\left(k^2 -\frac{z''}{z}\right)$ as follows:
\begin{align}
	&\left(k^2 -\frac{z''}{z}\right)\equiv \omega_k^2(t)\;, \label{eq:TD_frequency}
\end{align}
so the MS equation can be expressed as:
\begin{align}
&u''_k +\omega_k^2(t)u_k=0\; .\label{eq:MS_k-space_TDHO}
\end{align}
In this form it is easily recognized as a Time-Dependent Harmonic Oscillator (TDHO).
\section[Primordial power spectrum]{The primordial power spectrum - a signature of our history}
The patterns we currently see in the CMB, are remnants of the time of last scattering. The CMB photons that are just now detected in our instruments were last scattered when the universe was approximately $1000\sim 3000$ times smaller and hotter. At that time the different constituents of the universe were in thermal equilibrium. Photons were scattered off electrons, and in turn, electrons were constantly bombarded with photons. In such a state matter is in a plasma phase, the optical depth is very small and whatever information a photon carries is erased after a few interactions. The current working assumption is that the optical depth increased suddenly, due to a sudden temperature drop that enabled electrons to be captured into previously ionized hydrogen atoms. At that point, the photons were no longer tightly coupled to electrons and were free to escape. It is these photons that first broke free that we now collect in our CMB instruments.
\subsection{Matter power spectrum}
However, since the time of last scattering several mechanisms have affected photons over time. It is customary to decompose the CMB radiation to modes, each corresponding to a different physical scale. The smallest $k$-mode we can hope to detect now corresponds to the scale of the current universe. The mode that is just now becoming causally connected corresponds to the current light horizon size. It can be shown that any region separated by more than 2 degrees in the sky today would have been causally disconnected at the time of decoupling \cite{Liddle:1999mq}. Modes that are not yet causally connected are 'frozen' and are roughly measured "as-is". Conversely, causally connected modes can now discharge energy, dissipate, or otherwise interact with other energy constituents of the universe. As such the CMB we currently detect differs from the radiation at the time of last scattering. 
\subsection{Transfer function}
In order to reflect the different physical processes sub-horizon and super-horizon modes undergo, as well as the overall expansion of the universe, which affect all modes, the Transfer function is called for. Given an initial power spectrum at the time of last scattering: 
\begin{align}
	&P(k)_{\text{matter}}\propto P(k)_{prim}T^{2}(k)\left(\frac{D(a)}{D(a=1)}\right)^{2},
\end{align}
where here $P(k)_{\text{matter}}$ is the matter power spectrum, $P(k)_{prim}$ is the PPS, $T(k)$ is the transfer function, and $D(a)$ is the growth function, which is dependent only on $a$, and reflects the change in modes due to the overall increase in the scale factor of the universe.
\subsection{Scalar power spectrum}
The primordial power spectrum (PPS) is traditionally characterized by its spectral index $n_s$ and the index running $\alpha_s$, which are given by the first and second logarithmic derivatives of the logarithm of the PPS:
\begin{align} 
&n_{s}=1+\left.\frac{\partial \ln\left(P_{s}\right)}{\partial \ln\left(k\right)}\right|_{aH=k},
\end{align}
\begin{align} 
&\alpha_s=\left.\frac{\partial^{2} \ln\left(P_{s}\right)}{\partial \ln\left(k\right)^{2}}\right|_{aH=k}=\left.\frac{\partial n_{s}}{\partial \ln\left(k\right)}\right|_{aH=k},
\end{align}
where $aH=k$ denotes the CMB scale.
Extending this Taylor expansion, we can write the PPS as:
\begin{align}
	&P_s=A_s\left(\frac{k}{k_0}\right)^{n_s-1 +\frac{\alpha_s}{2}\ln{\left(\frac{k}{k_0}\right)}+\frac{\beta_s}{6}\ln^2{\left(\frac{k}{k_0}\right)}+...},
\end{align}
where $k_0$ is the scale around which we Taylor expand, and is called the pivot scale. $\beta_s$ is the next term in the Taylor expansion and is called the running of running. It is given by:
\begin{align}
	&\beta_s=\left.\frac{\partial \alpha_s}{\partial \ln{k}}\right|_{aH=k}\; .
\end{align}
In most places the subscript denoting the assessment of the power spectrum at $aH=k$ is suppressed.
\section{Stewart-Lyth power spectrum}\label{sec:SL}
\subsection{Derivation}
Here we retrace the procedure of deriving the most widely used analytical expression for $n_s$. Recalling the definition for the pump field $z$, and the MS equation (Eqs. \eqref{Z},\eqref{eq:MS_k-space}). The parameter $\nu$ is properly defined as:
\begin{align}
	&\nu=+\sqrt{\frac{z''}{z}\tau^{2}+\frac{1}{4}}.\label{TrueNu}
\end{align}
However, in the Stewart-Lyth (SL) formulation, the approximations made lead to the definition of $\nu$ as the somewhat different:
\begin{align}
	&\nu_{\text{\tiny{SL}}}=\frac{3+2\delta_H +\epsilon_H}{2\left(1-\epsilon_H\right)}\; .
\end{align}
Then the MS equation becomes:
\begin{align}
	&u_{k}'' +\left(k^{2}-\frac{\left(\nu^{2}-\frac{1}{4}\right)}{\tau^{2}}\right)u_{k}=0.
\end{align}
For a constant $\nu$ this becomes the Bessel equation, with known solutions.
The value of $\tfrac{z''}{z}$ is formally given by:
\begin{align}
	&\frac{z''}{z}=2a^{2}H^{2}\left(1+\frac{3\delta_{H}}{2}+\epsilon_{H} +\frac{\delta_{H}^{2}}{2}+\frac{\epsilon_{H}\delta_{H}}{2} +\frac{1}{2H}\left(\dot{\epsilon_{H}}+\dot{\delta_{H}}\right)\right).\label{AccurateZoZ}
\end{align}
In many cases, one assumes that the time derivatives are small and can be neglected. However, these derivatives yield second-order terms that can significantly affect the value of $\tfrac{z''}{z}$. The full expression is given by:
\begin{align}
	&\frac{z''}{z}=2a^{2}H^{2}\left(1+\frac{3\delta_{H}}{2} +\epsilon_{H}+\epsilon_{H}	^{2} +2\epsilon_{H}\delta_{H} +\frac{1}{2}\frac{\delta_{H}\dddot{\phi}}{H\ddot{\phi}}\right), \label{AccurateZoZ-2}
\end{align}
which may  differ from Eq.~(\ref{AccurateZoZ}) after removing the time derivative terms, when $\delta^2_{H}$ and/or $\frac{\delta_{H}\ddot{\phi}}{H\dot{\phi}}$ are non-negligible. $\epsilon^2_H$ is usually of the order of $10^{-5}$ or less, even for models with high $r$.

Applying boundary conditions and taking the small arguments limit we are left with a power spectrum of:
\begin{align}
	\ln\left(P_{R}\right)&=-\ln(32\pi^{2}\Gamma^{2}(\tfrac{3}{2}))&   \\ \nonumber  &+ 2\nu \ln(2) +2 \ln(k) +2 \ln\left(\Gamma(\nu)\right) \\ \nonumber &+(1-2\nu) \ln(-k\tau),&
\end{align}
which yields the scalar index of:
\begin{align}
	n_{s}=4-2\nu +2\left(\ln(2) +\psi(\nu)\right)\frac{\partial \nu}{\partial \ln(k)},\label{eq:TrueNs}
\end{align}
with the digamma function $\psi(x)\equiv \frac{\Gamma'(x)}{\Gamma(x)}$.
The final expression is heavily dependent on the value and time derivative of $\nu$. This is a possible source of inaccuracy. 
In the original text the author now define:
\begin{align}
	&\begin{array}{ccc}
	\alpha=\left(\frac{V_{,\phi}}{V}\right)^{2}& \beta =\frac{V_{,\phi\phi}}{V} & \gamma=\frac{V_{,\phi^{3}}}{V_{,\phi}},\\
	\end{array} .
\end{align}
Having defined these, usually one connects the original slow-roll parameters with the above quantities by \cite{Stewart:1993bc}
\begin{align}
	&\begin{array}{c}
		\epsilon	_{H}\simeq \frac{\alpha}{2}-\frac{\alpha^{2}}{3}+\frac{\alpha\beta}{3}\\ 
		\\
		\delta_{H}\simeq \frac{\alpha}{2} -\beta -\frac{2\alpha^{2}}{3}+\frac{4\alpha\beta}{3} -\frac{\beta^{2}}{3}-\frac{\alpha\gamma}{3}\\ 
		\\
		\frac{\delta_{H}\dddot{\phi}}{H\ddot{\phi}}\simeq \alpha^{2}-\frac{5\alpha\beta}{2}+\beta^{2}+\alpha\gamma.
	\end{array} \label{eq:approximated_SR_SL}
\end{align}
With these relations, one can substitute the slow-roll parameters in Eq. \eqref{eq:TrueNs}, for the quantities in Eq. \eqref{eq:approximated_SR_SL}, to get the most commonly used analytical expression for the scalar index \cite{Lyth:1998xn}:
\begin{align}
	\nonumber n_{s}&\simeq 1-6\varepsilon_{V}+2\eta_{V} &\\ \nonumber&+2\times\bigg[\frac{\eta_{V}^2}{3}-(8b+1)\varepsilon_{V}\eta_{V}&& \\&-\left(\frac{5}{3}-12b\right)\varepsilon_{V}^2+\left(b+\frac{1}{3}\right)\xi_{V}^2\bigg],& \label{eq:SLR}
\end{align}
where $\varepsilon_{V}=\tfrac{\alpha}{2}\;;\;\eta_{V}=\beta\;;\;\xi^2_{V}=\alpha\gamma$, and $b=2-\ln{2}-\gamma_{\mathrm{\scalebox{.4}{Euler}}}$, $\gamma_{\mathrm{\scalebox{.4}{Euler}}}$ being the Euler number.
\subsection{Analytical term for index running $\alpha_s$}
Following the same procedure for the running of the scalar index, gives, to second-order:
\begin{align}
	\alpha_s \simeq & -16\varepsilon_{V}\eta_{V} +24\varepsilon_{V}^{2}+2\xi_{V}^{2}, \label{eq:Analytical_alpha}
\end{align}
where both \eqref{eq:SLR}, and \eqref{eq:Analytical_alpha}, are evaluated at the CMB point if one wishes to probe the CMB point. Equivalently if one wishes to compare to observables as derived from MCMC analyses at some pivot scale $k_{\mathrm{pivot}}$, one needs to find the correct $\phi_{\mathrm{pivot}}$, and evaluate these expressions at the pivot scale.
\section[Less common formulations for $n_s$]{Other formulations for the power spectrum observables}
In this section, attention should be paid to different formulations and notations. Every subsection is self-contained to encapsulate the derivations fully. Wherever possible we use the notation in the source material.
Over the years, our observations have become more accurate and spanned a longer interval of physical scales \cite{Bennett:2003ba,Leitch:2001yx,Halverson:2001yy,Pryke:2001yz,Netterfield:2001yq,deBernardis:2001mjh,Lee:2001yp,Stompor:2001xf,Tegmark:2000qy,Hannestad:2001nu}. Even well before the Planck mission \cite{Planck:2006aa} was proposed and eventually launched, the prospects of probing not just the slope of the PPS but the running of the slope was compelling. As a result, several forays into calculating the running of the spectral index were made. It was understood early on, that Eq. \eqref{eq:Analytical_alpha}, was put forth as a second order derivation of an expression that by definition vanishes. As a result different approaches were taken, at the level of analytic expression for the PPS itself, rather than calculating derivatives of the $n_s$ it predicts. All of these attempts were made around 2011. 
\subsection{Dodelson-Stewart}\label{sec:DS}
Twin papers written by Ewan Stewart and Scott Dodelson were published in September and October of 2011. Stewart published a detailed calculation of the PPS \cite{Stewart:2001cd}, dropping a previous assumption, that $|\alpha_s|\ll |n_s-1|$. Dodelson and Stewart published a joint paper, based on the calculations made in \cite{Stewart:2001cd}, that studies the ramifications of these formulae on inflationary potentials \cite{Dodelson:2001sh}. This treatment uses the slow-roll parameters in the efold-flow formulation, such that:
\begin{align}
	&\varepsilon=\frac{1}{2}\left(\frac{\dot{\phi}}{H}\right)\hspace{20pt}\textrm{and}\hspace{20pt} \delta_p\equiv \frac{1}{H^p \dot{\phi}}\left(\frac{d}{dt}\right)^p \dot{\phi}, \label{eq:Stewart_dodelson_slow-roll}
\end{align} 
and their evolution is governed by:
\begin{align}
	&\frac{d\varepsilon}{d\ln{a}}=\frac{d\varepsilon}{d N} =2\left(\varepsilon +\delta_1\right)\varepsilon\\
	&\nonumber \\
	&\frac{d\delta_p}{d \ln{a}}= \frac{d\delta_p}{d N}=\delta_{p+1}+\left(p\varepsilon -\delta_1\right)\delta_p .\label{eq:Stewart_dodelson_slow-roll2}
\end{align}
In this approximation the scalar index is given by:
\begin{align}
	&n_{s,\mathrm{\scalebox{.5}{DS}}}= 1-4\varepsilon -2\delta_1 -2\sum_{p=1}^{\infty}d_p \delta_{p+1} +\mathcal{O}(\xi^2), \label{eq:Dodelson_Stewart_ns}
\end{align}
where $d_p$ are numerical coefficients of order unity, and the remainder includes the terms $\varepsilon^2 ,\delta_1 \varepsilon$ and $\delta_1^2$. These $d_p$ terms can be calculated by using the following generating function:
\begin{align}
	&\sum_{p=0}^{\infty}d_p x^p =2^x \cos{\left(\frac{\pi x}{2}\right)} \frac{\Gamma(2-x)}{1+x}. \label{eq:DS_generator}
\end{align}
In this formulation the scalar index running is given by:
\begin{align}
	&\alpha_{s,\mathrm{\scalebox{.5}{DS}}}=-2\sum_{p=0}^{\infty}d_p\delta_{p+2}+\mathcal{O}(\xi^2). \label{eq:Dodelson_Stewart_alpha}
\end{align}
In terms of inflationary potential and derivatives thereof, the scalar index and the index running are given by:
\begin{align}
	&n_{s,\mathrm{\scalebox{.5}{DS}}}= -3\left(\frac{V'}{V}\right)^2 +2\sum_{p=0}^{\infty}q_p \left(\frac{V'}{V}\right)^{p}\frac{V^{(p+2)}}{V}\\
	&\nonumber \\
	&\alpha_{s,\mathrm{\scalebox{.5}{DS}}}=-2\sum_{p=0}^{\infty}q_p \left(\frac{V'}{V}\right)^{p+1}\frac{V^{(p+3)}}{V},
\end{align}
where here $q_p$ are generated by the following generating function:
\begin{align}
	&\sum_{p=0}^{\infty}q_p x^p \equiv 2^{-x} \cos{\left(\frac{\pi x}{2}\right)} \frac{3\Gamma(2+x)}{(1-x)(3-x)}.
\end{align}
As we shall see this formulation is close, but only in the 'horseshoes and hand grenades' sense.
\subsection{Stewart-Gong}\label{sec:SG}
The Stewart-Gong formulation \cite{Gong:2001he} is, in essence, similar. They first set up the Green's function formalism to tackle the problem of inflation. They rewrite the MS equation Eq. \eqref{eq:MS_k-space} as:
\begin{align}
	&\frac{d^2 y}{dx^2}+\left(1 -\frac{1}{z}\frac{d^2 z}{dx^2}\right)y=0,
\end{align}
where $x$ and $y$ are defined as\footnote{Note that in this derivation $\eta$ denotes conformal time}:
\begin{align}
	&x=-k\eta\\
	&\nonumber \\
	&y=\sqrt{2k}u_k .
\end{align}
They then choose the ansatz $z=\frac{1}{x}f(\ln{x})$, where $f$ is some function. Defining a new function $g$ such that:
\begin{align}
	&g=\frac{-3f' +f''}{f}, 
\end{align}
the equation of motion now transforms to:
\begin{align}
&\frac{d^2 y}{dx^2}+\left(1 -\frac{2}{x^2}\right)y=\frac{1}{x^2}g(\ln{x})y.	
\end{align}
Now we can take the Green's function approach to get the eigenfunction $y_{0}=\left(1+\frac{i}{x}\right)e^{ix}$, and yield the full solution:
\begin{align}
	&y(x)=y_0(x) +\frac{i}{2}\int_x^{\infty} du \frac{1}{u^2}g(\ln{u})y(u)\left[y_0^*(u)y_0(x) -y_0^*(x)y_0(u)\right].
\end{align}
The authors continue to write the full expression for $n_s$:
\begin{align}\label{eq:n_s-Stewart-Gong}
	n_{s,\mathrm{\scalebox{.5}{SG}}}&=1-4\varepsilon_1 -2\delta_1 +(8b-8)\varepsilon_1^2 +(10b-6)\varepsilon_1\delta_1 -2b\delta_1^2 +2b\delta_2&\\
	\nonumber & +\left(-16b^2 +40 b -108 +\frac{28\pi^2}{3}\right)\varepsilon_1^3 +\left(-31b^2 +60b -172 +\frac{199\pi^2}{12}\right)\varepsilon_1^2\delta_1 \\
	\nonumber & +\left(-3b^2 +4b -30 +\frac{13\pi^2}{4}\right)\varepsilon_1\delta_1^2 +\left(-2b^2 +8 -\frac{5\pi^2}{6}\right)\delta_1^3\\
	\nonumber & +\left(-7b^2 +8b -22 +\frac{31\pi^2}{12}\right)\varepsilon_1\delta_2 +\left(3b^2 -8 +\frac{3\pi^2}{4}\right)\delta_1\delta_2 +\left(-b^2 +\frac{\pi^2}{12}\right)\delta_3,  
\end{align}
where the hierarchy of slow-roll parameters is defined as:
\begin{align}\label{eq:Slow-roll_a_la_Stweart-Gong}
	&\varepsilon_1=\frac{-\dot{H}}{H^2}\; ; \; \delta_n=\frac{1}{H^n \dot{\phi}}\frac{d^{n+1}\phi}{dt^{n+1}}. 
\end{align}
At the end of this formulation, as we shall see, lay no cigars as well.\\
It is, however, worth mentioning, that Dvorkin \& Hu \cite{Dvorkin:2009ne}, used this method iteratively to calculate the PPS to $\sim 0.3\%$ accuracy.
\subsection{Schwarz-Escalante-Garcia}\label{sec:SEG}
The formulation in \cite{Schwarz:2001vv} claims not to rely on the slow-roll expansion, but as we shall see this is not accurate. We first define the horizon flow parameters as:
\begin{align}
	&\varepsilon_0=\frac{d_H(N)}{d_{H_i}},
\end{align}
where $d_H\equiv\frac{1}{H}$ is the Hubble distance, and $d_{H_i}$ is the Hubble distance at the start of inflation (at $t_i$, where $N(t_i)=0$). The hierarchy of functions is defined as:
\begin{align}
	&\varepsilon_{m+1}\equiv \frac{d\ln{|\varepsilon_m|}}{dN},\hspace{20pt}\; m>0.
\end{align}
According to this definition, $\varepsilon_1=\frac{d \ln{d_H}}{dN}$, and it can be shown that 
\begin{align}
	&\varepsilon_1\equiv \varepsilon_H=-\frac{\dot{H}}{H^2}.
\end{align}
Upon closer inspection, one finds this is nothing but a rewriting of the slow-roll parameters of Eq. \eqref{eq:Slow-roll_a_la_Stweart-Gong}, as shown in table~.\ref{table:SEG_to_SG}.
\begin{table}[!h]
	\begin{center}
		\begin{tabular}{|c|c|}
			\hline
			Schwarz-Escalante-Garcia&Stewart-Gong\\
			\hline
			\hline
			$\varepsilon_1$&$\varepsilon_1$\\
			\hline
			$\varepsilon_2$&$2\varepsilon_1+2\delta_1$\\
			\hline
			$\varepsilon_2\varepsilon_3$&$4\varepsilon_1^2 +6\varepsilon_1\delta_1 -2\delta_1^2 +2\delta_2$\\
			\hline
			\hline
		\end{tabular}
	\end{center}
	\caption[{\bf Comparison of slow-roll parameters SEG to SG}]{Slow-roll parameters in Schwarz-Escalante-Garcia formulation, translated to Stewart-Gong formalism. \label{table:SEG_to_SG}}
\end{table}\\
As such their analysis is not so different from previous ones made. However, they provide a taxonomy of models which divide models into two categories: Constant-Horizon (CH) models, and Growing-Horizon (GH) models. The CH category is reserved for models that have a very small time derivative of the Hubble distance. This implies that $\varepsilon_1 \ll 1$, but it doesn't necessarily mean that $\varepsilon_m$ is small for arbitrary $m$. Thus the authors define $\mathrm{CH}n$ as models in which $|\varepsilon^{n}_2|<\mathrm{max}\left(|\varepsilon_1\varepsilon_2|,|\varepsilon_2\varepsilon_3|\right)$. In this case, the following terms are allowed in the approximation: $1,\varepsilon_1,\varepsilon_2,...,\varepsilon_2^n$. In the GH treatment, the approximation scheme is different. This scheme is inspired by the power-law inflation case in which $\varepsilon_1=\frac{1}{p}$, for inflation with $a\propto t^p$, and the other slow-roll parameters strictly vanish. Thus this scheme defines $n$ as the maximal integer in which $\varepsilon_1^n>\mathrm{max}\left(|\varepsilon_1\varepsilon_2|,|\varepsilon_2\varepsilon_3|\right)$ holds true. So the slow-roll terms allowed in this approximation scheme are $1,\varepsilon_1,...,\varepsilon_1^n,\varepsilon_2$. \\
The power spectra given in the CH case are given by:
\begin{align}
	&k^3 P_s&=&\frac{H^2}{\pi\varepsilon_1}\big[a_0 +a_1 \ln{(k/k_0)} +a_2\ln^2{(k/k_0)} +a_3\ln^3{(k/k_0)}&+...\big],\\
	&a_0&=& 1-2(1-b)\varepsilon_1 +b\varepsilon_2 +\frac{1}{8}\left(4b^2 +\pi^2 -8\right)\varepsilon_2^2&\\
	&\nonumber && -\frac{1}{24}\left(3b(8-\pi^2)-4b^3 +14\zeta(3) -16\right)\varepsilon_2^3,&\\
	&a_1&=& -2\varepsilon_1 -\varepsilon_2 -b\varepsilon_2^2 -\frac{1}{8}\left(4b^2 +\pi^2 -8\right)\varepsilon_2^3,&\\
	&a_2&=&\frac{1}{2}\varepsilon_2^2 +\frac{b}{2}\varepsilon_2^3,&\\
	&a_3&=& -\frac{1}{6}\varepsilon_2^3,&
\end{align}
with the same $b$ as before: $b=2-\ln{2}-\gamma_{\mathrm{\scalebox{.4}{Euler}}}$, and where $\zeta(3)\approx 1.2021$. 
The power spectra of GH models are given by:
\begin{align}
	&k^3 P_s&=&\frac{H^2}{\pi\varepsilon_1}\big[c_0 +c_1 \ln{(k/k_0)} +c_2\ln^2{(k/k_0)} +c_3\ln^3{(k/k_0)}&+...\big],\\
	&c_0&=& 1-2(1-b)\varepsilon_1 +\frac{1}{2}\big[4b(b-1)+\pi^2 -10\big]\varepsilon_1^2\\
	&\nonumber &&+\frac{1}{3}\big[4b^3 +3b(\pi^2-12)-14\zeta(3) +19\big]\varepsilon_1^3 +b\varepsilon_2,\\
	&c_1&=& -2\varepsilon_1 +2(1-2b)\varepsilon_1^2 -(4b^2+\pi^2 -12)\varepsilon_1^3 -\varepsilon_2,\\
	&c_2&=& 2\varepsilon_1^2 +4b\varepsilon_1^3,\\
	&c_3&=& -\frac{4}{3}\varepsilon_1^3
\end{align}
with the same values of $b$ and $\zeta(3)$.\\
While this approach at least gives us consistent conditions for the voracity of different approximation schemes, it is no closer to the mark than the previous ones.
\section[Lyth bound]{The Lyth bound, original and extended versions}
One of the most useful heuristics out there, to build and evaluate inflationary models, is the so-called Lyth bound. In the original paper \cite{Lyth:1996im}, there are several conclusions. However we concentrate on the following argument: Given that the PPS can be written as
\begin{align}
	P_s\simeq \left.\frac{8\pi G H^2}{9\varepsilon_H k^3}\right|_{k=aH},
\end{align}
and the tensor power spectrum is given by: 
\begin{align}
	P_T\simeq\left.\frac{8\pi G H^2}{k^3}\right|_{k=aH},
\end{align}
we can define the \textit{tensor-to-scalar ratio}, $r$, as the power spectra ratio at some scale $k_0$ of our choosing. 
Thus $r$ is set as:
\begin{align}
	r=9\varepsilon_H.
\end{align}
The numerical coefficient is subject to some finer detail, but the order of magnitute is always the same. In \cite{Lyth:1996im} we have $r=2\cdot 6.9 \varepsilon_h$, whereas other sources state $r=12\sim 16\varepsilon_H$. This statement, along with the (extreme) slow-roll assumption in which $\varepsilon_H=\varepsilon_V$ yields the following:
\begin{align}
	\left(\frac{r}{6.9}\right)^{\frac{1}{2}}=\left|\frac{V'}{V}\right|. \label{eq:Lyth_to_V}
\end{align} 
In order to complete the argument, we must look at the quantity
\begin{align}
	\frac{d\phi}{dN}=\frac{d\phi}{Hdt}=\frac{\dot{\phi}}{H}=\sqrt{2\varepsilon_H}.
\end{align}
So, plugging in the former equation we get:
\begin{align}
	\frac{d\phi}{dN}=\left(\frac{r}{6.9}\right)^{\frac{1}{2}},
\end{align}
where we adhere to the numerical coefficient in \cite{Lyth:1996im}.\\
The scales on which the tensor power spectrum is most prominent, and therefore is more likely to be discovered at is $1<l\lesssim 100$. With $k\sim\frac{H_0 l}{2}$, the interval in $k$-space over which we expect to find tensor modes is $\Delta \ln{k}\simeq 4.6$. Taking $H$ to be roughly constant, when we evaluate $k=aH$ we have
\begin{align}
	&\ln{k}=\ln{aH}=\ln{H}+\ln{a}=\ln{H}+N,
\end{align}
thus \begin{align*}
		&d\ln{k}=dN.
\end{align*}
If we additionally take $d\phi/dN$ to be approximately constant along the aforementioned $k$-interval we have:
\begin{align}
	\frac{d\phi}{dN}\simeq \frac{\Delta \phi}{\Delta N}.
\end{align}
Putting these together yields:
\begin{align}
	\Delta \phi \simeq 4.6\left(r/6.9\right)^{1/2}=0.46\left(r/0.07\right)^{1/2}. \label{eq:Lyth_bound_original}
\end{align}
This means that given some $r$ we can approximate the field excursion, in Planck units, during the first $\sim 4.6$ efolds of inflation. For example, suppose we find $r\lesssim 0.03$, which is possible in the foreseeable future. In that scenario we have:
\begin{align}
	\Delta \phi \sim 0.3\;.
\end{align}
Applying Eq. \eqref{eq:Lyth_to_V}, we can also get a one-to-one relation between $r(k)$ and the slope of the potential at a chosen $k_0$, as long as $k_0$ is inside the $1<l\lesssim 100$ window. This relation is formalized in \cite{BenDayan:2009kv} as
\begin{align}
	\frac{V'}{V}=-\sqrt{\frac{r}{8}},
\end{align}
where the numerical coefficient is subject to slight changes, and the minus sign comes from the notion of the field rolling towards the larger positive $\phi$ values.\\

~This original argument was extended in \cite{Easther:2006qu,Efstathiou:2005tq}. In \cite{Easther:2006qu}, the argument was amended to include higher slow-roll parameters, to allow for models that change by a significant amount during the tensor mode window. This extended Lyth bound is given by:
\begin{align}
	\Delta \phi \gtrsim \sqrt{\frac{r}{4\pi}}\left[2-n_s -\frac{r}{8}\right], \label{eq:Lyth_small_fields}
\end{align}
and in the case of $r=0.03$ we have $\Delta\phi\gtrsim 0.05$.
In \cite{Efstathiou:2005tq} however, some assumptions were added, in order to extend the bound to the full efold extent of inflation, i.e. $N=55\sim 60$. In this case the authors extend the Lyth bound such that:
\begin{align}
	\Delta \phi \approx 6r^{1/4}. \label{eq:Lyth_extended_efolds}
\end{align} 
This would imply that models that predict $r=0.03$, should incur a field excursion of $\Delta\phi\sim 2.5$.
\section{Small and Large field models taxonomy}
Current nomenclature distinguishes between 'Large field models' and 'Small field models'. This distinction is based on \cite{Dodelson:1997hr}, where several models were studied and a taxonomy of models was presented. In the interest of brevity and accessibility, we will first reiterate the key concepts needed to understand this distinction. First, the slow-roll condition on the potential is given by:
\begin{align}
	\varepsilon_V\equiv \frac{1}{2}\left(\frac{V'}{V}\right)^2<1, \label{eq:Taxonomy_epsilon}
\end{align}
where $V'=\frac{dV(\phi)}{d\phi}$, and usually it is understood as a strong requirement i.e. we prefer $\varepsilon_V\ll 1$, in order to be in the slow-roll regime.  The number of efolds generated within a certain inflation is given by:
\begin{align}
	N=\int_{t_{\mathrm{init}}}^{t_{\mathrm{fin}}} H dt \simeq \int_{\phi_{\mathrm{init}}}^{\phi_{\mathrm{fin}}} -\frac{V}{V'}d\phi,\label{eq:Taxonomy_efolds}
\end{align}
where the minus sign is due to our interpretation of the field $\phi$ rolling down the potential. Finding $\phi_{\mathrm{fin}}$ given the potential is usually straightforward, by setting $\varepsilon_V=1$ we identify the end of slow-roll inflation, thus:
\begin{align}
	\left(\frac{V'}{V}\right)=\pm \sqrt{2}, \label{eq:Taxonomy_phi_end}
\end{align}
from which we can extract $\phi_{\mathrm{fin}}$.\\
In \cite{Dodelson:1997hr}, the authors divide the models studied into 3 categories:
\begin{itemize}
	\item[1. ] Large field models: in which the inflaton field $\phi$ is displaced far from its minimum, and rolls down the potential towards a minimum at the origin. The models $V(\phi)=\Lambda^4 \left(\phi^p\right)\;;\; p>1$ and $V(\phi)=\Lambda^4 \exp\left(-\sqrt{2\phi^2/p}\right)\;;\;p>0$ are classic examples of large field models.
	\item[2. ] Small field models: in which the field is initially near the origin, and rolls down towards a minima that is removed from the origin ($<\phi>\neq 0$), these types of models would be expected as a result of spontaneous symmetry breaking.
	\item[3. ] Hybrid models of inflation: in which the field evolves towards a minima with a non-zero vacuum energy. Usually, hybrid models are realized with a number of dynamic fields. However during inflation, in most cases one can identify a dominant field, and thus treat this type of models as single field inflation models, at least a posteriori. We do not discuss hybrid models of inflation within the context of this work. 
\end{itemize}
\subsection{Large field models}
Since these type of models, as defined in \cite{Dodelson:1997hr}, usually have initial $\phi$ values of $\sim 10\; m_{pl}$ or more, the taxonomy had evolved to the following simplified meaning. Large field models are models in which the field excursion during inflation is more than a few $m_{pl}$. Let us examine the class of Large field polynomial potentials to understand why that is:
\begin{align}
	V(\phi)=\Lambda^4 \left(\phi^p\right)\;;\; p>1.
\end{align}
By virtue of Eq.~\eqref{eq:Taxonomy_phi_end} we have $\phi_{\mathrm{fin}}=\pm \frac{p}{\sqrt{2}}$, and by applying Eq.~\eqref{eq:Taxonomy_efolds}, we recover the following condition on the initial $\phi$ value:
\begin{align}
	\sqrt{p\left(2N+\frac{1}{\sqrt{2}}\right)}=\phi_{\mathrm{init}}.
\end{align}
With $N\gtrsim 50$, and even when $p=1$, this produces $\phi_{\mathrm{init}}\sim 10\; m_{pl}$ and $\Delta\phi$ approximately the same. Thus in these models $\Delta \phi\gtrsim 10\; m_{pl}$. With the exponential models the end of inflation occurs near $\phi_{\mathrm{fin}}=\sqrt{p}$. The relation between the starting and end positions of the field is given by:
\begin{align}
	\sqrt{p}e^{\frac{2N}{p}}=\phi_{\mathrm{init}},
\end{align}
and a quick inspection of this as a function of $p$ reveals a minimum at $p=4N$, which means the minimum field excursion is given by:
\begin{align}
\Delta \phi \gtrsim 2\sqrt{N}\left(e^{\frac{1}{2}}-1\right) \approx 9 \;m_{pl}
\end{align}
when $N=50$. This is enough to justify the notion of large field models, as models in which the field excursion during inflation is of several Planck masses.
\subsection{Small field models}
The methodology of evaluating small field models is the same as the above, with the only difference of $\phi_{\mathrm{init}}$ starting near the origin:
\begin{align}
	<\phi_{\mathrm{init}}>\simeq 0.
\end{align}
The usual representative of this class of models is the "small-field polynomial model"
\begin{align}
	V(\phi)=\Lambda^4\left(1-(\phi/\mu)^p\right)\; \phi\ll\mu\ll m_{pl}\; \text{and}\; p>2.
\end{align}
This model is not to be confused with the models we study, as it is actually a monomial, meaning there is only one term in which $\phi$ appears. Note that since this is the most usual representative of the small field models, there is a tendency to study this class and apply these conclusions to the entire class. This is evident in \cite{Martin:2013tda} for example. In these types of models it is easily discerned that the field excursion can be of the order of $\Delta \phi \lesssim  1\sim 2 \; m_{pl}$, and usually much smaller.\bigskip \\
A different more observable-oriented classification of models can be found in \cite{Schwarz:2004tz}.
\clearpage


\graphicspath{{Chapter2_5/}}

\chapter{Our Models}\label{Models}
Small field models of inflation in which inflation occurs near a flat feature, a maxima, or a saddle point are studied (see \cite{Boubekeur:2005zm} for a review). This class of models is interesting because they appear in many fundamental physics frameworks, effective field theory, supergravity \cite{Yamaguchi:2011kg} and string theory \cite{Baumann:2014nda} in successive order of complexity. Our focus on such models is also motivated by the expected properties of the moduli potentials in string theory. More generally speaking these type of models can be viewed as a Taylor expansion approach to other models \cite{Dodelson:1997hr}. A different more observable-oriented classification of models can be found in \cite{Schwarz:2004tz}, in which analysis our models fall into the toward-exit class.

In general, inflation will occur in a multi-dimensional space. However, the results for multifield inflation cannot usually be obtained simply. In many known cases, it is possible to identify a-posteriori a single degree of freedom along which inflation takes place. To gain some insight about the expected typical results effective single field potentials can be used.

Generic small field  models predict a red spectrum of scalar perturbations,  negligible spectral index running and non-gaussianity. They also predict a characteristic suppression of tensor perturbations \cite{BenDayan:2008dv}. Hence, they were not viewed as candidate models for high-$r$ inflation. Large field models of inflation are thus the standard candidates for high-$r$ inflation. For more detailed model building considerations one can review \cite{Makarov:2005uh} and \cite{Lesgourgues:1998mq}.

In \cite{BenDayan:2009kv}, a new class of more complicated single small field models of inflation was considered (see also \cite{Hotchkiss:2011gz}) that can predict, contrary to popular wisdom \cite{Lyth:1996im,Martin:2013tda}, an observable GW signal in the CMB (see also \cite{Cicoli:2008gp}.) The notion that observable signal GW precludes small field models partly stems from \cite{Martin:2013tda} and similar analyses that study monomial potential models as small field models. The spectral index, it's running, the tensor to scalar ratio and the number of e-folds were claimed to cover all the parameter space currently allowed by cosmological observations. The main feature of these models is that the high value of $r$ is accompanied by a relatively strong scale dependence of the resulting power spectrum. Another unique feature of models in this class is their ability to predict, again contrary to popular wisdom \cite{Easther:2006tv}, a negative spectral index running.  The single observable consequence that seems common to all single field models is the negligible amount of non-gaussianity.
In \cite{Lesgourgues:2007gp} the inflationary potential was Taylor-expanded up to order $4$. The approach applied in \cite{Lesgourgues:2007gp} is similar to ours, however only potentials that are monotonic in the entire CMB window were considered. The family of potentials we study can easily contain members that have a shallow minimum point followed by an equally shallow maximum. As long as there is enough kinetic energy to clear that interval in $\phi$, we will clear the potential `valley' and not be trapped in a scenario where inflation doesn't end. A concrete example would be a third degree small field monomial of the form $V(\phi)=V_0(1-M\phi^3)$, after a shift such that $\phi \rightarrow \phi-\tilde{\phi}$. In this case $\phi$ and $\phi^2$ terms will appear in the shifted potential. Following this shift it is an easy matter to add some small terms of degree four, five and six to the potential, such that we get a six degree polynomial member with a shallow minimum. Though we haven't targeted potentials with minima especially, we have not ruled these out a-priori, thus some potentials as the one in the example were manufactured by the computerized model building package, and were tested for compliance.\\

The current work yields corrected predictions of this class of models by a systematic high-precision analysis, thus providing a viable alternative to the large field-high $r$ option. The analysis of \cite{BenDayan:2009kv} is extended, in preparation for a subsequent detailed comparison of the models to data. This is done in order to simplify the parametrization of the potential and facilitate a comprehensive numerical study.
\subsection{Inflaton potentials with $r=0.001$}
\begin{figure}[!ht]
\includegraphics[width=1\textwidth]{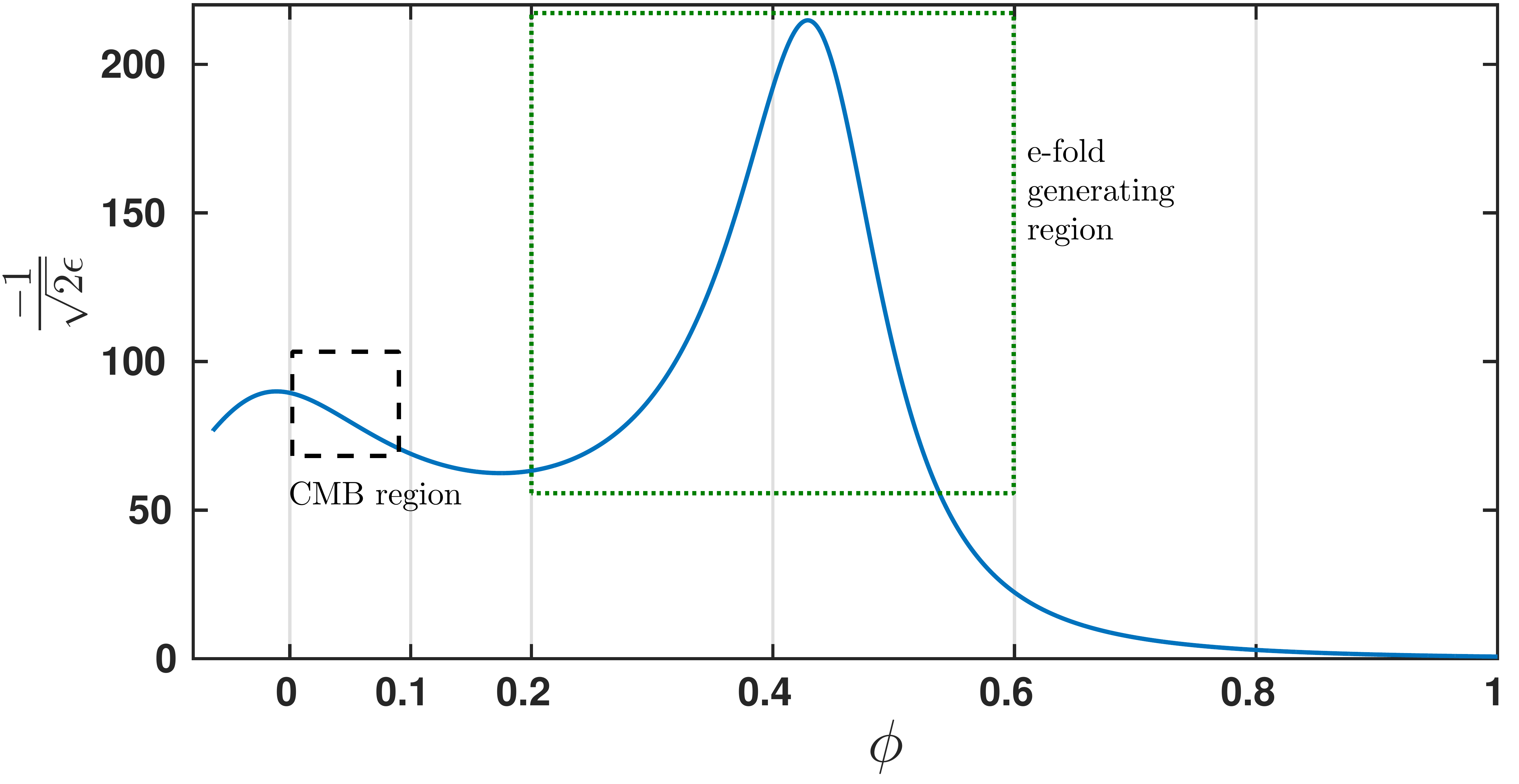}
\caption[{\bf Efold generation regimes}]{A graph depicting $-1/\sqrt{2\epsilon}$ as a function of the inflaton $\phi$ for a model for which $r_0=0.001$. The CMB interval is covered by $\sim 8$ e-folds generated while the field changes by about $\Delta\phi\sim 0.1$. Most of the e-folds are generated when $\phi$ reaches $\sim 0.4$.}
\label{fig:1OverSqrt(2eps)}
\end{figure}

The following class of polynomial inflationary potentials proposed in \cite{BenDayan:2009kv} is:
\begin{align}
V(\phi)=V_{0}\left(1+\sum_{p=1}^{5}a_{p}\phi^{p}\right)\;. \label{Full_form}
\end{align}
The virtue of these models from a phenomenological point-of-view is the ability to separate the CMB region from the region of large e-fold production. Hence, these potentials can produce  a very different spectrum early on, than in the later stages of inflation. Fig.~\ref{fig:1OverSqrt(2eps)} illustrates this point, with separate CMB region and e-fold generation region.
In the context of both classification systems mentioned, current observational data weakly support these \cite{Martin:2014lra,Vennin:2015eaa}. However the small field model studied in \cite{Martin:2014lra} are monomial potential models of the form $V\propto 1-a_p\phi^p$, which are different from many of our models.\bigskip

In many models, $\varepsilon_V\sim 1/N^2$, $\eta_V\sim 1/N^2$, and the time derivative $\frac{d}{Hdt}$ can approximately be replaced with a factor of $\frac{1}{N^2}$ \cite{Kosowsky:1995aa}. In the above models, this standard hierarchal dependence is broken, they have a more complicated dependence while obeying the slow-roll conditions $\epsilon_H$, $\delta_H \ll 1$.
In \cite{BenDayan:2009kv} it was shown that these models can be written as:
\begin{align}
V(\phi)=V_{0}\left(1-\sqrt{\frac{r_{0}}{8}}\phi +\frac{\eta_{0}}{2}\phi^{2} +\frac{\alpha_{0}}{3\sqrt{2 r_{0}}}\phi^{3} +a_{4}\phi^{4}+a_{5}\phi^{5}\right).\label{BB-potential}
\end{align}
Here  $r_0$, $\eta_{0},\alpha_{0}$ are defined as $r= 8 \left(\frac{V'}{V}\right)^2$,  $\eta=\frac{V''}{V}$, $\alpha=-2\xi^{2}$, respectively. The subscript $0$ means that these are the values at the CMB point.

Specifically for a potential of the form $V\propto 1+\sum_{p=1}^5a_{p}\phi^p$, the SL analytic expressions for the scalar index and its running (Eqs.~(\ref{eq:SLR},\ref{eq:Analytical_alpha})) are given by
\begin{align}
    n_{s}\simeq&1-3a_{1}^2+4a_{2} \\
    &\nonumber +2\Bigg[\frac{4a_{2}^2}{3}-\left(\frac{5}{3}-12b\right)\frac{a_{1}^4}{4}\\ \nonumber &-\left(8b+1\right)a_{1}^2 a_{2} +\left(6b+2\right)a_{1}a_{3}\Bigg] ,\\ \nonumber \\
    \alpha_s \simeq & 16a_{1}^2 a_{2}-6a_{1}^4-2a_{1}a_{3}.
\end{align}
\subsection{Reduced parameter space}
The potential in Eq.~\eqref{BB-potential} is a small field candidate, which after some scaling and normalization, depends on four free parameters. One parameter is used for setting  $r_0$ at the CMB point, and thus the predicted amplitude of the GW signal produced, while the other two parameters are used to parametrize the $n_s-\alpha_s$-plane. The fourth parameter determines the number of e-folds from the CMB point to the end of inflation. $\phi_{\text{\tiny{end}}}$ is set to $1$ to simplify the analysis. It follows that
 \begin{align}
  \frac{1}{2}\left(\frac{V'}{V}\right)^2{|_{\phi=1}}=1. \label{CONDITION}
\end{align}
Suppose we want inflation to end at $\phi=\zeta$, we can rescale $\phi$:
	\begin{align}
		\phi\rightarrow \tilde{\phi}=\frac{\phi}{\zeta}.
	\end{align}
	In this formulation,
	\begin{align}
		V=V_0\left(1+\sum_{p}a_p \zeta^{p} \tilde{\phi}^{p}\right)=V_0\left(1+\sum_{p} \tilde{a_p}\tilde{\phi}^p\right),
	\end{align}
	where $\tilde{a_p}=a_p \zeta^{p}$. Since this is the same potential, it follows the same CMB observables are produced. Thus, applying condition \eqref{CONDITION} can be viewed as a scaling scheme for the different terms in the potential which does not limit the generality of our results. \\
	
~ Substituting the expression for the potential and its derivative at $\phi=1$ we get:
\begin{align}
	-\sqrt{2}=\frac{\sum_{p=1}^{5}p\cdot a_p}{1+\sum_{p=1}^{5}a_p}.
\end{align}
 $a_4$ is now given in terms of the other coefficients:
 \begin{align}	
a_4=\frac{-1}{4+\sqrt{2}}\left(\sqrt{2} +\sum_{p\in(1,2,3,5)}\left(p+\sqrt{2}\right)a_p\right)
 \label{sola4}
 \end{align}

Using the standard definition for the number of e-folds  $N=\int_{t_{\text{\tiny{CMB}}}}^{t_{\text{\tiny{end}}}}Hdt\simeq -\int_{\phi_{\text{\tiny{CMB}}}}^{\phi_{\text{\tiny{end}}}}\frac{V}{V'}d\phi$, and the approximation $V(\phi)=1+\sum_{p=1}^{5}a_{p}\phi^{p}\simeq 1$ yields a rough estimate for  $a_5$ as a function of $N$,
\begin{align}
N\simeq-\int_{0}^{1}\frac{V(a_{1},a_{2},a_{3},a_{5})}{V'(a_{1},a_{2},a_{3},a_{5})}d\phi \simeq -\int_{0}^{1}\frac{d\phi}{V'(a_{1},a_{2},a_{3},a_{5})}.
\end{align}
This estimate is then used as a starting point to refine $a_5$ by solving the background equations iteratively thereby obtaining the accurate coefficient $a_5$ that yields the correct $N$. Thus a 4-dimensional parameter space $r_0$, $a_2$, $a_3$, $N$ is defined. The parameters  $a_{2},a_{3}$ are constrained by the requirement $|a_{2}|,|a_{3}|\ll 1$, $a_{1}$ is constrained by the observable value of $r$ and $a_{5}$ is determined by the other parameters and by the number of e-folds (taken to be in between $50\sim 60$).  The PPS considered is in the range of the first $\log(2500)\sim 8$ e-folds of inflation.

\subsection{Inflaton potentials with $r=0.01$}
\begin{figure}[!h]
\includegraphics[width=1\textwidth]{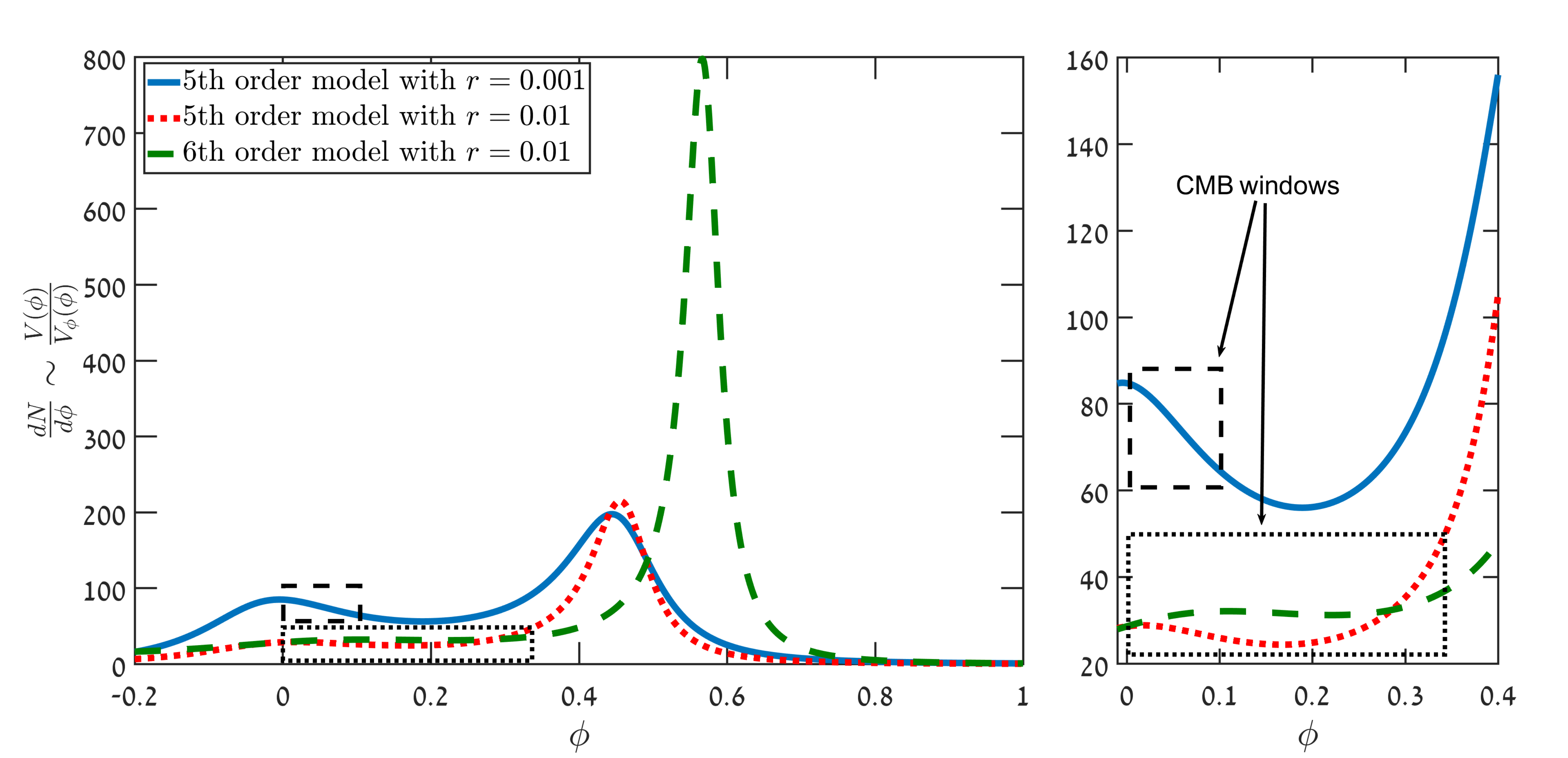}
\caption[{\bf Efold generation as $r$ changes}]{A graph depicting $-1/\sqrt{2\epsilon}\sim\frac{V}{V_{\phi}}$ as a function of the inflaton $\phi$ for two fifth-order polynomial models, and a sixth-order polynomial model. For a fifth-order polynomial model with $r=0.001$ (Blue line) the CMB window width is $\sim 8$ e-folds, while the field changes by about $\Delta\phi\sim 0.1$. Most of the e-folds are generated when $\phi$ reaches $\sim 0.4$.  When $r$ is increased the CMB window widens and approaches the e-fold generating peak (Red dots). While marginally affecting the CMB window width, the introduction of an additional coefficient, $a_6$, allows shifting the peak to higher values of $\phi$, thereby decoupling the CMB window and the e-fold generating peak (Green dash).}
\label{fig:CMBwindow}
\end{figure}

In \cite{Wolfson:2016vyx}, a class of small field inflationary models which can reproduce the currently measured CMB observables, while also generating an appreciable primordial GW signal was studied. The existence of such small field models provides a viable alternative to the large field models that generate a high Tensor-to-Scalar ratio. Our exact analysis was shown to give accurate results \cite{Wolfson:2016vyx}. Models which yield Tensor to Scalar ratio, of less than $r\lesssim 0.003$ were previously studied in \cite{Wolfson:2016vyx}. The initial study additionally demonstrated a significant difference between analytical Stewart-Lyth \cite{Stewart:1993bc,Lyth:1998xn} estimates and the exact results. This result should be confronted with analyses such as in \cite{Martin:2013tda} where the Stewart-Lyth expression is relied upon, and \cite{Dodelson:2001sh} in which the authors use a Green's function approach and perturbation theory, but assume the log of the input is well behaved. Our method extends and improves the method of the model building technique employed in \cite{BenDayan:2009kv,Hotchkiss:2011gz}. Previous analytical work \cite{Choudhury:2014kma,Chatterjee:2014hna,Choudhury:2015pqa} has shown that a fourth-order polynomial potential is sufficient to generate a high tensor-to-scalar ratio, even up to $r\gtrsim 0.1$. However, it was hard to realize this numerically. It was discovered in \cite{Wolfson:2016vyx}, that a fifth-order polynomial potential was required for generating $0.001\lesssim r\lesssim 0.003$. Furthermore a sixth-order polynomial seems to be required for a tensor-to-scalar ratio greater than $r\gtrsim 0.003$. A simple explanation is offered by observing (see Fig.~\ref{fig:CMBwindow}) that increasing $r$ by factor $\sim 10$, causes the e-folds per field excursion generated at the CMB window to decrease by a factor of $\sim 3$. This means widening the CMB window and losing the decoupling between the CMB window and the e-fold generating peak. Adding the 6th coefficient pushes the peak from $\phi\in[0.4,0.5]$, to higher values of $\phi$ and decouples these regions.
\subsection{Inflaton potentials with $r=0.03$}
We continue our investigations \cite{Wolfson:2016vyx,Wolfson:2018lel} of a class of inflationary models that were proposed by Ben-Dayan and Brustein  \cite{BenDayan:2009kv} and were followed by  \cite{Hotchkiss:2011gz,Antusch:2014cpa,Garcia-Bellido:2014wfa}. This class of models is compatible with several fundamental physics considerations. Recently, interest in this class of models was revived by the discussion about the ``swampland conjecture", \cite{Lehners:2018vgi,Garg:2018reu,Kehagias:2018uem,Ben-Dayan:2018mhe} which suggests that small field models are favoured by various string-theoretical considerations (see \cite{Palti:2019pca} for a recent review).

In addition, for this class of inflationary models, high values of $r$ result in a scale dependence of the scalar power spectrum. Future experiments such as Euclid \cite{Amendola:2012ys}, and SPHEREx \cite{Dore:2014cca} aim to probe the running of the scalar spectral index $\alpha_s$ at the level of $1\sigma \simeq 1\times 10^{-3}$. This is a major improvement  in comparison to the Planck bounds on $\alpha_s$ which are currently at the level of $1\sigma \simeq 17 \times 10^{-3}$. Such future measurements could provide additional constraints on our models.

\begin{figure}[!h]
\includegraphics[width=1\textwidth]{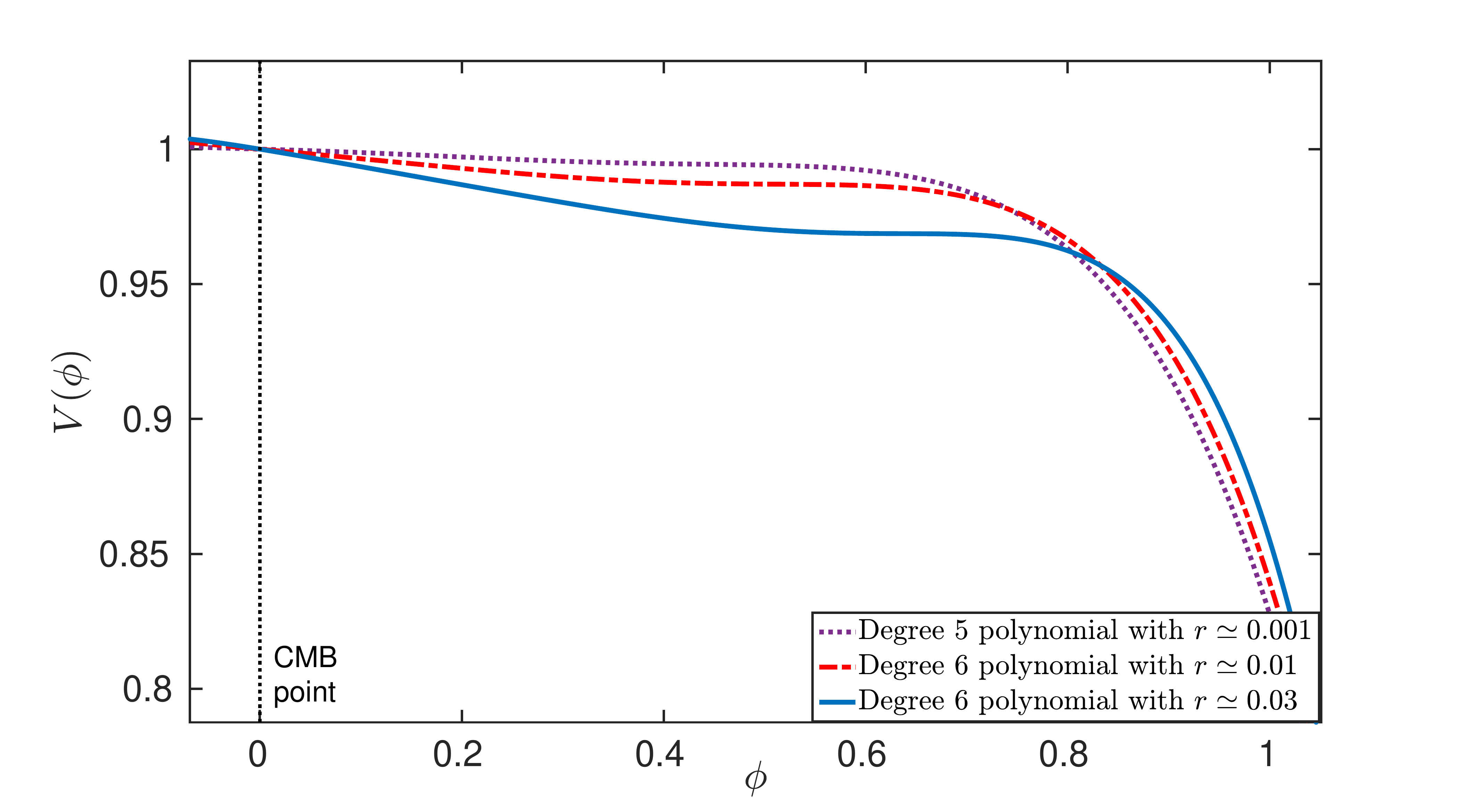}
\caption[{\bf Polynomial potentials}]{Polynomial potentials. The blue (solid) line depicts a potential of a model that predicts $r\simeq 0.03$, while the red (dash-dotted) line depicts a potential that predicts $r\simeq 0.01$. The purple (dotted) line depicts a degree five polynomial potential of a model that predicts $r\simeq0.001$. All models are variants of the hilltop model, with a  flatter region in which most e-folds are generated.
\label{potentials}}
\end{figure}

The small field models that we study are single-field models. The action of such models is given by
\begin{align}
	S=\int d^4 x \sqrt{-g}\left[\frac{R}{2}-\frac{1}{2}\partial^{\mu}\phi\partial_{\mu}\phi -V(\phi)\right].
\end{align}
The metric is of the FRW form and the potential given by
\begin{align}
	V(\phi)=V_0\left[1+\sum_{p=1}^{6}a_p\phi^p\right].
\end{align}
Previously, in \cite{BenDayan:2009kv,Wolfson:2016vyx} this class of models was discussed from a phenomenological and theoretical points of view. In \cite{Wolfson:2016vyx}, the technical details of model building and simulation methods were discussed, while in \cite{Wolfson:2018lel}, the analysis and the extraction of the most probable model were discussed. Additionally, in \cite{Wolfson:2018lel}, the most likely model which yields $r=0.01$ was identified.

\begin{figure}[!h]
\includegraphics[width=1\textwidth]{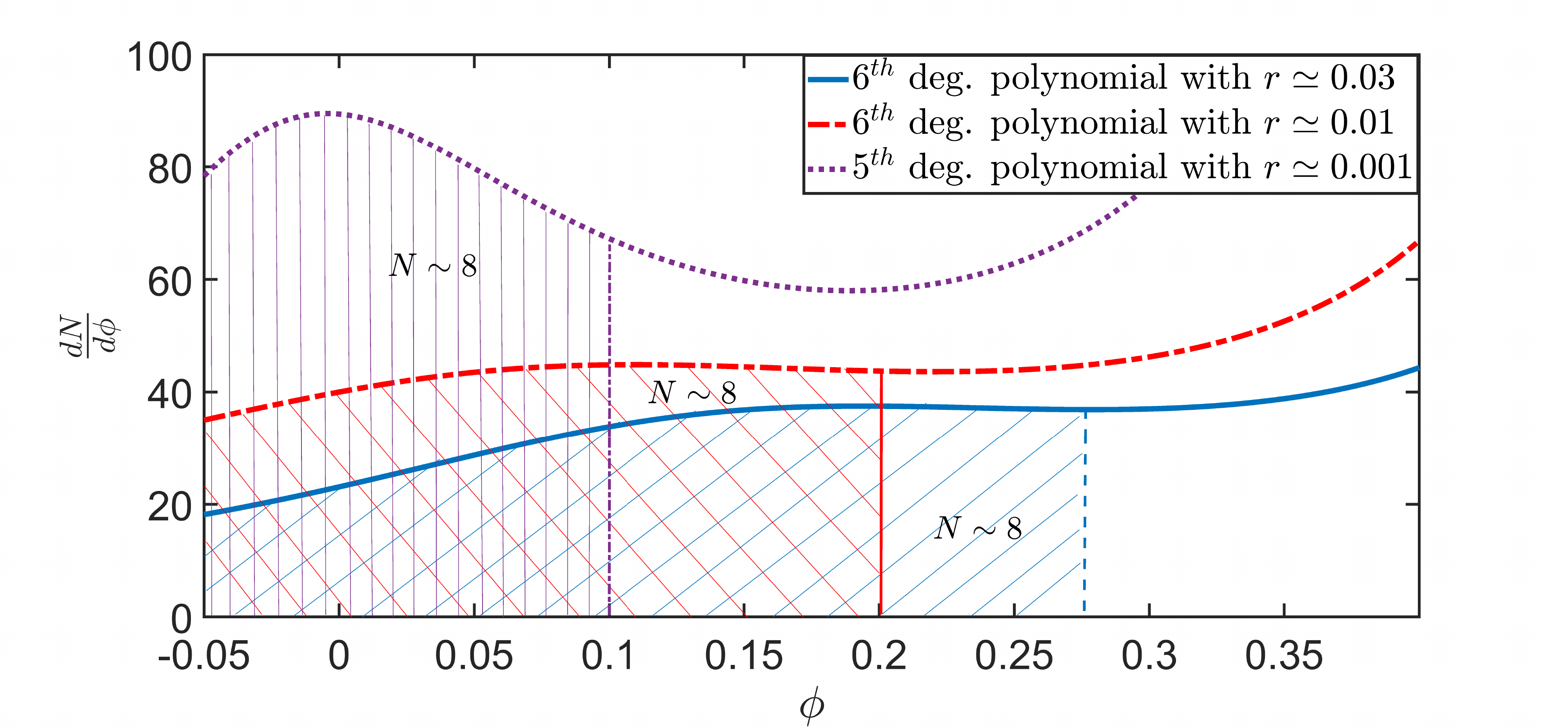}
\caption[{\bf Efold generation for models with $r\simeq 0.001,0.01,0.03$ }]{Field excursions in (reduced) Planck units. Different predicted values of $r$ require different field excursions to generate the $\sim 8$ e-folds probed by the CMB. The model predicting $r\simeq 0.03$ (blue line) requires an excursion of $\left(\Delta\phi\right)_{CMB}\simeq 0.28$ to generate the same amount of e-folds which the model predicting $r\simeq 0.01$ model (red dash-dot) generates in $\left(\Delta\phi\right)_{CMB}\simeq 0.2$. This means more tuning is required for models that predict $r\simeq 0.03$ . The model predicting $r\simeq0.001$  (purple dots) requires only $\left(\Delta\phi\right)_{CMB}\simeq 0.1$ to generate the CMB window. \label{dNdphi}}
\end{figure}

The small field models previously studied in \cite{Wolfson:2016vyx} yielded results that are consistent with observable data up to values of $r\simeq 0.003$. While these values agree with the current limits on $r$ set by Planck \cite{Ade:2015tva,Ade:2015xua}, we are interested in studying models with higher $r$. For models with $r\gtrsim 0.003$, significant running of running is found. This means that while three free parameters (corresponding to $n_{s},\alpha_s,N$) were previously needed, we now need an additional free parameter. Therefore we turn to a model of a degree six polynomial potential. Obviously considering higher degree models complicates the analysis by adding other tunable parameters.
The potential is given by the following polynomial:
\begin{align}
	V=V_{0}\left(1+\sum_{p=1}^{6}a_p \phi^p\right)\;.
\end{align}
It has been shown \cite{BenDayan:2009kv} that the potential can be written as:
\begin{align}
	V=V_0\left(1-\sqrt{\frac{r_0}{8}}\phi +\frac{\eta_0}{2}\phi^2 +\frac{\alpha_0}{3\sqrt{2 r_0}}\phi^3 +a_4\phi^4+a_5\phi^5+a_6\phi^6\right).
\end{align}
However, for simplicity, we express the potential as follows:
\begin{align}
	V=V_0\left(1-\sqrt{\frac{r_0}{8}}\phi +\sum_{p=2}^{6}a_p \phi^p\right)\;,
\end{align}
with the subscript $0$ denoting the value at the CMB point. By setting $\phi_0=0\; ; \;\phi_{end}=1$ we limit ourselves to small field models in which $\Delta\phi=1$ in Planck units, with little effect on CMB observables. 
According to the Lyth bound \cite{Lyth:1996im,Easther:2006qu}, given a Tensor-to-Scalar ratio of $r\simeq 0.01$, the lower bound on the field excursion is approximately given by $\Delta\phi_{4}\gtrsim 0.03\;m_{pl}$. Here $\Delta\phi_{4}$ is the field excursion while the first $\sim 4$ efolds are generated. Our models satisfy this strict bound, as the first 4 efolds or so typically result in $\Delta\phi_4 \sim 0.15$ which is well above $0.03$. The Lyth bound was further extrapolated \cite{Efstathiou:2005tq} to cover the entire inflationary period. Applying this approach to models with $r\sim 0.01$ yields $\Delta\phi\simeq 2\; m_{pl}$. However, in \cite{Hotchkiss:2011gz}, it was shown that in models such as the ones we study, the value of $\Delta\phi$ can be smaller because $\epsilon_H$ is non-monotonic. In this case, $\Delta\phi=1\; m_{pl}$ from the CMB point to the end of inflation is consistent with the Lyth bound.

When the coefficients $\{r_0,a_2,a_3,a_4\}$ are fixed, the remaining coefficients are related by:
\begin{align}
	a_5 =f_1(r_0,a_2,a_3,a_4,a_6),\\
	a_6 = f_2(r_0,a_2,a_3,a_4,N).
\end{align}
The procedure of finding $f_1$ and $f_2$ was explained in detail for the degree 5 polynomial models in \cite{Wolfson:2016vyx}, and here we follow a similar procedure for the degree 6 models. So, ultimately, the model is parametrized by 5 parameters: the two physical parameters $r_0$ and $N$ and the three other parameters ($a_2,a_3,a_4$) that are used to parametrize the $n_{s},\alpha_s,\beta_s$ parameter space. It should be pointed out that $N$ is not an observable, rather $N\sim 50-60$ is a 'soft' constraint. Strictly speaking, $N$ depends on the reheating temperature and only its maximum value can be determined. However, for simplicity, we treat $N$ as an observable, in order to facilitate the study of a large sample of models.

\clearpage


\graphicspath{{Chapter2_75/}}

\chapter{Methods}
\subsection{Coefficient extraction methods\label{methods}}
In this section, we explain the two methods for calculating the most likely coefficients $\{a_2,a_3,a_4\}$, given a large number of simulated models and the likelihood data for the CMB observables. This data is available through CosmoMC \cite{Lewis:2002ah} analysis of CMB data, such as the Planck data \cite{Ade:2015tva}.
\begin{figure}[!h]
\includegraphics[width=1\textwidth]{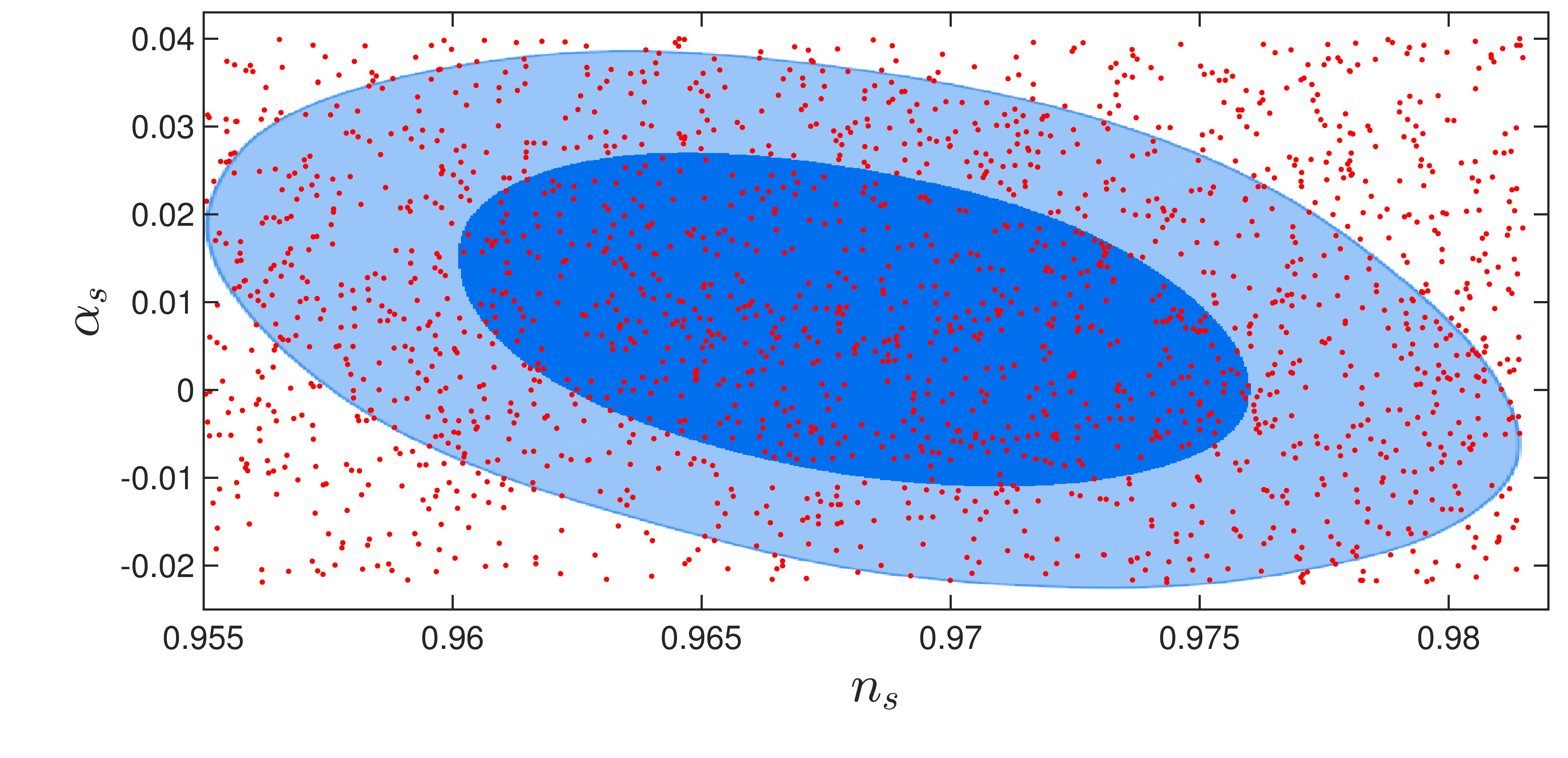}
\caption[{\bf Covering $n_s-\alpha_s$ with $r=0.01$ models.}]{Small field inflationary potentials which yield $r=0.01$, as well as PPS observables within 68\% and 99\% confidence levels. Every $(n_s,\alpha_s)$ pair is accessible using these models. The likelihood curves are results of a CosmoMC \cite{Lewis:2002ah} run with the latest BICEP2+Planck \cite{Ade:2015tva} data. \label{main_001} }
\end{figure}
\subsubsection{Likelihood assignment method - Gaussian extraction\label{Probability}}
To each potential, after calculating the observables $n_{s},\alpha_s,\beta_s$, we assign a likelihood. For each observable, we calculate the likelihood according to the MCMC likelihood analysis of the data sets used. We then assign the product of the likelihoods $L_{(n_s)}\times L_{(\alpha_s)}\times L_{(\beta_s)}$ to the potential. A concrete example is the following:
suppose we extract the trio $(n_{s},\alpha_s,\beta_s)=(0.96,0.011,0.024)$, we look up the likelihoods: $(L_{\left(n_{s}=0.96\right)},L_{\left(\alpha_s=0.011\right)},L_{\left(\beta_s=0.024\right)})$. We now multiply them, and so the likelihood attached to that specific model which yielded these observables is given by $
L_{potential}=L_{\left(n_{s}=0.96\right)}\times L_{\left(\alpha_s=0.011\right)}\times L_{\left(\beta_s=0.024\right)}\;.$ We proceed to extract likelihoods for the different coefficients by process of marginalization. The expectation is that this method will yield a (roughly) Gaussian distribution for each of the values of $a_2,a_3,a_4$. The advantage of this method is in yielding not only the most likely value but also the width of the Gaussian. This width can then be used as an indication for the level of tuning that is needed in these models.
\begin{figure}[!h]
\includegraphics[width=1\textwidth]{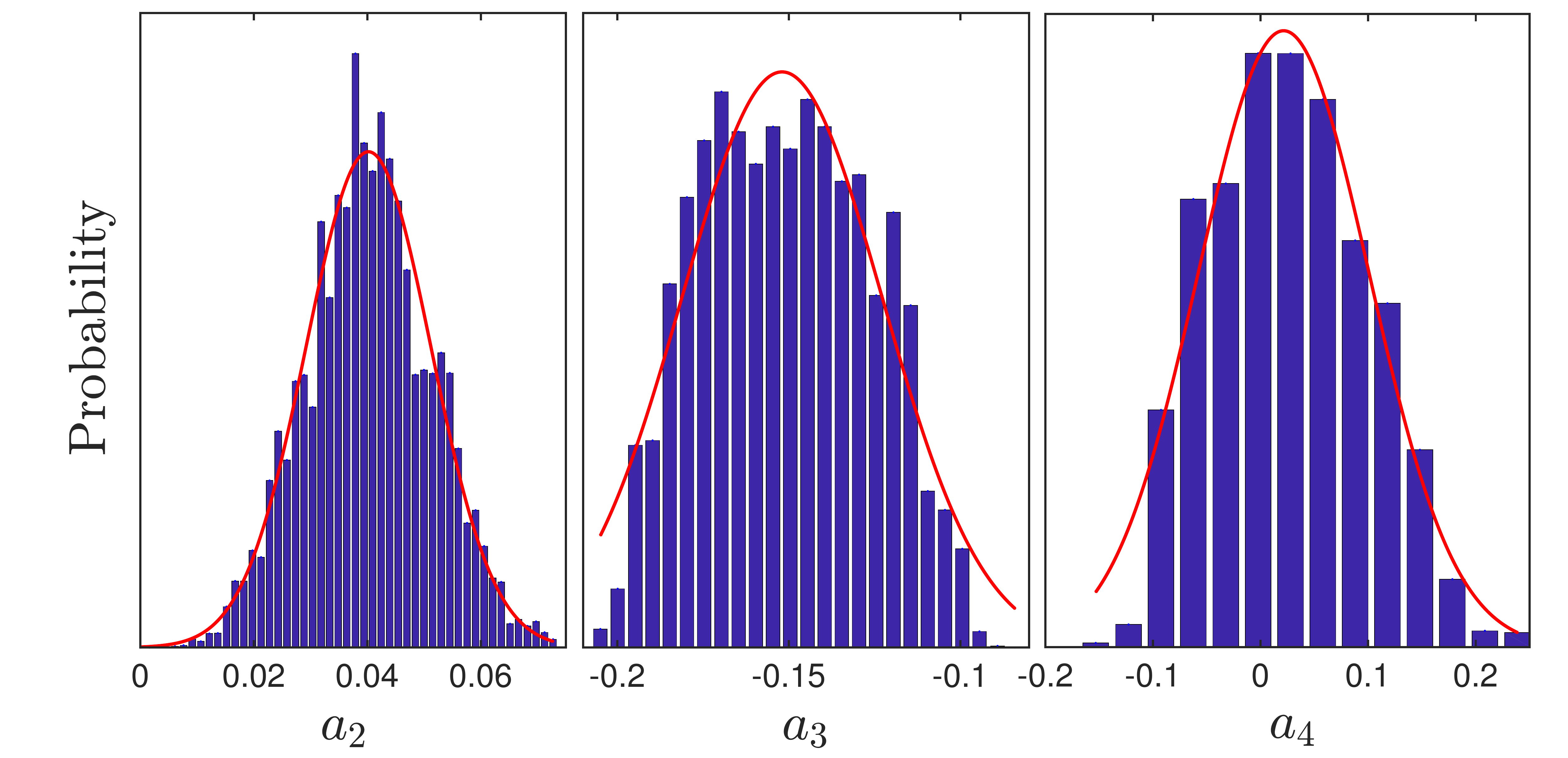}
\caption[{\bf Gaussian fitting of coefficients}]{The calculated likelihoods for the coefficients $\{a_2$,$a_3$,$a_4\}$, in models with $r=0.01$. The most likely coefficients are given by: $a_2=0.04$, $a_3=-0.15$, $a_4=0.02$. The tuning level for each coefficient is given by Barbieri-Giudice measure \cite{Barbieri:1987fn} and is $(0.375,0.27,5.5)$. \label{Prob_001} }
\end{figure}
\subsubsection{Possible pitfalls}
This method of likelihood assignment is vulnerable in two ways:
\begin{itemize}
	\setlength{\topsep}{0pt}
    \setlength{\itemsep}{0pt}
    \setlength{\parskip}{0pt}
    \setlength{\parsep}{0pt}
\item[a)] To be valid, this method requires a uniform cover of the relevant parameter space, by the potential parameters. If the cover significantly deviates from uniform, the results might be skewed by overweighting areas of negligible likelihood, or underweighting areas of significant likelihood. Fig. \ref{main_001} shows a mostly uniform cover.
	\item[b)] Since $ L_{n_s,\alpha_s,\beta_s} \simeq L_{(n_s)}\times L_{(\alpha_s)}\times L_{(\beta_s)}$ only if the paired covariance is small, we must make sure that this is the case. In our underlying MCMC analysis this is indeed the case. The covariance terms are, in general, one to two orders of magnitude smaller than the likelihoods at the tails of the Gaussian.
	\item[c)] We also run the risk of false results if the fit we apply to the data points produced by the numerical analysis yields a large fitting error. However, the fitting error of the polynomial function to the $\log PPS - \log k$ data is usually of the order of $10^{-6}$. This fitting is done over 30 data points generated by the MS equation numerical evaluation, for each potential. The error is calculated as $\Delta=\sqrt{\sum_1^{30}\left((\log PPS)_i-\mathrm{fit}((\log k)_i)\right)^2}$, thus the error per data point is of the order of $10^{-6\sim 7}$. We conclude that the $\log PPS$ function is well fitted.
\end{itemize}
\subsubsection{Multinomial fit\label{multi}}
Another method for calculating the most likely coefficients is by fitting the simulated data with a multinomial function of the CMB observables. We aim to find a set of functions $F_i$ such that, for example, $a_2=F_2(n_{s},\alpha_s,\beta_s)$. We assume that this function is smooth and thus can be expanded in the vicinity of the most likely CMB observables. Hence, we can find a set of multinomials $(F_2,F_3,F_4)$, such that:
\begin{align}
	\begin{array}{ccc}
	a_2&=& F_2(n_{s},\alpha_s,\beta_s)\\
	a_3&=& F_3(n_{s},\alpha_s,\beta_s)\\
	a_4&=& F_4(n_{s},\alpha_s,\beta_s).
	\end{array}
\end{align}
We have found that a quadratic multinomial is sufficiently accurate and that using a higher degree multinomial does not improve the accuracy significantly. Thus we may represent these by a symmetric bilinear form  plus a linear term, as follows:
\begin{align}
	F_i=OB_iO^{\dag} + A_iO^{\dag} +p_{0,i}\; ,
\end{align}
where $O=(n_{s},\alpha_s,\beta_s)$, $B_i$ is the bilinear matrix, and the linear coefficient vector is $A_i$.
\begin{figure}[!h]
\includegraphics[width=1\textwidth]{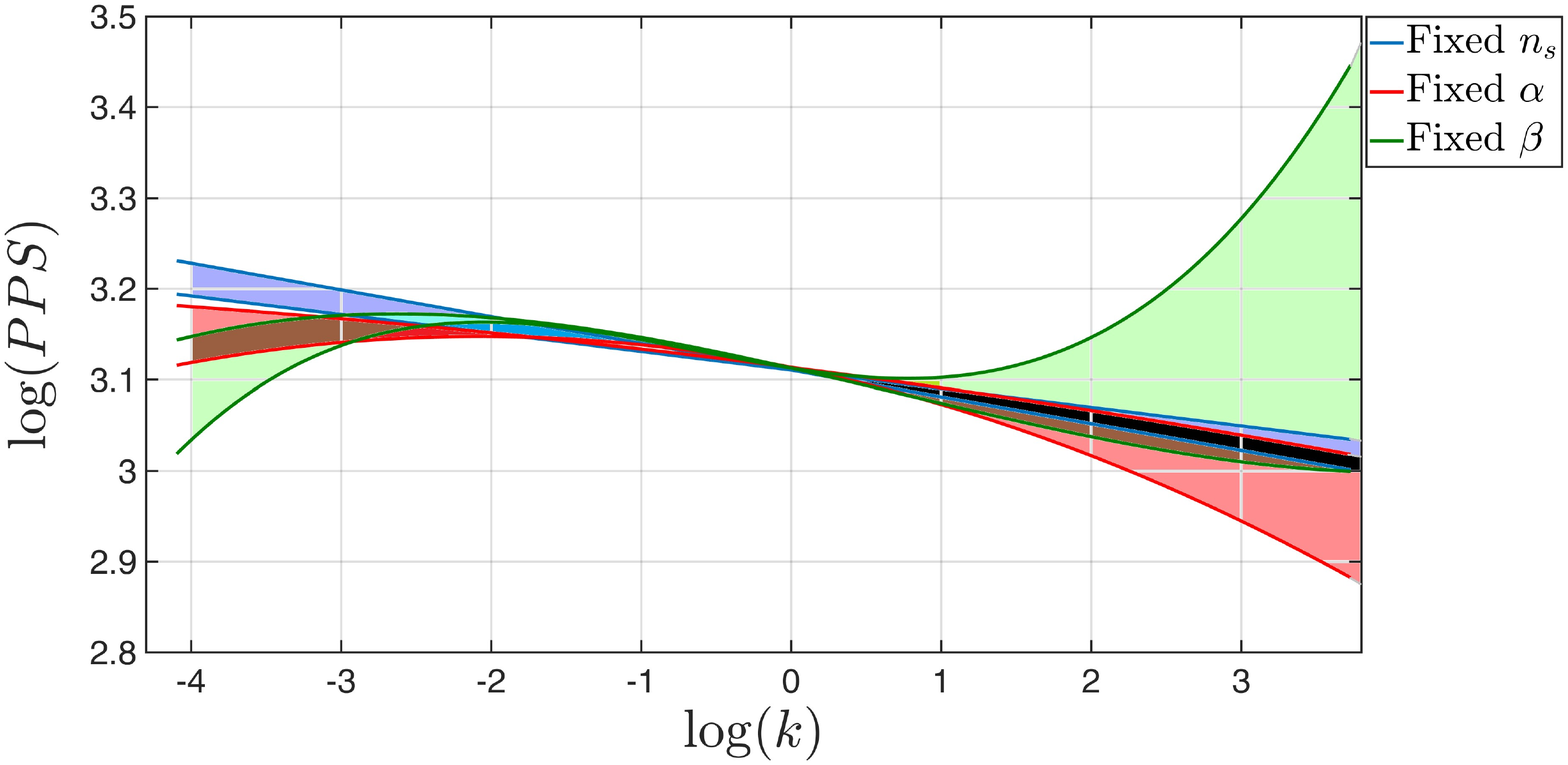}
\caption[{\bf Power spectra in $k$-space}]{Power spectra as recovered using CosmoMC \cite{Lewis:2002ah} analysis with latest BICEP2+Planck data \cite{Ade:2015tva}. Allowed area (68\% CL) for a fixed $n_s$ analysis is shown (blue). Similarly, a fixed $\alpha_s$ analysis (red), and  a fixed $\beta_s$ (green) are shown. The other colors are intersection areas. The pivot scale in this graph is at $\log\left(\frac{k}{k_0}\right)=0$, where $k_0=0.05\;h Mpc^{-1}$. The apparent divergence in high k's is due to the inability of Planck to constrain these k's. This is also shown in Fig. \ref{Power_Spectra_l}. With more data, it will be possible to differentiate between the three possibilities.  \label{POWER_SPECTRA}}
\end{figure}
\subsubsection{Pivot scale}
\begin{figure}[!h]
\hspace*{-2.3cm}
\includegraphics[height=14cm, width=1.65\textwidth]{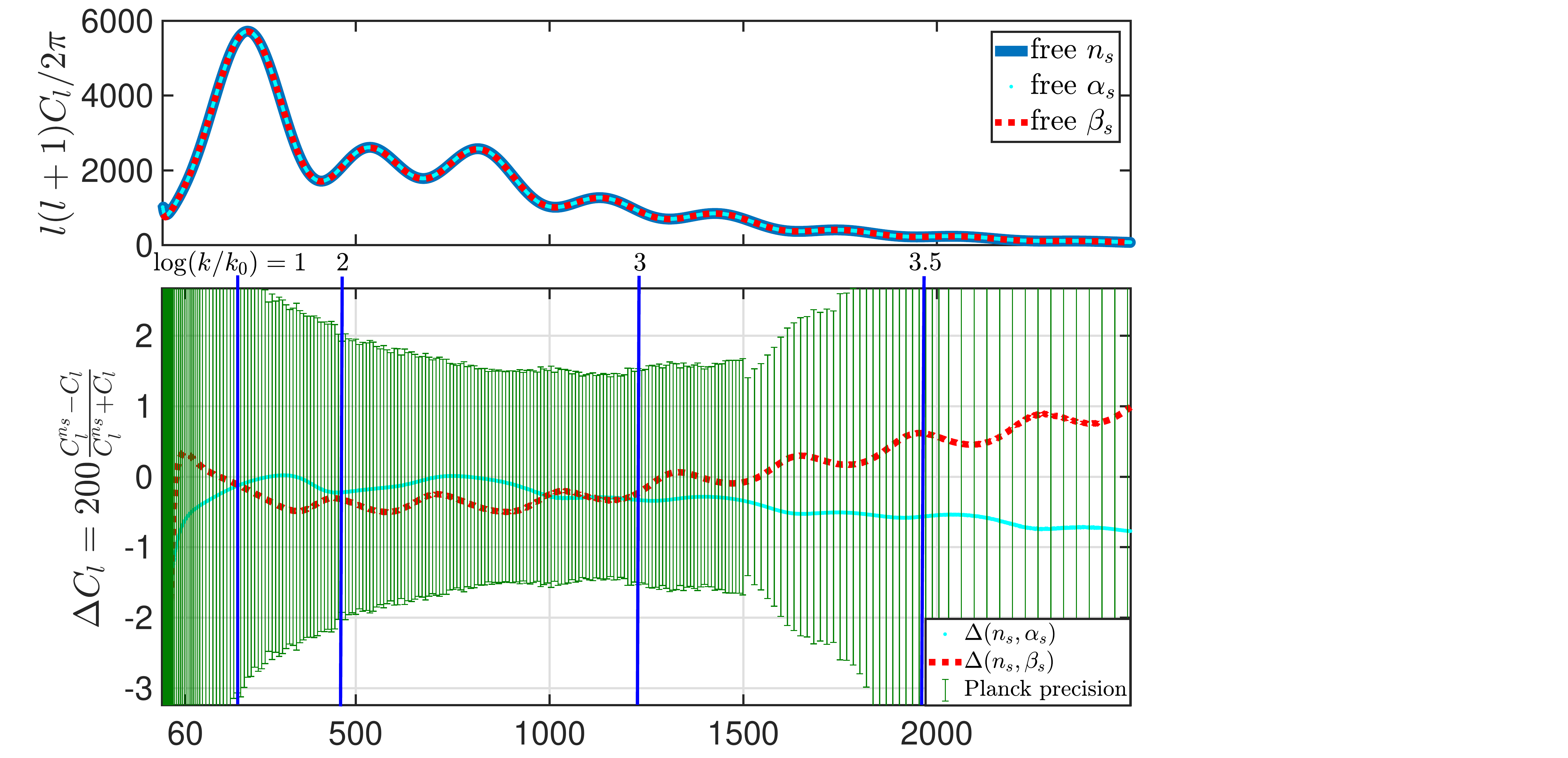}
\caption[{\bf Power spectra in $C_l$ representation.}]{Power spectra in the $C_l$'s decomposition (upper panel), with a free $n_s$ (thick blue line), free $\alpha_s$ (thin cyan dots), and free $\beta_s$ (medium red dash). The lower panel shows the relative difference between the different cases. The relative difference (lower panel) is bound from above by $\sim 1\%$. Additionally, the Planck observation error bars are shown.\label{Power_Spectra_l} }
\end{figure}
So far, we discussed matching potentials and their resulting PPS around the CMB point. However, in order to correctly compare the results of the PPS to observables, one has to take into account the pivot scale at which the CMB observables are defined. Since, in this case, the pivot scale is given by $k_0=0.05 \;h Mpc^{-1}$, and the CMB point is at $k\sim 10^{-4}\; hMpc^{-1}$ , the observables in the CMB point and $k_0$ should be related in a simple way only if the spectrum varies slowly with $k$. This is not true for the case at hand.  Two potentials can yield very different power spectra near the CMB point, and nevertheless yield the same observables at the pivot scale. These degeneracies, stem from our limited knowledge of the power spectra on small scales, and at the CMB point. For concreteness take two PPS functions, one that is well approximated by a cubic fit near the pivot scale, and the other that is well approximated only when we consider a quartic fit. Suppose, additionally, that these two PPS functions have the exact same first three coefficients, it follows that they yield the exact same observables $\{n_s,\alpha_s,\beta_s\}$. However, if we go to sufficiently small scales or high enough $k$ values, these functions will diverge. This is also true at the large scale end, where the CMB point is set. Hence the degeneracy.\\
A possible solution to this problem is classifying the resulting power spectra by the level of minimal good fit. We define a good fit as one in which the cumulative relative error $\Delta=\sqrt{\sum_k \left(\log(PPS(\log(k)))-fit(\log(k))\right)^2}$, is less than $10^{-7}$.  Given a single power spectrum, we fit our result with a polynomial fit, increasing in order until the accumulated relative error is sufficiently small. The minimal degree polynomial fit that approximates the $\log(PPS)$ function to the aforementioned accuracy is called the minimal good fit. We then study separately power spectra that are well fitted by cubic polynomials, quartic polynomials etc. In this way we make sure that we compare non-degenerate cases.
\subsection{Monte Carlo analysis of Cosmic Microwave Background with running of running}
In \cite{Cabass:2016ldu} it was shown that the inclusion of additional parameters, i.e., the running of the spectral index ($\alpha_s$), and the running of the running ($\beta_s$) resolves much of the tension between different data sets. In this section, we briefly discuss the effect of considering non-vanishing $\alpha_s$ and $\beta_s$ on the most likely shape of the PPS. First, we find $n_s$ when it is the only free parameter. We then use $n_s$, and $\alpha_s$ as the free parameters, and finally we conduct an analysis with $n_s,\alpha_s$ and $\beta_s$ as the free parameters. The shape of the power spectrum changes significantly when running of running is considered.\\
The data sets that were used are the latest BICEP2+Planck baseline \cite{Ade:2015tva}, along with the low $l$'s \cite{Bennett:2012zja}, low TEB and lensing likelihoods.  The results of these analyses are given in Table \ref{Table_cosmomc}, as well as in Fig. \ref{POWER_SPECTRA}. As expected the resulting power spectra converge at the pivot scale $k_0=0.05\;h Mpc^{-1}$. However, for lower $k$'s, the resulting spectra diverge considerably, consistent with cosmic variance. Notably, the spectra also diverge at higher $k$'s. This indicates the inability of current observational data to constrain  the models in this range of $k$'s. This inability is also demonstrated in Fig. \ref{Power_Spectra_l} where, for $l>1500$, the most restrictive data cannot rule out models with significant running, or running of running. Figure \ref{Power_Spectra_l} also shows that the three models are virtually indistinguishable in terms of the observed $C_l$'s.\\
The conclusion is that we will need additional accurate data from smaller cosmic scales to be able to differentiate between the three scenarios. These extra e-folds might come from future missions such as Euclid \cite{Amendola:2016saw}, or $\mu$-type distortion data \cite{Diacoumis:2017hhq,Abitbol:2017vwa}.
\begin{table}
\begin{center}
\begin{tabular}{|c||c|c|c|}
\hline
Parameter (68\%)& free $n_s$ & free $\alpha_s$&free $\beta_s$\\
\hline
$\log(10^{10}A_s)$&$3.1047\pm0.0057$ &$3.1073\pm0.006$  &$3.1061\pm0.0065 $\\
$n_s$&$0.9751\pm0.0045$&$0.973\pm 0.0057$  &$0.9687^{+0.0051}_{-0.006}$\\
$\alpha_s$& N/A&$-0.009\pm0.0067$  &$0.008\pm 0.013$\\
$\beta_s$& N/A & N/A&$0.020\pm 0.013$\\
\hline
\end{tabular}
\end{center}
\caption[{\bf MCMC results}]{Results from 3 analyses of the latest BICEP2+Planck dataset, each adding a free parameter in the power spectrum. The results shown are best fits, within the 68\% confidence level for each analysis. \label{Table_cosmomc}}
\end{table}

\clearpage


\graphicspath{{Chapter3/}}

\chapter{The INSANE code}
In order to assess the primordial power spectrum given an inflationary potential, a stand-alone simulator was built. The code was given the name INflationary potential Simulator and ANalysis Engine (INSANE) and is a fully numerical simulator that solves the background and MS equations fully and precisely, for a wide variety of symbolic potentials. Of course this is by no means the first foray into numerical cosmology. Salopek, Bond and Bardeen \cite{Salopek:1988qh} have calculated power spectra resulting from different potentials as early as 1989. Adams, Cresswell and Easther \cite{Adams:2001vc} have studied PPS responses to features in the potential, while Peiris et. al. have utilized such codes to analyse the first-year results from WMAP mission \cite{Peiris:2003ff}. Mortonson, Dvorkin, Peiris and Hu \cite{Mortonson:2009qv} have also utilized such codes to study features of inflation. There are of course many others who have solved such inflationary systems numerically, some have published their codes (e.g. for example \cite{Price:2014xpa} and \cite{Ringeval:2005yn,Martin:2006rs,Ringeval:2007am}). However the code presented here differs in two main features:
\begin{itemize}
	\item[(a)] Most if not all currently existing codes use the parameter flow equations, first introduced by Kinney \cite{Kinney:2002qn}, that replace time as the underlying quantity over which integration is performed, with the number of efolds. Since $N=\int H dt$, it follows that $\frac{d}{dt}=H\frac{d}{dN}$ thus, the equations of motion are now given by:
	\begin{align}
		\left\{\begin{array}{rcl}
			\frac{d^2\phi}{dN^2}&=&\left(\frac{1}{2}\left(\frac{d\phi}{dN}\right)^2 -3\right)\frac{d\phi}{N} -\frac{1}{H^2 }\frac{dV}{d\phi}\\
			\frac{dH}{dN}&=&\frac{-H}{2}\left(\frac{d\phi}{dN}\right)^2 .
		\end{array}\right. 
	\end{align}
		While this formulation does wonders to hasten the integration process, it is our experience that this flow formulation, while analytically sound, results in information loss in the computational implementation. For instance in a tilted natural inflation scheme, near the limit of tilting the model such that $\epsilon=1$ is accessible, there is a marked disagreement between the results of integration with the two formulations. This can be seen in figure \ref{figure_1}.\\
\begin{figure}[!h]
	\includegraphics[width=0.8\textwidth]{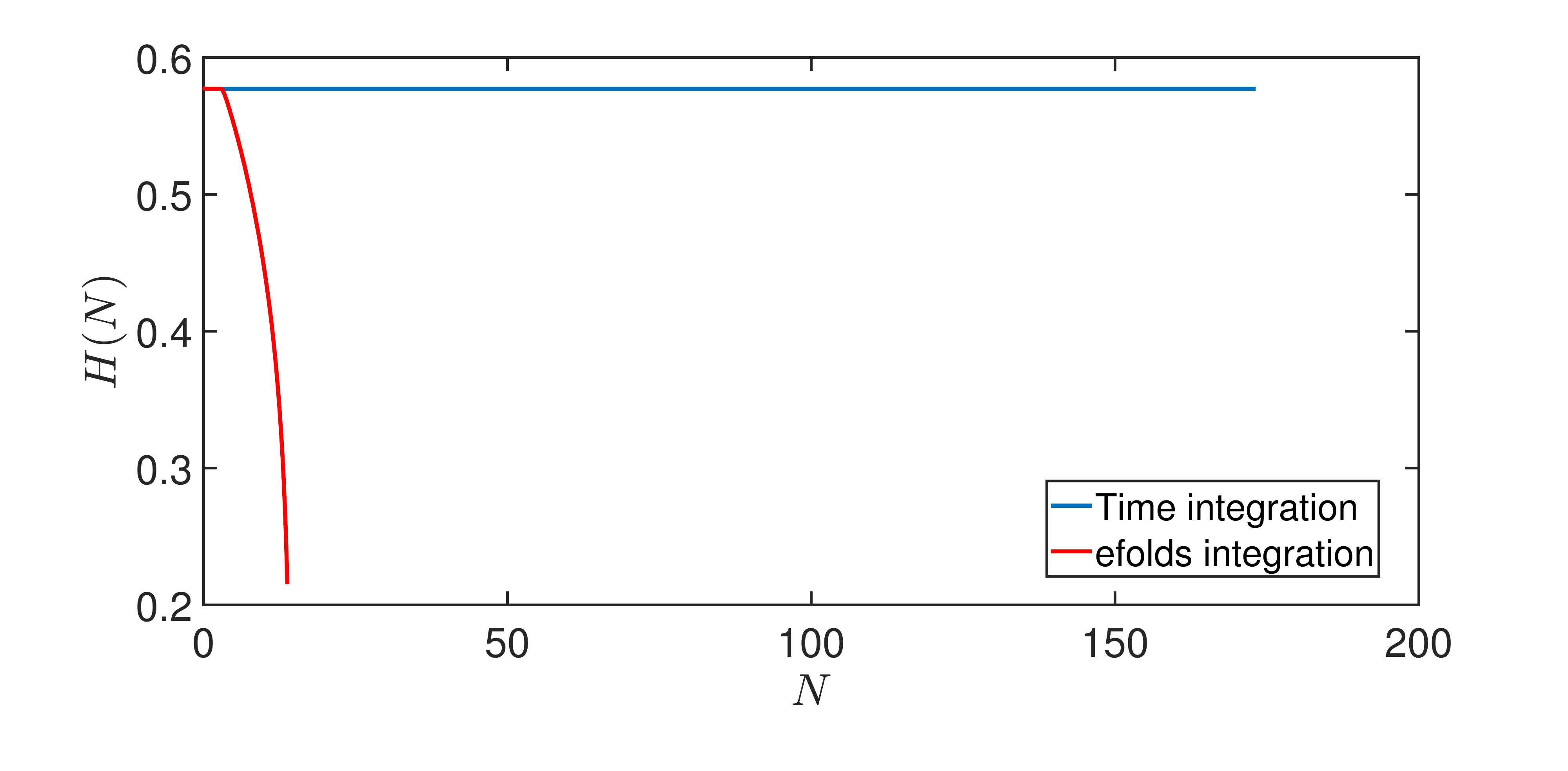}
	\caption[{\bf Difference between cosmic time and efold based integration}]{While analytic integration over efolds, cosmic or conformal time are equivalent, integration over conformal time is nigh impossible due to tendancy toward an 'exploding' solution. Numerical integration over efolds or cosmic time usually give near identical results, but in some cases, as shown above they diverge considerably. \label{figure_1}}
\end{figure}
To be perfectly clear, we do not state with absolute confidence that one is better than the other, simply that they may differ. Since neither cosmic time nor efold number are strictly observables, and we are unable to run an experiment to validate one version over the other it is hard to prefer one over the other. However we contend that this does merit some additional discussion within the computational cosmology community. Further differences, especially in potentials which sport a plateau, were discussed by Coone, Roest, and Vennin \cite{Coone:2015fha}.
\item[(b)] At the time of writing the INSANE package, no computing language had both an embedded symbolic math capability as well as good numerical solvers (the one exception to this is Mathematica which was, at the time, opaque to `under-the-hood' scrutiny). Thus by writing this package using Python, we were able to include fully symbolic parsers.
\end{itemize}
\section{Code architecture}
The code is organized into two main components each with its routines and sub-routines. The chosen computing language was Python, since at the time it was the only language to support both numerical integration and symbolic math, along with an object-oriented-programming (OOP) paradigm. The OOP compatibility was paramount for the construction of code that is reusable and can accept any inflationary potential. Initially, the plan was to have the code accept any physical action along with any underlying metric, construct the appropriate Friedmann equations and then develop and solve the background and MS equations to yield the full PPS. However, since we decided on studying small field scalar potential models, this was deemed a wild overshoot and was relegated to possible future projects. Additionally, the inclusion of a Boltzman code, pyCAMB \cite{Lewis:1999bs}, and a sky-map realization, HEALpy \cite{Gorski:2004by}, is a recent implementation and as such have yet to be fully debugged.
\subsection{Background evolution}
This part of the code is given several parameters, for the evaluation of the background geometry given an inflationary potential.
The equations of background evolution are given by the Friedmann equation, along with the Klein-Gordon for the inflaton field:
\begin{align}
\left\{\begin{array}{ccc}
	\dot{H}&=&-\frac{\dot{\phi}^2}{2}\\
	&&\\
	\ddot{\phi}&=&-3H\dot{\phi}-\frac{dV(\phi)}{d\phi}
\end{array}\right.
\end{align}
It is implicitly assumed that we are dealing with a FRW metric such that the scale factor $a(t)$ can be gleaned directly from knowing $H(t)\equiv\frac{\dot{a}}{a}$.
The parameters supplied are the following:
\begin{table}[!h]
	\renewcommand{\arraystretch}{1.2}
	\centering
	\begin{tabular}{||l|c|c|l||}
	\hline
	\hline
	Param.&Type&Default value& Description\\
	\hline
	\hline
	\code{V0}&\code{double}&$1$&$V_0$ the scalar potential at $\phi=0$\\
	\hline
	\code{H0}&\code{double}&$\sqrt{\frac{V_0}{3}}$&$H_0$, Initial Hubble parameter\\
	\hline
	\code{phi0}&\code{double}&$0$&Initial $\phi$ value\\
	\hline
	\code{phidot0}&\code{double}&$0$&Initial $\dot{\phi}$ value\\
	\hline
	\code{efolds}&\code{double}&$\log{2500}$& Number of efolds to simulate \\ 
	& & & in the CMB window\\
	\hline
	\code{efoldsAgo}&\code{double}& 60 & Number of efolds in the simulated \\
	& & & inflation, minimum is 50\\
	\hline
	\code{Tprecision}&\code{double}& 0.01 & Precision of time integration\\
	 & & &for the Background geometry solver\\
	\hline
	\end{tabular}
	\renewcommand{\arraystretch}{1}
	\caption[{\bf Background evolution parameters}]{Possible parameters to input into the code, for background evolution simulation.}
\end{table}
\subsubsection{Potential}
The inflationary potential that is supplied to the simulator can be in the form of a single variable potential, for instance, $V(\phi)=V_0\cos\left(c\phi\right)$ with some set of predefined $V_0,c$, or a set of numerical coefficients $\{V_0,a_0,a_1...a_n\}$ such that the potential is a finite polynomial representation of some function:
\begin{align}
	V(\phi)=V_0\sum_{p=0}^{n} a_p\phi^p.
\end{align}
Currently, the code accepts polynomials up to degree 5, degree 6 and more polynomials should be inserted as symbolic functions.
The code then parses this expression, and decides whether to manufacture a lookup table as a stand-in for the potential at different values of $\phi$, or use the potential as is, in the case of a polynomial representation.
\subsubsection{Background Geometry}
The Background geometry solver class receives the potential, parses it, and integrates the Friedmann equations of the zero-order scalar field $\phi^{(0)}(t)$. The design choice was to use the cosmic time equations, since the conformal solution explodes to infinity faster than exponentially, close to the end of inflation, which makes the integration step go down to zero. Thus using an integration scheme that can handle exponential integration over a large stretch of time was called for. The equations we integrate are given by:
\begin{verbatim}
H_rhs=(-1/float(2*Mpl**2))*phidot**2 
phi_rhs=phidot  
phidot_rhs=-3*H*phidot -vd  ,
\end{verbatim}
where the \code{rhs} suffix means this is the coordinate derivative such that \code{phi\_rhs} means $\dot{\phi}$, and \code{vd} is the numerical equivalent of $dV(\phi)/d\phi$.\\
In order to solve the background equations, initial values must be supplied. The \code{BackgroundGeometry} initiator method \code{\_\_init\_\_} takes the following arguments:
\begin{verbatim}	
(l1=0,l2=0,l3=0,l4=0,l5=0,Hubble0=None,Phi0=None,PhiEnd=1,
Phidot0=None,Tprecision=0.01,endTime=200,Planck_Mass=1,
physical=True,selfCheck=False,poly=True,
Pot=(sym.Symbol('x'))**0,mode='silent') ,
\end{verbatim}
where \code{l1} through \code{l5} are a degree 5 polynomial coefficients such that $V(\phi)=V_0\left(1+\sum_{p=0}^5 l_p\phi^p\right)$. \code{Hubble0,Phi0,Phidot0} are initial values for $H_0,\phi_0,\dot{\phi}_0$ correspondingly. We can set the integration precision as well as several other quantities. In general in the Python syntax, where an equality sign appears, this is the default value, such that \code{Tprecision=0.01} means the integration step is $0.1$ `seconds', unless otherwise specified.
\subsection{Cosmic Perturbation}
As outlined in \ref{sec:scalar_perturbations}, the equations for modes of scalar perturbation are given by the set of MS equations \eqref{eq:MS_k-space}, one for each $k$-mode. In order to solve these, we first need to construct the pump field \eqref{eq:Pump_field}. The Background solver contains, among others, the fields in Table \ref{table:Background_Solver}, which are used to construct the pump field $z$. One important note - the field \code{a} in the \code{BackgrounGeometry} class denotes $\log{a(t)}$ instead of $a(t)$, and needs to be exponentiated before construction of $z$.
\begin{table}[!h]
\begin{center}
	\begin{tabular}{||l|c|c||}
	\hline
		Class field& type & Description\\
	\hline
	\hline
		\code{BG.a}      & \code{ndarray(float128)}& Instead of $a(t)$, this is $\log{a(t)}$ \\
					    &						 & or equivalently the number of efolds $N$\\
					    &						 & from the start of inflation.\\
	\hline
		\code{BG.phi}	&\code{ndarray(float128)}& $\phi(t)$\\
	\hline
		\code{BG.phidot}	&\code{ndarray(float128)}& $\dot{\phi}(t)$\\								
	\hline	
		\code{BG.H}	&\code{ndarray(float128)}& $H(t)$\\
	\hline
	\end{tabular}
\end{center}
\caption[{\bf Background evolution output}]{Some of the outputs of the \code{BackgrondGeometry} class, which are used to construct the pump field $z=\frac{a\dot{\phi}}{H}$. The instance of \code{BackgrondGeometry} in this table is  \code{BG}. \label{table:Background_Solver}}
\end{table}
After constructing the pump field it is possible now to solve the MS equation for each mode.
\subsubsection{The Mukhanov-Sasaki solver}
The class responsible for solving the MS equations for each mode, and construct the PPS is \code{MsSolver}.
In general, the class is fed the quantities $a,H,\phi,\dot{\phi},t$ at initiation as well as other parameters. \code{MsSolver} then splines everything, finds the analytic expressions for the observables in several approximations (Eqs.~(\ref{eq:SLR},\ref{eq:Analytical_alpha},\ref{eq:Dodelson_Stewart_ns},\ref{eq:n_s-Stewart-Gong}, etc.)), and finds the CMB point by calculating a user-specified number of efolds back from the end of inflation. It then finds the $k$ corresponding to the CMB point. This is done by finding the pump field value at $t=t_{CMB}$, by setting the time-dependent frequency in Eq.~\eqref{eq:TD_frequency} $\omega_k^2(t)=0$. we thus find the $k$ number which is just now entering the horizon and call it $k_0$. This corresponds to the scale of the universe and is, therefore, the physical $k=1.2\cdot 10^{-4}\;\mathrm{h\;Mpc^{-1}}$. If we are interested in physical inflation, this scale will be the default pivot scale, unless some other scale was set as the pivot scale. 
~ A preparatory phase is immediately initiated after the class is built, in which an interpolation table is created for the pump field and all quantities needed for a solution of the MS equation to arbitrary precision. The precision is predefined by the user, usually due to memory and running time considerations vs. precision demands. 
The solver then proceeds to integrate the MS equation for each $k$-mode, build a two variable function $U(k,t)$ which is  a stack of eigenfunctions $u_k(t)$ along the $k$ axis. The solver finally builds the quantity 
\begin{align}
	P_s(k,t)= \frac{k^3}{2\pi^2} \left|\frac{U(k,t)}{z(t)}\right|^2,
\end{align}
and evaluates this function at a time $t_{ps}$ much later than the freeze-out time of the last mode of interest, to yield:
\begin{align}
	P_s(k)=\frac{k^3}{2\pi^2} \left|\frac{U(k,t_{ps})}{z(t_{ps})}\right|^2\; .
\end{align}
This concludes the calculation phase, after which we have an array of $k$-numbers, corresponding to the $k$-modes, and the PPS as a function of these $k$-numbers.
The analysis phase now begins, in which we assign the pivot-scale specified. We first reassign $k_0$ to the correct physical scale:
\begin{verbatim}
	k=(k/k0)*1.2*10**(-4) .
\end{verbatim}
If the user specified a pivot scale $k_{pivot}$, we further assign:
\begin{verbatim}
	k=pivotScale*k/(1.2*10**(-4)) .
\end{verbatim}
This effectively sets the point around which we will Taylor expand as the pivot scale. The analysis phase continues by fitting the power spectrum with a polynomial fit around the pivot scale. The polynomial fit degree starts at $1$, i.e. linear fitting, and if the SSE value is over a set threshold, the degree is increased. This is done until the SSE is lower than the set threshold or the polynomial degree reaches 20. We assume that if polynomial of degree 20 does not accurately fit the power spectrum, we either have some error along the way, or we are describing a manifestly non-physical scenario.\\
In order to correctly initialize the \code{MsSolver} class the following  input needs to be supplied:
\begin{verbatim}	
(a,H,phi,phidot,t,epsilon,eta,xisq,Vd4V,Vd5V,V,
eh,delh,pp,Tprecision=0.0064,Kprecision=15,
log=True,pivotK=0.05,efoldsAgo=60,efoldsNum=8,
check=False,physical=True,mode='aprox') ,
\end{verbatim}
where \code{a,H,phi,phidot,t,epsilon,eta,xisq,Vd4V,Vd5V,V,eh,delh,pp} are all fields manufactured in the \code{BackgroundGeometry} class, and held as internal fields within it. \code{Tprecision} and \code{Kprecision} are precision parameters, for integration time-step, and for the number of $k$-modes to solve and include in the construction of the PPS. If \code{log=True} the code draws $k$-numbers such that the distribution of $k$'s will be uniform in a log-log scale, to have equal fidelity across the CMB window. \code{pivotK,efoldsAgo,efoldsNum} are the pivot scale, number of efolds to consider in the inflation as `visible' inflation, and the number of efolds in the window to analyse as the CMB window, correspondingly. We allow for non-physical inflationary scenarios in which slow-roll is not held along the entire CMB window, in which case we will set \code{physical=False}. Finally we consider the  exact number of inflationary efolds by setting  \code{mode='exact'}, otherwise, setting \code{mode='aprox'} means the CMB point will be set at $\phi=0$. The difference between both cases is very slight, but may be of import in some cases.
\subsection{Usage}
In order to correctly use INSANE, one must specify several arguments.
These arguments are conveniently edited in the \code{params.in} file. The full file along with example inputs is given in Appendix \ref{INSANE_CODE_APPENDIX}.
\subsection{Benchmarking}
Several tests are now in order to asses the voracity of this code. We first test the code against the only known relevant analytical solution - the power law inflation\footnote{There are other analytical solutions (c.f. \cite{Martin:2000ei}) but they are not relevant since they produce a very blue tilt, not in the vicinity of the observable regime.}.
\begin{figure}
	\includegraphics[width=0.9\textwidth]{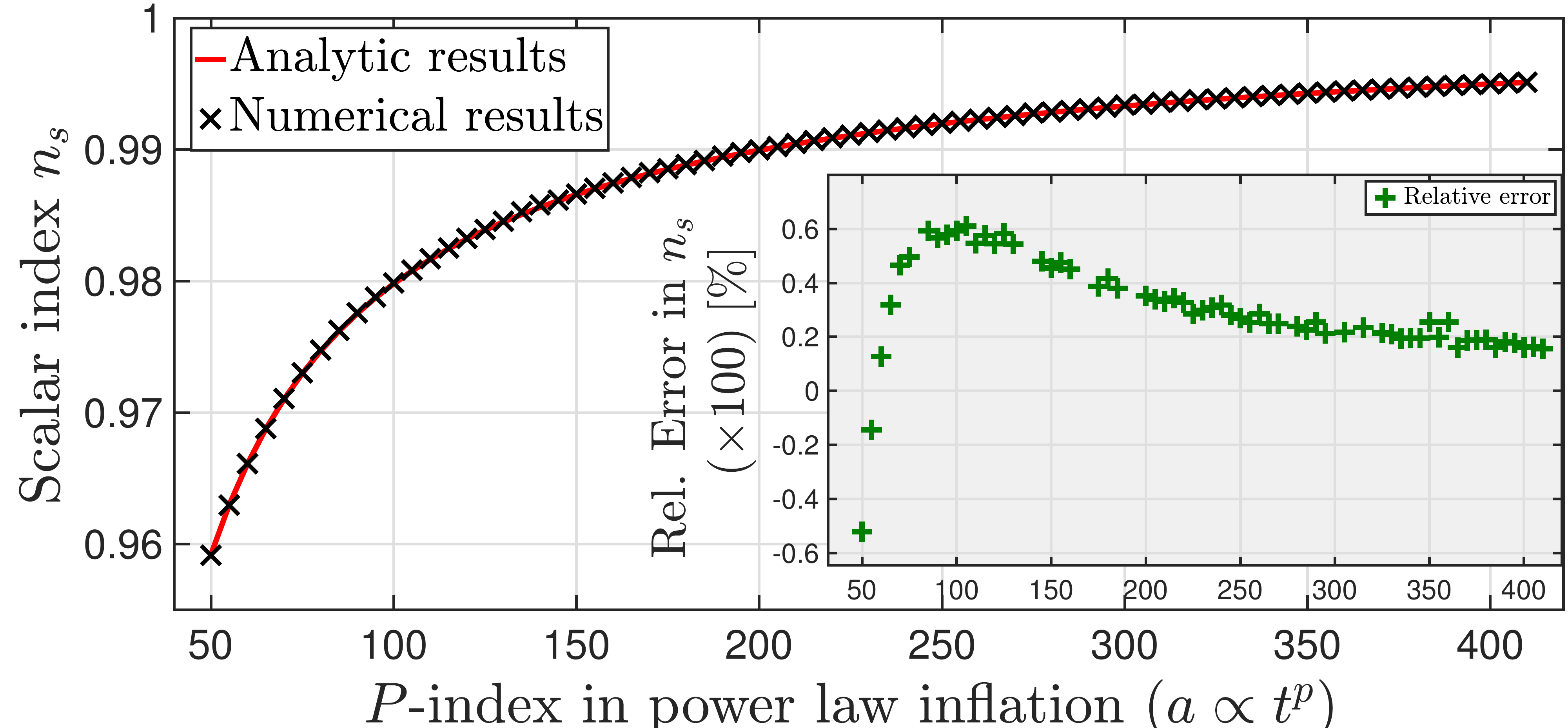}
	\caption[{\bf $n_s$ benchmarking in power law inflation}]{Benchmarking our code against the analytical case of power-law inflation reveals relative errors of order $10^{-3}\%$, which we take as a sign of agreement between analytical and numerical cases. \label{fig:Benchmark1_powerLaw}}
\end{figure}
\subsubsection{The power law case}
The power law inflation is the case where the Hubble parameter $H$ scales as $H\propto t^p$ during inflation. The scalar potential that results in such a case is given by:
\begin{align}
	V(\phi) = A\exp{\left(-\sqrt{\frac{2}{p}}\phi\right)},
\end{align}
in which $p$ is the same as the power law index. The resultant scalar index $n_s$ of such an inflationary scenario is given by:
\begin{align}
	n_s=1-\frac{2}{p},
\end{align}
and is scale independent, such that the index running $\alpha_s$ vanishes by definition.
We have run the INSANE code against these scenarios to compare analytic and numeric solutions. The results are shown in Fig.~ \ref{fig:Benchmark1_powerLaw}, and reveal an agreement between analytical and numerical analysis to a relative error of order $10^{-3}\%$. In evaluating the fidelity of $\alpha_s$ there is an intrinsic problem since in the analytic case it vanishes by definition. Thus no intrinsic scale is available for us to relate to. We, therefore, took the heuristic approach of assessing $\Delta n_s / \Delta \log{k}$. If $\alpha_s$, as recovered by the code, is lower than $\Delta n_s / \Delta \log{k}$, we are satisfied that the code yields a correct $\alpha_s$.
\begin{figure}
	\includegraphics[width=0.9\textwidth]{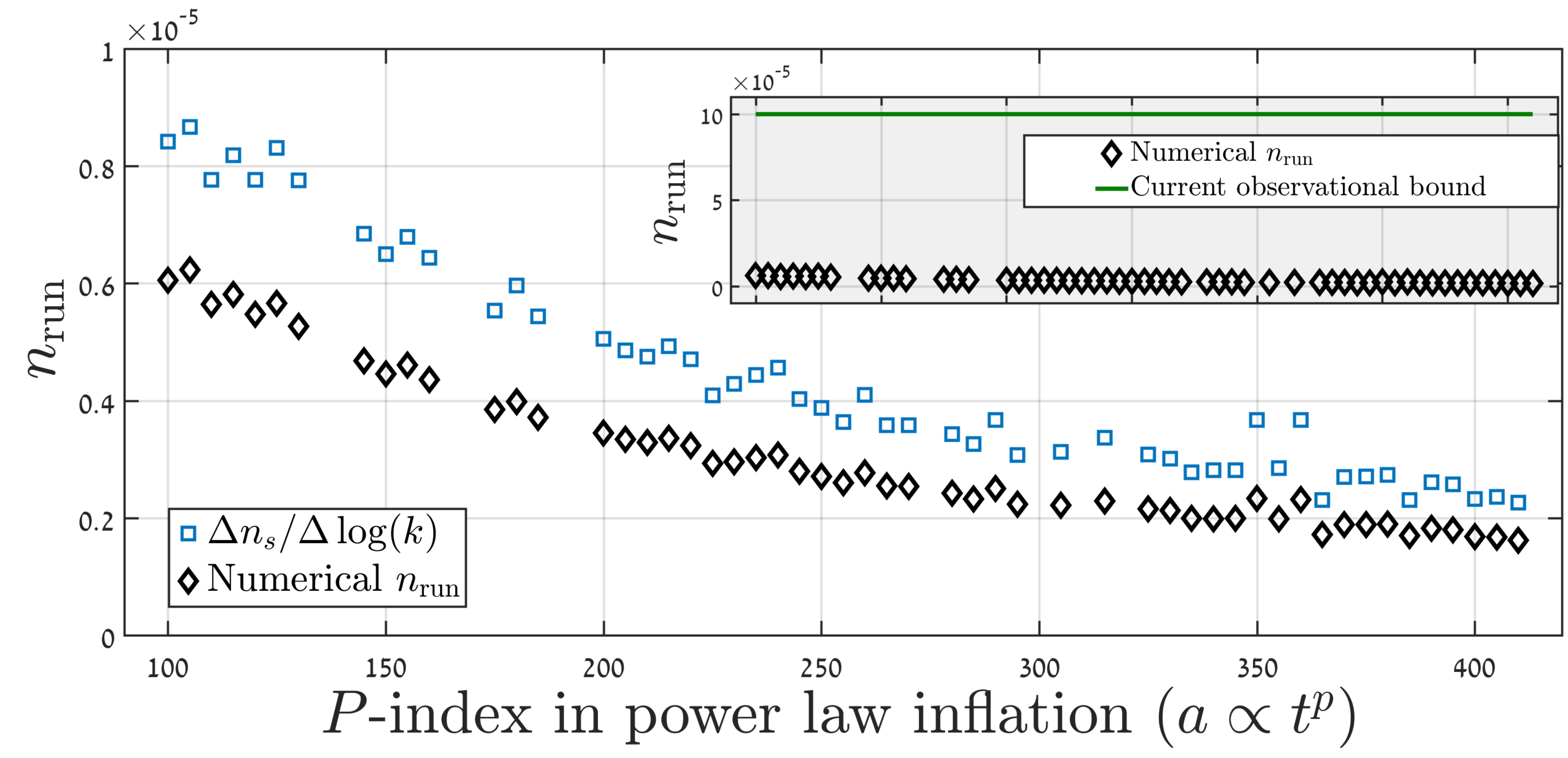}
	\caption[{\bf Heuristic $\alpha_s$ benchmarking in power law inflation}]{A heuristic study of $n_{run}$ (or $\alpha_s$), reveals the values recovered numerically are consistently below the error threshold for $n_s$. We take this as a sign of $n_{run}$ being at the level of numerical error, hence it is equivalent to a vanishing running. \label{fig:Heuristic_nrun}}
\end{figure}
We find that the $\alpha_s$ values recovered by the numerical calculation are always below the numerical error in $n_s$ divided by the overall $\log{k}$. We take this as a sign that the recovered $\alpha_s$ is at the level of numerical error, thus equivalent to a vanishing $\alpha_s$. Fig.~\ref{fig:Heuristic_nrun} shows these findings. Additionally, we show the current observational threshold of $\alpha_s$.
\subsubsection{The quadratic case}
\begin{figure}
	\includegraphics[width=0.9\textwidth]{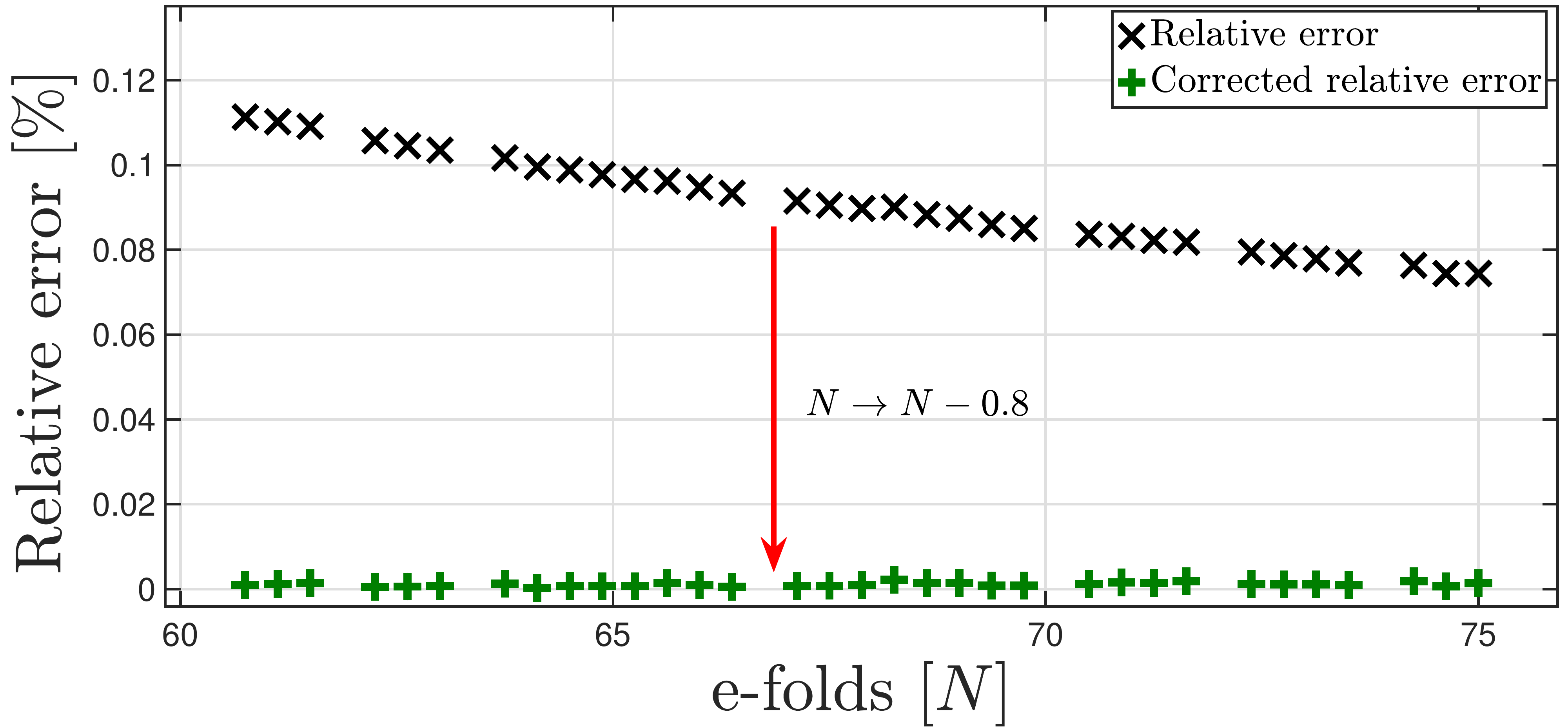}
	\caption[{\bf Benchmarking of quadratic potentials}]{Relative error (in percent) between numerical results and the SL analytical expression (black X's). The errors converge to $0$ for large values of $N$. Shifting the number of the efolds by $N\rightarrow N-0.8$ yields a relative error of the order of $10^{-3} \sim 10^{-4} \%$ (green pluses).  \label{fig:quad1} }
\end{figure}
As we aim to study models that produce slow-roll parameters which are time-dependent, we need to check the precision of the numerical code against such models.
Consequently, we tested the accuracy of our calculations for quadratic potentials of the type
\begin{align}
V=\frac{1}{2}m^{2}\phi^{2}.
\end{align}
In these cases, the analytic expression for the scalar index is given by,
\begin{align}
n_{s}=1-\frac{8}{4N+2}+\frac{32b}{\left(4N+2\right)^2},\label{SL-quad}
\end{align}
Here $N$ is the number of efolds and $b$ is the same as in \eqref{eq:SLR}.
Fig.~\ref{fig:quad1} presents the results of this study, as relative errors between precise calculations and the SL analytic expressions. These results are accurate to $\sim 0.1 \%$. However, there is a systematic error that is traced back to the inaccuracy of the approximation:
\begin{align}
	N=\int_{t_{\text{\tiny{CMB}}}}^{t_{\text{\tiny{end}}}}Hdt \simeq -\int_{\phi_{\text{\tiny{CMB}}}}^{\phi_{\text{\tiny{end}}}}\frac{V}{V'}d\phi.
\end{align}
A shift $N\rightarrow N-0.8$ is sufficient to reduce the systematic error such that the relative error is of the order of $10^{-3}\sim 10^{-4}\%$. Additional types of simple potentials, which yield time-dependent slow-roll parameters were also studied. In all cases, the relative error between calculated results and the traditional SL expression Eq.~\eqref{eq:Lyth_Riotto_ns} is bounded from above by $\sim 0.1\%$. Furthermore, a more careful analytical treatment leads to better accuracy, bounded from above by about 0.02\% relative error. Additionally, we were able to recover the ``Cosmic ring'' phenomenon, that is the PPS response to a step function in the potential. This response feature in the PPS was first studied in \cite{Adams:2001vc}.\\
We take all these results as a strong indication of sufficient accuracy of our calculations.
\section[Code improvements]{Improvements - Teaching a PC some physics}
In order to speed up the calculation, and enable the study of a class of inflationary potentials rather than a solitary example, some adaptations of the code were called for. The most crucial and time-consuming element of calculating the power spectrum is the numerical integration of the different $k$-modes that make up the PPS. We, therefore, identified several mechanisms that, while technically might be interesting, produce little to no effect in the context of yielding the PPS.
\subsection{Finding the mode `freeze-out' point}
The first and most influential improvement was achieved by using the physical understanding that each mode of quantum oscillation `freezes out' of the horizon at some point in its evolution. That is, at some point of the inflationary evolution the light horizon becomes smaller than the mode's wavelength. If we take the viewpoint of the classical harmonic oscillator such that
\begin{align}
U_k=A_ke^{i\omega t},
\end{align}
with $\omega^2 = k^2 -\frac{z''}{z}$, which at some point becomes negative, it is immediately apparent that $\omega$ becomes purely imaginary around
\begin{align}
	k^2=\frac{z''}{z}.
\end{align}
And so, since we care only about the modes after the freeze-out time, we can in theory start the integration approximately around the freeze-out time of each mode. 
We call this 'finding the "knee"' since the modes oscillate around an overall similar amplitude all throughout their evolution inside the horizon, but around the freeze-out time their amplitude begins to grow quasi-exponentially and the phase becomes locked. Thus if we look at the square of the eigenfunction the 'knee' for each mode is approximately at the freeze-out point. 
\begin{figure}[!ht]
\includegraphics[width=1\textwidth]{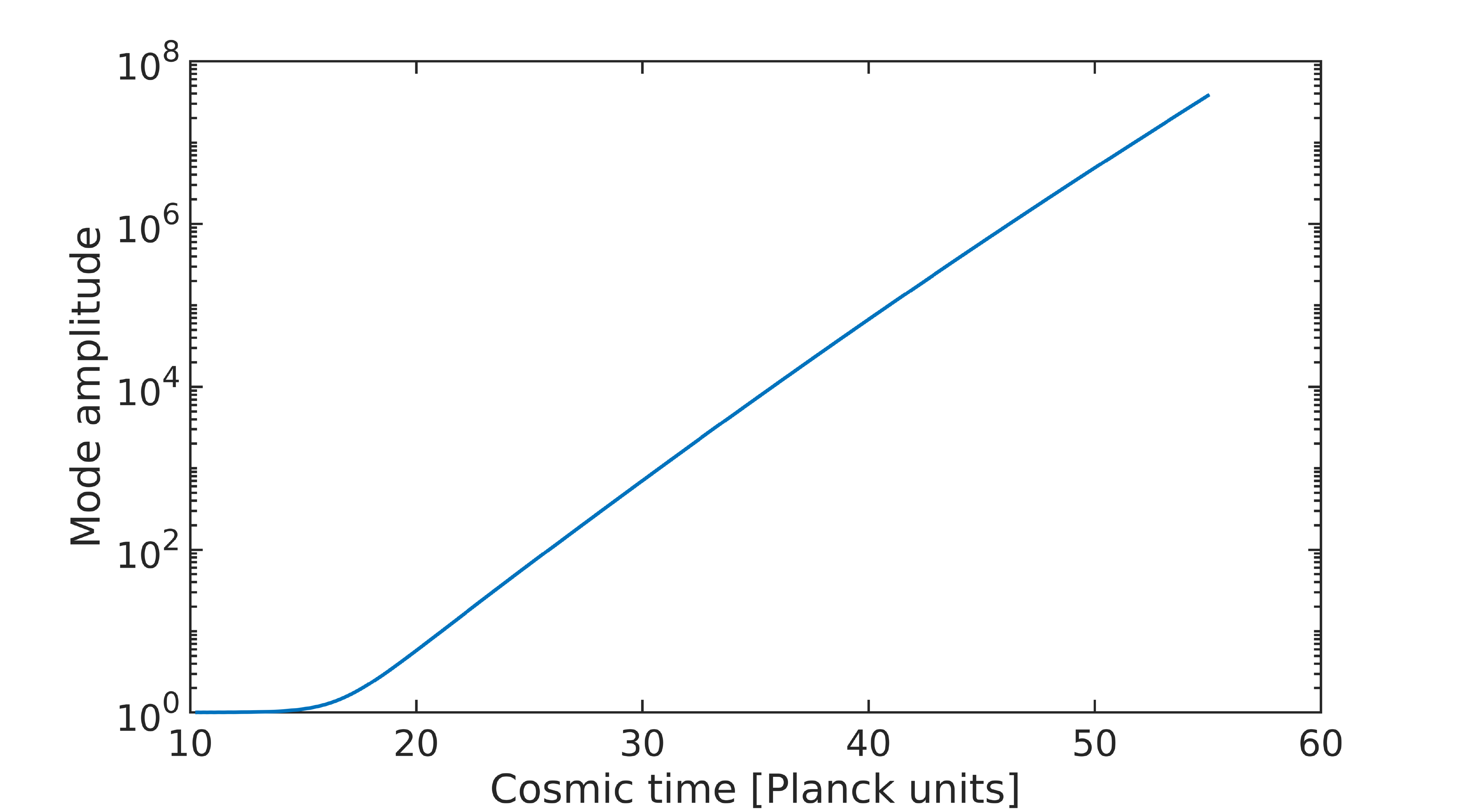}
\caption[{\bf Freeze-out point}]{A single mode's amplitude as a function of time (half-log graph). The amplitude starts at some value and is approximately constant, up to $t\sim 15$. At that time the amplitude starts growing, and around $t\sim 17$ the mode leaves the horizon and the amplitude becomes exponential. Finding this point for each mode enables faster integration.}
\label{fig:mode_knee}
\end{figure}
\subsection{Forgetting the phase}
While the former subsection discussed the amplitude, we are also interested in the phase. Since we are dealing with TDHO the phase and amplitude are coupled to some degree. In theory, then, one cannot simply forget about the phase and integrate the amplitude alone. In fact, this is a common mistake that is done in most numerical schemes to date. However, it is true that the phase gets "locked" when the mode is frozen out of the horizon. In this case, the information of the phase does not evolve any more and provided we know the amplitude and phase at the freeze-out time we no longer have to take them into account when integrating the mode.
Several efolds prior to the freeze-out scale of a mode the associated frequency $\omega$ satisfies $|\omega|\gg 1$, thus the mode oscillates wildly, and in effect the mode 'forgets' the initial phase. This is an oversimplification since some of this information is still carried in the evolution if the amplitude. Be that as it may, the longer prior to the freeze-out time we start the integration this added correction to the amplitude becomes negligible.
\subsection{Pre-Integration}
With the combination of these two last insights, it is evident that it is enough to start the numerical integration some 2.5 to 3.5 efolds prior to the freeze-out time for each mode. 
In this fashion we both truncate the runtime for each mode while keeping the correct information. Another issue that is critical from which it is apparent that we should perform integration several efolds before freeze-out is that of initial conditions. The Bunch-Davies initial conditions are applicable for each mode only deep into the oscillatory phase. These conditions contain critical information, as they contain the seed amplitude which will define the evolution and ultimately the $k$ dependence of the PPS.

\ifpdf
    \graphicspath{{Chapter3/Figs/Raster/}{Chapter3/Figs/PDF/}{Chapter3/Figs/}}
\else
    \graphicspath{{Chapter3/Figs/Vector/}{Chapter3/Figs/}}
\fi


\graphicspath{{Chapter4/}}

\chapter{Fresh eyes on $n_s$}\label{Chap:Fresh_ns}
\section{The $n_s$ discrepancy}
One of the surprising results found during our studies is that a fully numerical treatment of inflation yields different results, from an analytical perturbative treatment. This should not surprise us, as the usual way perturbation theory is used, is by removing sub-leading terms of some order. However, the underlying assumption of this practice is that the sub-leading discarded terms issue tiny corrections relative to the leading terms. This might not always be true. Furthermore, the non-perturbed case should be chosen wisely, such that there is a natural scale to perturb over.\\
The surprising aspect is not the difference itself, but the scope of the difference. We show that in the type of models we study the relative difference in $n_s$ is usually well over $1\%$. We have also seen that even in the case of 'well behaved' models there is a marked difference in $n_s$ between analytical and numerical treatment. 
\begin{figure}
\includegraphics[width=0.9\textwidth]{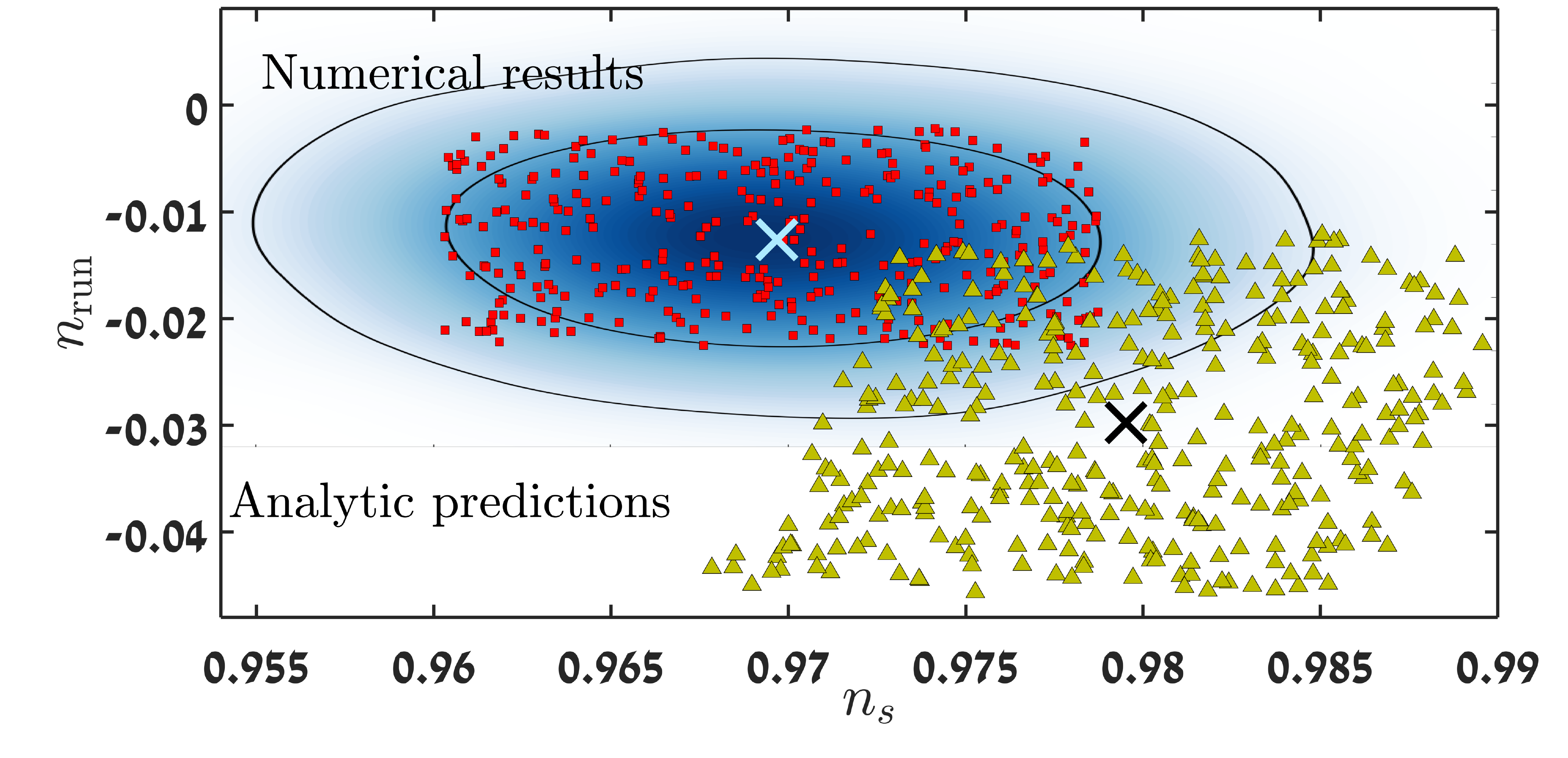}
\caption[{\bf $n_s$ errors}]{Shown are the results of a precise calculation of the cosmological parameters of $\sim$ 200 models (red squares), as well as the corresponding analytic predictions (yellow triangles) calculated according to the 2nd order SL term and the corresponding term for $\alpha_s$. The cyan and black x's  mark the mean value of the precise and analytic results (respectively).}\label{fig:Errors}
\end{figure}
\subsection{Slow-roll hierarchy}
In potentials which are so-called large field models, i.e. where inflation occurs over $\Delta \phi \ll 1$, the number of efolds per field excursion $\frac{dN}{d\phi}\simeq - \frac{V}{V_{\phi}}$, and is usually taken to be monotonously decreasing. The Lyth bound \cite{Lyth:1996im}, originally states that in accordance with current observations, there is a bound on the field excursion that creates the observable scales in the CMB. Several works since (c.f. \cite{Efstathiou:2005tq}) have expanded the Lyth bound to include the entire span of the inflationary potential, up to the end of inflation. However, these works always assume a generally monotonous $\varepsilon_V$. This yields a natural hierarchy for the slow-roll parameters. The relations between slow-roll and their temporal derivatives is also suggestive of this:
\begin{align}
	\dot{\varepsilon_H}=2H\varepsilon_H\left(\varepsilon_H+\delta_H\right)\\
	\dot{\delta_H}=H\delta\left(\frac{\dddot{\phi}}{H\ddot{\phi}}-\delta_H +\varepsilon_H\right),
\end{align}
and this relationship continues where in general \cite{Gong:2001he}:
\begin{align}
	\delta_1=\delta_H,& \\
	\nonumber \\
	\delta_{n+1}=\frac{\dot{\delta_n}}{H}&+\delta_n \left(\delta_1 -n\varepsilon_H\right)\\
\end{align}
However, this relation does not always hold. When $\frac{dN}{d\phi}$ is not monotonous, the derivative $\frac{d\delta_n}{dt}$ can be high enough to break the hierarchy. This is the case with the small field models we study. While the original Lyth bound still apply in these cases, the extended version is broken.
\begin{figure}[!ht]
 \includegraphics[width=1\textwidth]{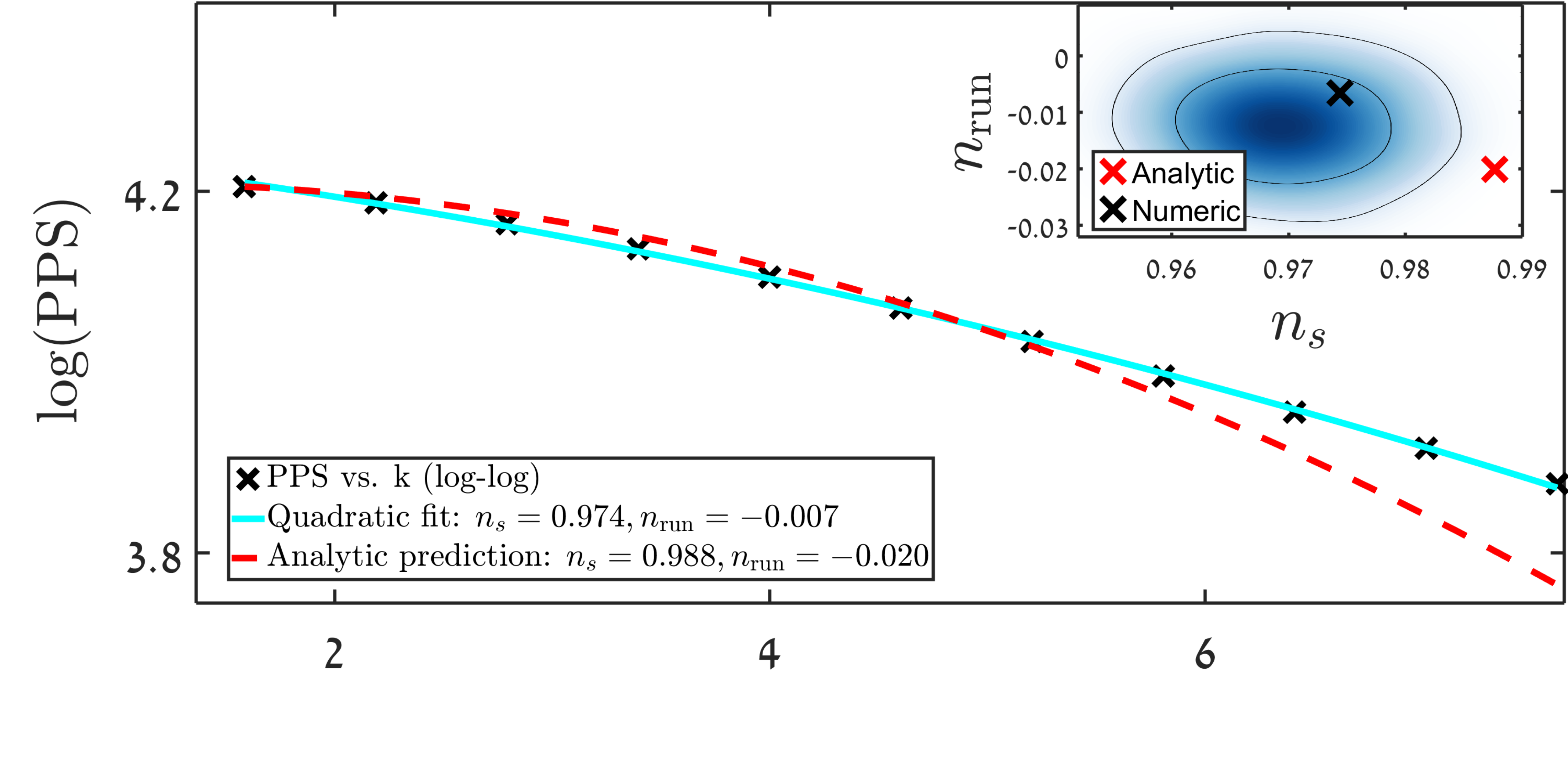}\\
  \includegraphics[width=1\textwidth]{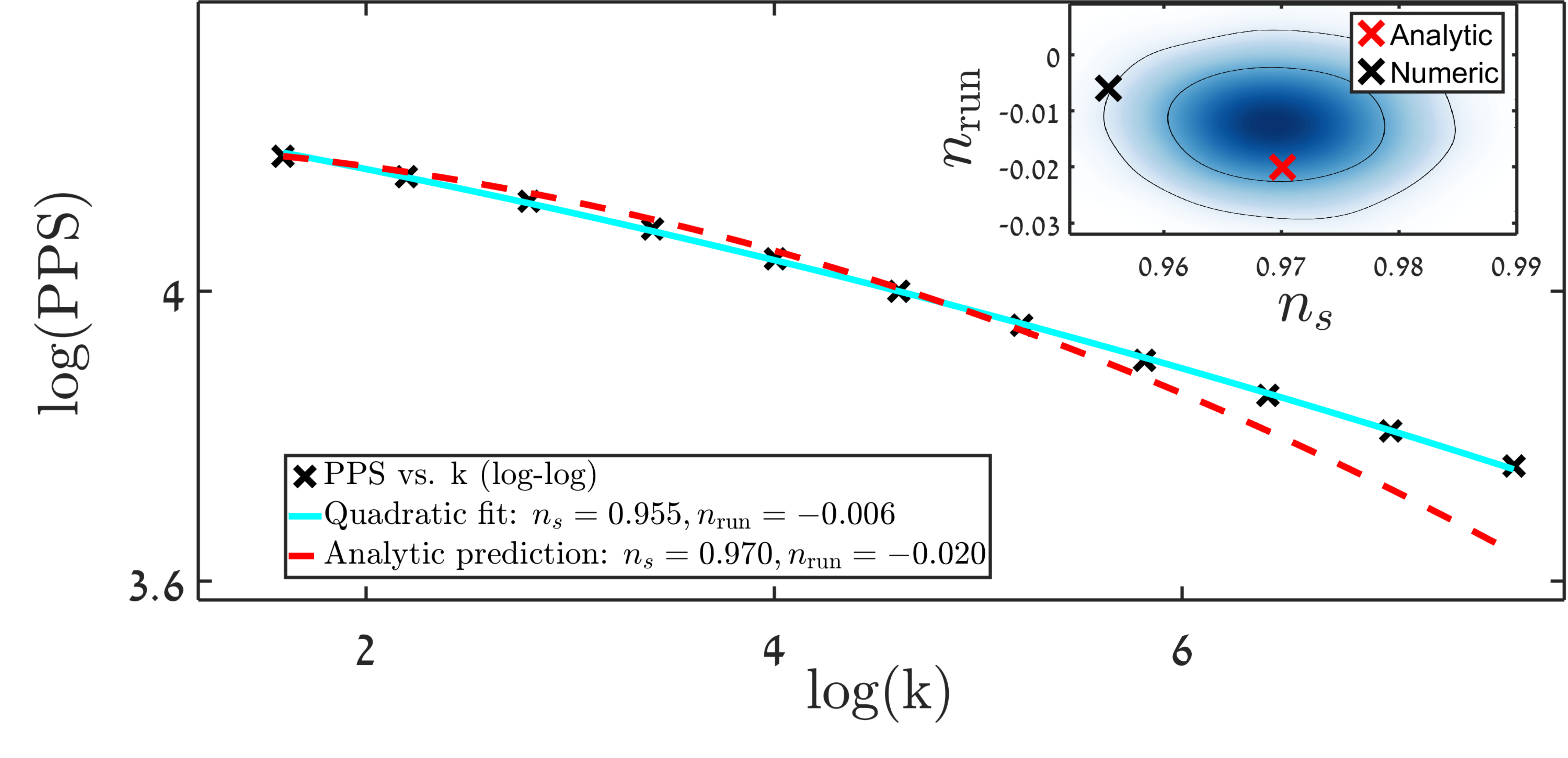}
  \caption[{\bf Examples of power spectra discrepancy}]{Comparison of the precise results and analytic predictions made with \eqref{eq:SLR}. Each panel shows the precisely calculated results, fitted by a quadratic polynomial to extract $n_{s}$ and $n_{\text{\tiny{run}}}$.  The curve predicted by Eqs.~(\ref{eq:SLR},\ref{eq:Analytical_alpha}) is plotted as a reference.  In the upper panel, we show a potential that would be excluded based on the analytic result, whereas the precise results is well within the $68\%$ probability curve. In the lower panel the exact opposite is the case, with an analytically accepted result, but an excluded precise one.}\label{fig:exemplars}
\end{figure}
\subsection{Another degree of separation - slow-roll in potential language}
The original formulation that ties the slow-roll parameters to the characteristics of the PPS uses the slow-roll parameters per their Taylor expansion definition. In \cite{Stewart:1993bc} it was shown, that assuming $\frac{\phi^{(4)}}{H\phi^{(3)}}$ is small, where $\phi^{(n)}\equiv \frac{d^n\phi}{dt^{n}}$, the following relations apply:
\begin{align}
	\left\{
	\begin{array}{lcl}
	\varepsilon_H&\simeq&\varepsilon_V -\frac{4}{3}\varepsilon_V^2 +\frac{2}{3}\varepsilon_V\delta_V\\
	&&\\
	\delta_H&\simeq&\varepsilon_V-\delta_V +\frac{8\varepsilon_V}{3}\left(\delta_V-\varepsilon_V \right)-\frac{1}{3}\left(\delta_V^2 +\xi^2\right) \\
	&& \\
	\delta_H\frac{\dddot{\phi}}{H\ddot{\phi}}&\simeq& 4\varepsilon_V^2 -5\varepsilon_V\delta_V +\delta_V^2 +\xi^2,
	\end{array}
	\right.  ,\label{eq:slow_roll_H_to_V}
\end{align}
where the potential derivative terms are given by:
\begin{align}
	\varepsilon_V=\frac{1}{2}\left(\frac{V'}{V}\right)^2\;;\;\delta_V=\frac{V''}{V}\;;\;\xi^2=\frac{V'V'''}{V^2}\;;\; \sigma^{(n)}=\frac{V'V^{(n)}}{V^2} \label{eq:slow_roll_V}.
\end{align}
As was previously shown in section \ref{sec:slowRollParadigm} these relations are exact at the end of inflation as well as in the case of an exponential potential, that corresponds to the pure dS case. The way to derive these relations is the following. We begin with the set of exact relations:
\begin{align}
	\frac{V}{H^2}=3-\varepsilon_H \label{eq:V_over_Hsq} \\
	\nonumber \\
	\ddot{\phi}+3H\dot{\phi}+\frac{dV}{d\phi}=0\label{eq:KG_2}\\ 
	\nonumber \\
	\dot{H}=\frac{-\dot{\phi}^2}{2},\label{eq:Hdot}
\end{align}
and the definitions for $\delta_H=\frac{\ddot{\phi}}{H\dot{\phi}}$, and $\varepsilon_H=-\frac{\dot{H}}{H^2}$. We now derive the Klein-Gordon equation with regards to cosmic time to yield:
\begin{align}
	\dddot{\phi}+3\dot{H}\dot{\phi}+3H\ddot{\phi}+\frac{d^2 V}{d\phi^2}\dot{\phi}=0.
\end{align}
proceeding to substitute the original KG equation for $\ddot{\phi}$, and setting the highest derivative $\dddot{\phi}=0$, we get the following:
\begin{align}
	3\dot{H}\dot{\phi}-9H^2\dot{\phi} -3HV' +V''\dot{\phi}=0,
\end{align}
where applying the relations in (\ref{eq:Hdot},\ref{eq:V_over_Hsq}), yields the desired term. This procedure can be followed to arbitrary order; derive \eqref{eq:KG_2} w.r.t $t$, $n$ times and \eqref{eq:Hdot} $n-1$ times. Set the highest derivative $\phi^{(n+2)}=0$, substitute iteratively the previous derived relations $\phi^{(n+1)}=...$ and $H^{n}=...$, until we get a term that is a function of $V^{0},V^{(1)}...V^{(n+1)},H,\dot{H},\dot{\phi}$. Assign the slow-roll parameter definitions where applicable, and arrange the potential terms to combinations of \eqref{eq:slow_roll_V}, and a new term will arise of the form:
\begin{align}
	\frac{V'V^{(n+1)}}{V^2}.
\end{align}
A word of caution though, in order for complete equality between the $H$ formulation of the slow-roll parameters and the potential and derivatives formulation to be exact, the infinite tower of equations should be developed. This is the analogue of a Taylor expansion exactly agreeing with the underlying function only in the case of an infinite Taylor series. It can be shown that the remainder takes the form of
\begin{align}
	\Delta \propto \frac{\phi^{(n+2)}}{H \phi^{(n+1)}}.
\end{align}
Since a small $\phi^{(n+1)}$ does not assure us of a substantially smaller $\phi^{(n+2)}$, there is no real way of assessing this remainder without either having a fully analytical expression for $\phi(t)$, or a numerical value of $\phi(t)$ through the entirety of the CMB window. This is unfortunate since it means there is no assurance of these relations except in the case of either extremely controlled slow-roll or fully analytical cases. Of the latter there is only one case that approximates observed values, which is the pure dS case. We have no reason to believe a-priori that the former case is more likely than any other.
\begin{table}[!h]
\begin{center}
\begin{tabular}{||c|c|c|c|c|c||c||}
\hline
$a_2$&$a_3$&precise&analytic&precise&analytic&Fit error\\
&&$n_s$&$n_s$&$n_{\text{\tiny{run}}}$&$n_{\text{\tiny{run}}}$&  \\
&&&&&&($\times 10^{-4}$)\\
\hline
$0.0005$&$-0.3041$&$0.9777$&$0.9856$&$-0.0196$&$-0.0409$&$	1.8$\\
\hline
$-0.0013$&$-0.2795$&$0.9713$&$0.9796$&$-0.0175$&$-0.0373$&$	1.5$\\
\hline
$-0.0001$&$-0.2188$&$0.9780$&$0.9877$&$-0.0125$&$-0.0293$&$	1.1$\\
\hline
$-0.0042$&$-0.1538$&$0.9627$&$0.9748$&$-0.0067$&$-0.0203$&$	0.8$\\
\hline
$-0.0032$&$	-0.2923$&$0.9631$&$0.9711$&$-0.0185$&	$-0.0387$&$	1.9$\\
\hline
$-0.0002$&$-0.2709$&$0.9760$&$0.9843$&$-0.0168$&	$-0.0363	$&$	1.6$\\
\hline
$-0.0026$&$-0.1342$&$0.9710$&$0.9820$&$-0.0055$&	$-0.0178	$&$	0.6$\\
\hline
$-0.0031$&$-0.1517$&$0.9670$&$0.9793$&$-0.0066$&	$-0.0201	$&$	0.8$\\
\hline
$-0.0011$&$	-0.1563$&$0.9757$&$0.9868$&$-0.0072$&	$-0.0209	$&$	0.7$\\
\hline
$-0.0024$&$	-0.2808$&$0.9662$&$0.9752$&$-0.0174$&	$-0.0373	$&$	1.9$\\
\hline
\end{tabular}
\caption[{\bf $n_s$ discrepancy for a small sample of models.}]{Shown is a table of 10 potentials constructed such that $r_0=0.001$, and $N=60$. The parameters $a_2$ and $a_3$ are constructed by randomly drawing from a uniform distribution as explained in Section 4. The discrepancy in $n_s$ is around $0.8\%\sim 1.25\%$, while the $n_{\text{\tiny{run}}}$ discrepancy is much more pronounced.  } \label{table:PotTable}
\end{center}
\end{table}
\subsection{Does the $\alpha_s$ formulation work?}
The Lyth-Riotto formulation in \cite{Lyth:1998xn} for the running of the scalar index, $\alpha_s$, is derived by using the analytical expression for $n_s$, up to second order:
\begin{align}
	n_s=1-6\varepsilon_V +2\delta_V -2\left[\left(\frac{5}{3}+12C\right)\varepsilon_V^2 +\left(C-1\right)\varepsilon_V\delta_V+\frac{\delta_V^2}{3}-\left(C-1\right)\xi^2\right], \label{eq:Lyth_Riotto_ns}
\end{align}
with $C=-2+\log{2} +b\simeq -0.73$, where $b$ is the Euler-Mascheroni constant. 
By deriving Eq.\eqref{eq:Lyth_Riotto_ns} with respect to $\log{k}$, keeping terms up to second order in slow-roll parameters the term for $\alpha_s$ in \cite{Lyth:1998xn} is given: \footnote{Note that in \cite{Lyth:1998xn} our $\delta_V$ corresponds to $\eta$}
\begin{align}
	\alpha_s=-16\varepsilon_V\delta_V +24\varepsilon_V^2 +2\xi^2. \label{eq:Lyth-Riotto-alpha}
\end{align}
However, the original expression for the scalar index was derived by \cite{Stewart:1993bc} for the exponential inflationary potential, which yields a strictly linear $n_s$. Since $0$ has no natural scale, even the slightest perturbation around it may be considered large. This means that this perturbative formulation for $\alpha_s$ is by definition impossible.
In addition, almost all such formulations assume an unbroken hierarchy of slow-roll parameters. This assumption is based on the faulty logic of:
\begin{align}
	a(x)\simeq b(x) \Rightarrow \frac{da(x)}{dx}\simeq \frac{db(x)}{dx} .
\end{align}
This assumption is only correct if both $a(x)$ and $b(x)$ are strictly smooth, non-oscillatory, and the relation $a(x)\simeq b(x)$ holds for a long interval on the $x$ axis. These conditions are in no way guaranteed, especially since the natural eigenfunctions of the PPS are Hankel functions which are by nature oscillatory. Furthermore, even a slight step-like feature in the inflationary potential yields a non-proportional response of an oscillatory nature. In fact, the motivation for looking at the numerical results for $n_s$ as compared to analytical predictions first came up from looking at the numerical results for the index running $\alpha_s$, as compared to the analytical expression in Eq.~\eqref{eq:Lyth-Riotto-alpha}. We found that the disparity between these can be of the order of $500\%$ sometimes, and there doesn't seem to be any consistency in the comparison between analytic predictions and numerical results. 
\begin{figure}
\includegraphics[width=0.85\textwidth]{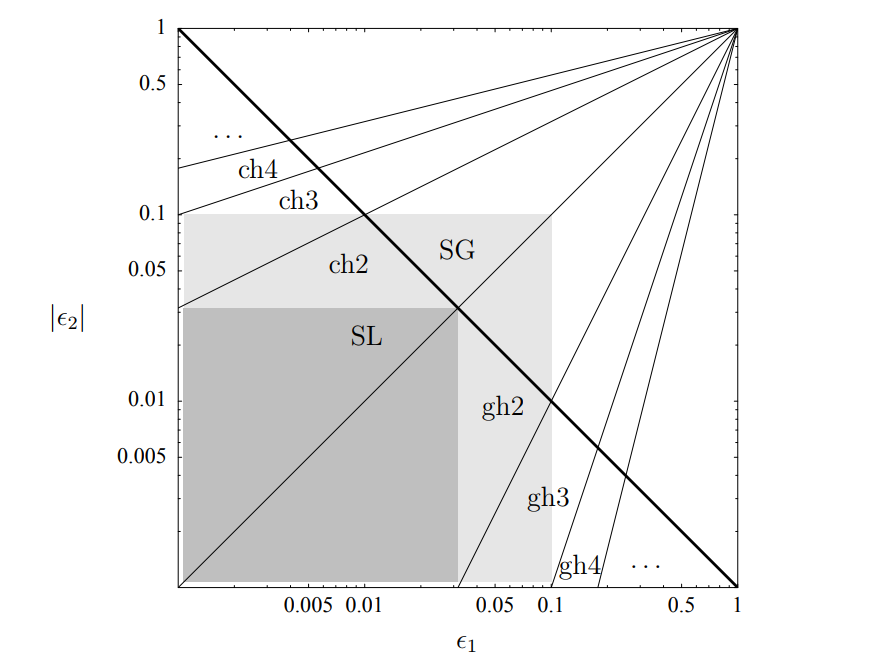}
\caption[{\bf SEG $n_s$ evaluation map}]{Regions in the $\epsilon_1$-$|\epsilon_2|$ parameter space where the 
spectral amplitudes could be calculated with an accuracy better than $1\%$, according to the analysis presented in \cite{Schwarz:2001vv}.
In the dark shaded region the Stewart-Lyth (SL) approximation \cite{Stewart:1993bc}, as 
well as all other approximations are supposedly sufficiently accurate. Second-order corrections,
as calculated by Stewart and Gong (SG) \cite{Gong:2001he}, extend that region to the 
light shaded region. The constant horizon approximation at order $n$ (ch$n$), 
and the growing horizon approximation at order $n$ (gh$n$), do well below
the thick line. The rays indicate where the corresponding higher order 
corrections are necessary.
The thick line itself is the condition $\epsilon_1 |\epsilon_2| < 
(A/100\%)/\Delta N$, with $\Delta N =10$ and $A = 1\%$. We study these approximations and others, and find that our models defy these analyses.
Figure and caption adapted, with permission, from \cite{Schwarz:2001vv}.}\label{Different_regions}
\end{figure}  
\section{A Green's function approach}
There are two analytic approaches we discuss, the first is the so-called 'raw' Green's function approach in which we present the Green's function for inflation. The second is to use the same mathematical 'trick' used by Stewart in \cite{Gong:2001he}, and use the adjusted Green's function.
\begin{figure}[!ht]
\includegraphics[width=0.95\textwidth]{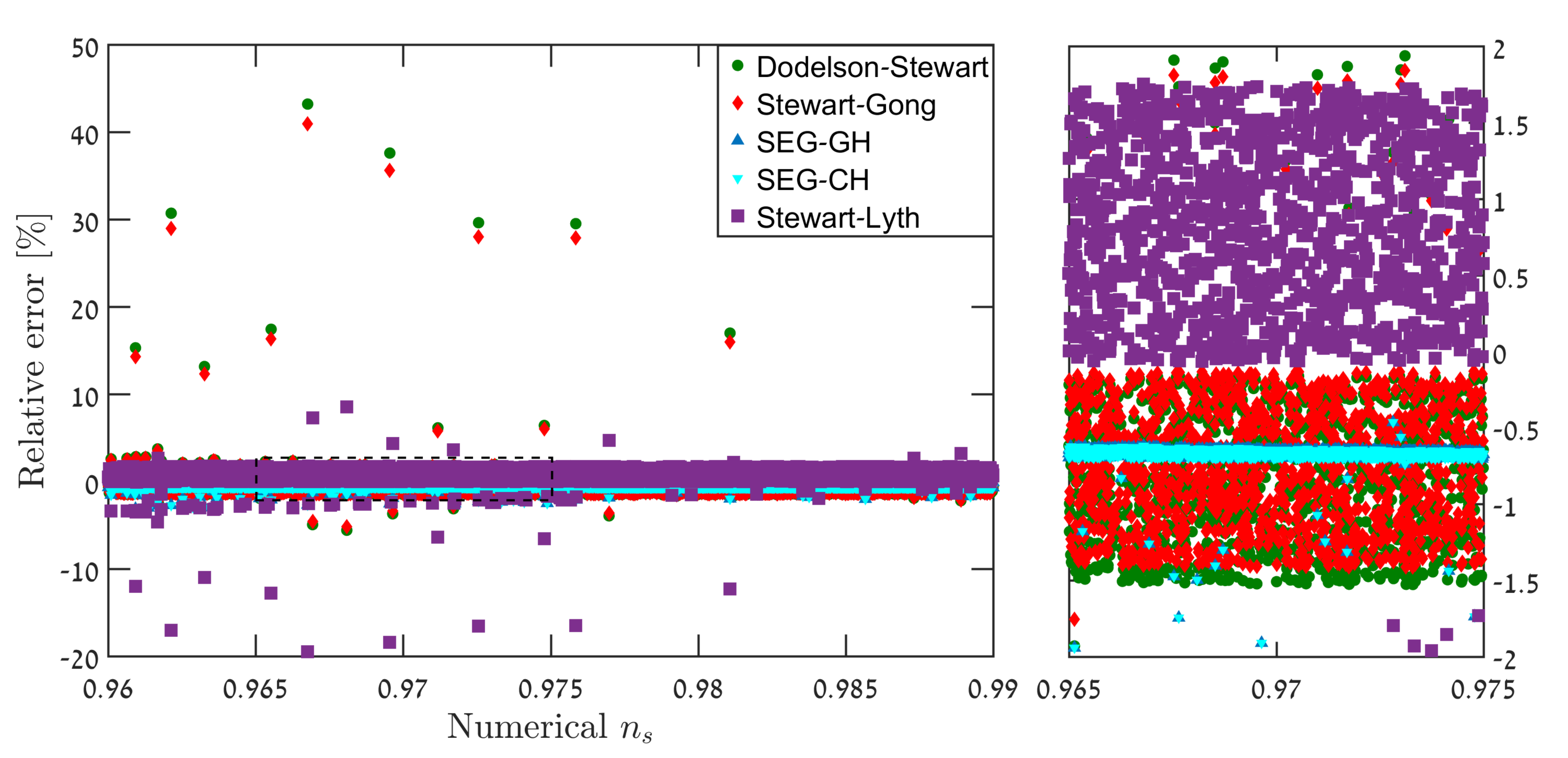}
\caption[{\bf $n_s$ discrpency with different formulation}]{Around 50,000 of our models numerically simulated and compared to different analytical expressions reveals a varying level of accuracy in predicting the correct scalar index. The figure shows only a partial sample of $\sim 8000$ restricted to $\epsilon_1 < 0.0275$, $|\epsilon_2| < 0.0275$ and $0.96<n_s<0.99$. Each data point is a relative error between the numerical result of a model and an analytical expression from \cite{Dodelson:2001sh} (DS,green circles), \cite{Gong:2001he} (SG,red diamonds),\cite{Schwarz:2001vv} (SEG-GH, growing horizon variant - blue triangle, and SEG-CH, constant horizon variant - inverted cyan triangle), and the usual SL \cite{Stewart:1993bc} expression (purple squares).}
\label{different_ns_SEG}
\end{figure}
\subsubsection{Raw Green's function approach}
Over the years some attempts have been made to analytically derive the PPS in a more accurate way. The most promising of which was the Green's function approach along with perturbative corrections. The Green's function for scalar field driven dS inflation is given by:
\begin{align}
	\label{green}\\ \nonumber
	G(x,x')=\left\{\begin{array}{lr}
		\frac{\sqrt{\pi}}{2}e^{\tfrac{i\pi}{2}(\nu+\tfrac{1}{2})}\sqrt{x}H^{(1)}_{\nu}(kx)& x'<x\\
		&\\
		\frac{\sqrt{\pi}}{2}e^{\tfrac{i\pi}{2}(\nu+\tfrac{1}{2})}\sqrt{x}H^{(1)}_{\nu}(kx)& x<x'\\
		+\frac{i\pi k \sqrt{xx'}}{4}\left(H^{(2)}_{\nu}(kx')H^{(1)}_{\nu}(kx)-H^{(1)}_{\nu}(kx')H^{(2)}_{\nu}(kx)\right)&
	\end{array}\right. , 
\end{align}
where $H^{(1)}_{\nu}$ is the first Hankel function of order $\nu$.
In order to derive the correct $u_k$ eigenfunctions given some other inflationary potential, one decomposes the pump field term to a dS pump field term $(z''/z)_{dS}$ + a correction term
\begin{align}
	F(\tau)=(\frac{z''}{z}-\left.\frac{z''}{z}\right|_{dS})
\end{align} 
In this manner, we get the perturbed eigenfunction by:
\begin{align}
	u^{1}_k(\tau)=\int_{0}^{\infty}G(-\tau,-\tau')u^{0}_k(\tau)F(\tau)d\tau',
\end{align}
where $u^{0}_k$ is the set of dS k-eigenfunctions which are in general a linear combination of spherical Hankel functions of the first and second kind.\\
This means that there are closed-form analytical solutions only for a restrictive set of input functions, as were evaluated for instance in \cite{Dvorkin:2009ne}.\\
In the general case, there are no known analytic solutions to the problem of inflation with a scalar inflationary potential. Furthermore, the perturbative approach was not probed sufficiently for its validity limits. \\

One point of interest is that the preferred eigenfunction space for evaluating the PPS and its features is the Bessle function one. This is also demonstrated in \cite{Adams:2001vc}, where a step function feature in the potential has a compressed Bessel function response over the baseline PPS. Since the step function creates a 'double delta'-like feature in the pump field, one can consider the compressed Bessel response to be the natural decomposition basis of the PPS.
\subsubsection{Adjusted Green's function}
In \cite{Gong:2001he} the author used a mathematical 'trick' where we define:
\begin{align}
	\left\{ \begin{array}{lcr}
	y&\equiv& \sqrt{2k}u_k\\
	&&\\
	x&\equiv& -k\tau
	\end{array}\right. ,
\end{align}
along with the ansatz:
\begin{align}
	z=\frac{1}{x}f\left(\ln{x}\right) .
\end{align}
In this approach the eigenfunctions are simpler to integrate perturbatively since the equation takes the form:
\begin{align}
	\frac{d^2 y}{dx^2} +\left(1-\frac{2}{x^2}\right)y=\frac{g\left(\ln{x}\right)}{x^2 }y \; ; \hspace{20pt} g=\frac{f''-3f'}{f},
\end{align}
with a homogeneous solution of 
\begin{align}
	y_0(x)=\left(1+\frac{i}{x}\right)e^{ix}.
\end{align}
This approach amount to the following solution:
\begin{align}
	\lim_{x\rightarrow 0}y(x)= \frac{i}{x}\left\{1+\frac{\xi x_{\star}^{-\nu}}{(3-\nu)(\nu)}\left[\left(\frac{i}{2}\right)^{\nu}\frac{\Gamma(2+\nu)}{1-\nu}-x^{\nu}\right]\right\},
\end{align}
in which $\xi$ is the small perturbation parameter, $x_{\star}$ is some convenient time around horizon crossing, and $\Gamma$ is the generalized factorial Gamma function. 
While this formulation is sound, several assumptions are made. These include the first order evaluation of $\xi$ to be small and the validity and smoothness of the function $g(\ln x)$. It was shown in \cite{Dvorkin:2009ne}, that in this case, several iterations are needed to achieve a less than $1\%$ accuracy as compared to the exact solution. Thus this approach serves at best as a good guideline. However, in the age of precision cosmology, this is not enough.
\begin{figure}[!h]
\includegraphics[width=0.95\textwidth]{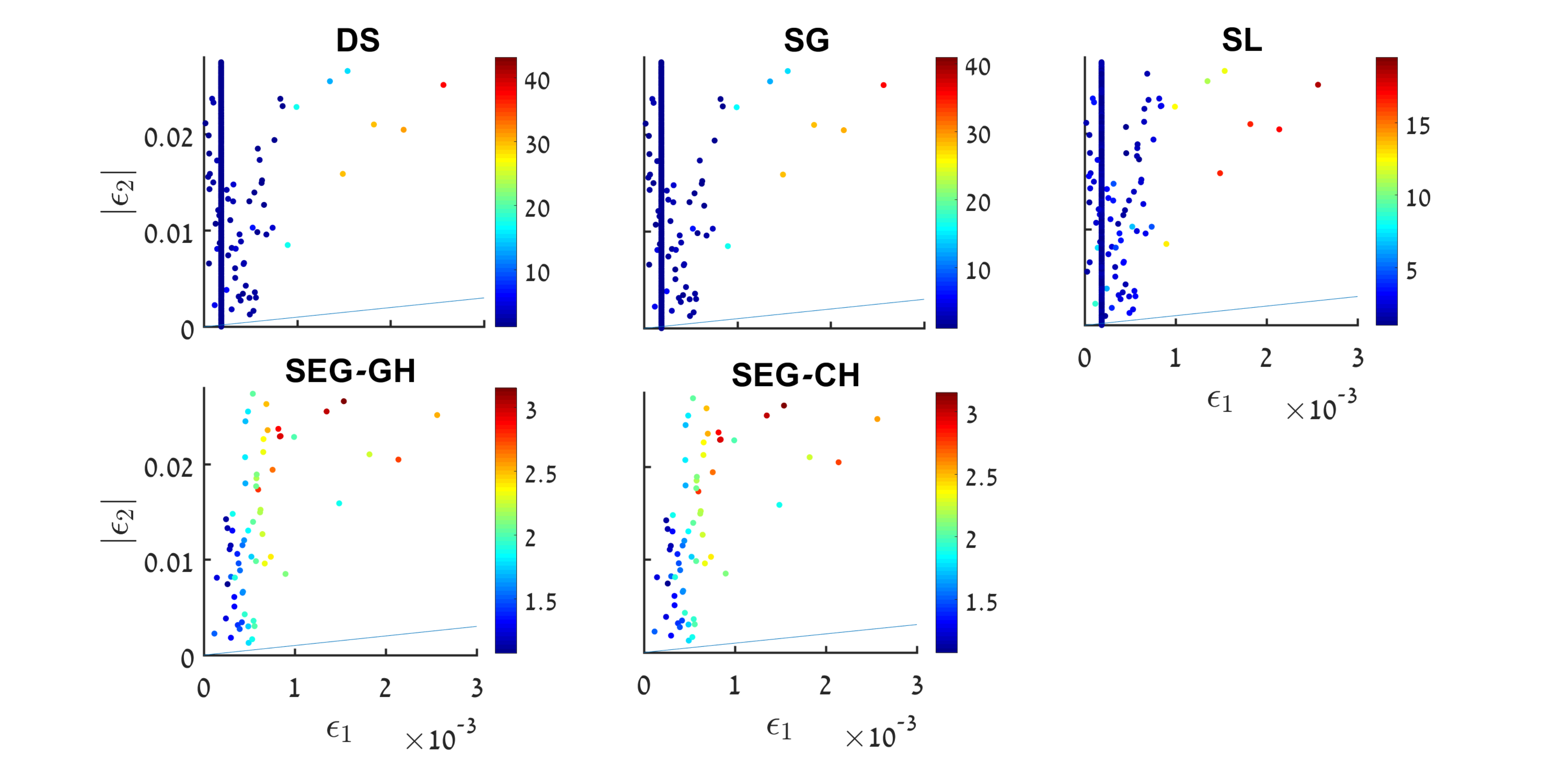}
\caption[{\bf Within SEG approximation still yield $n_s$ discrepancy.}]{Different analytical expressions and their errors relative to the exact numerical analysis, presented on the $\epsilon_1 - |\epsilon_2|$ plane. Each data point is the relative error between the analytic expression and the numerical result, and the color bars to the right of each panel indicate the percentage of relative error. The errors are filtered to show only errors above $1\%$, with numerical results $0.96<n_s<0.99$.}
\label{different_ns_METHODS2}
\end{figure}
\subsection{Possible explanations of the source of deviation between precise results and analytical estimates.}
From the discussion in section \ref{sec:SL} , one can easily see that the definition of $\nu$, is potentially the most significant discrepancy. The effect of this change in definition is an error of less than about $0.4\%$.

Table \ref{Examples} contains three examples of potentials. Two yield observables that are within acceptable limits and a third shows an excluded precise result with an allowed analytic prediction. Examples such as these are used to study the origin of discrepancy.
\begin{table}
	\begin{center}
		\begin{tabular}{|c||c|c|c|c|c|c|}
		\hline
		Ex. no.& $a_{1}$&$a_{2}$&$a_{3}$&$a_{4}$&$a_{5}$&$n_{s}$\\
		\hline
		1&$-0.01118$&$-0.0008$&$-0.2468$&$0.8726$&$-0.7825$&$0.9698$\\
		\hline
		2&$-0.01118$&$-0.0057$&$-0.2344$&$0.8631$&$-0.7804$&$0.9495$\\
		\hline
		3&$-0.01118$&$-0.0025$&$-0.1782$&$0.7100$&$-0.6916$&$0.9661$\\
		\hline
		\end{tabular}\caption[{\bf $n_s$ discrepancy of $1\%$, for likely models.}]{Shown are three examples for a degree 5 polynomial inflationary potentials. Examples no. 1 and 3 yield a precise result for $n_s$ which is well within the 68\% probability region. Example no.2 is the opposite case, with an analytic prediction within the 68\% region, but a precise result which is excluded. Tables \ref{Table_SlowVsPot} and \ref{ns of different methods}, refer to these potential examples. \label{Examples} }
		\end{center}
\end{table}
The differences between the slow-roll parameters defined via the potential vs. their definition in terms of time derivatives are also discussed in section \ref{sec:SL}.
We have found, that in the degree 5 polynomial potentials that were studied, small but significant departures from the relations in Eq. \eqref{eq:slow_roll_V} are detected.
For instance $\delta_{H}=-0.0016$ and $\delta_{V}=\left(\eta_V-\varepsilon_V\right)=0.001$ at the time when $n_{s}$ is evaluated. Table \ref{Table_SlowVsPot} contains values of the three quantities $\epsilon_{H},\delta_{H},\tfrac{\delta_{H}\dddot{\phi}}{H\ddot{\phi}}$ as precisely calculated and analytically approximated, for three potentials of the degree 5 polynomial class. Table \ref{ns of different methods} contains the scalar index for the corresponding potentials (examples 1,2 and 3).
\begin{figure}[!h]
\includegraphics[width=0.95\textwidth]{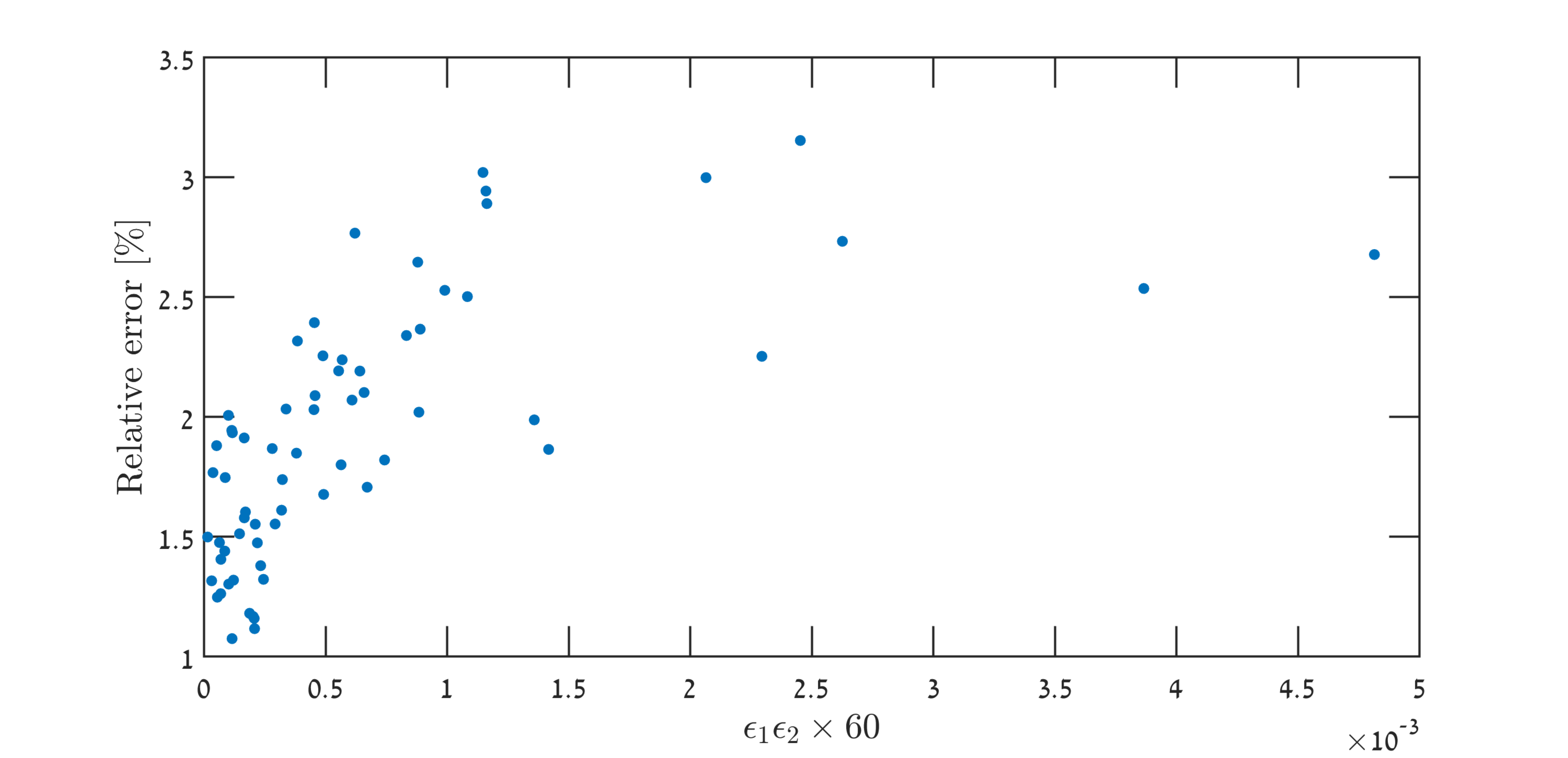}
\caption[{\bf Satisfying the condition $\epsilon_1 |\epsilon_2| \times \Delta N <10^{-2\sim 3}$ still yields discrepancy.}]{While satisfying the condition $\epsilon_1 |\epsilon_2| \times \Delta N <10^{-2\sim 3}$, for $\Delta N = 60$, one finds a relative difference of well over $1\%$ between analytical predictions and numerical results. This is in contrast to the analysis proposed in \cite{Schwarz:2001vv}.}
\label{e1e260}
\end{figure}

\begin{table}[!h]
	\begin{center}
	\begin{tabular}{|l||c|c|c|}
		\hline
		Ex. no. & Quantity & slow-roll value& pot. der. value\\
		\hline \hline
		 & $\epsilon$ & $6.28\cdot 10^{-5}$ & $6.24\cdot 10^{-5}$\\
		 1& $\delta$ & $-0.0068$ & $-0.0038$ \\
		 & $\frac{\delta \dddot{\phi}}{H\ddot{\phi}}$ &$0.0255$ & $0.0165$ \\
		 \hline
		  & $\epsilon$ & $6.23\cdot 10^{-5}$ & $6.20\cdot 10^{-5}$\\
		 2& $\delta$ & $0.0037$ & $0.0063$ \\
		 & $\frac{\delta \dddot{\phi}}{H\ddot{\phi}}$ &$0.0237$ & $0.0159$ \\
		 \hline
		  & $\epsilon$ & $6.26\cdot 10^{-5}$ & $6.23\cdot 10^{-5}$\\
		 3& $\delta$ & $-0.0016$ & $0.001$ \\
		 & $\frac{\delta \dddot{\phi}}{H\ddot{\phi}}$ &$0.0198$ & $0.0119$ \\
		 \hline
	\end{tabular}\caption[{\bf Leading slow-roll parameters, numerical vs. analytic.}]{A table containing the three leading slow-roll parameters, as precisely calculated, vs. the values evaluated by the analytic approximation in Eq.~\eqref{eq:SLR}. While the difference in value for $\epsilon_H$ is negligible, the difference in $\delta_H$ might already be substantial and the difference for $\frac{\delta_{H}\dddot{\phi}}{H\ddot{\phi}}$ is significant.\label{Table_SlowVsPot}}
	\end{center}
\end{table}
The overall effect of this discrepancy can sometimes amount to a $5\sim 8\%$ error towards higher values.

Finally, there is also a significant difference in the derivatives of $\nu$ and $\nu_{\text{\tiny{SL}}}$, $\nu_{\text{\tiny{SL}}}$ being $\nu$ in the SL formulation:
\begin{align}
	\nu_{\text{\tiny{SL}}}=\frac{3+2\delta_H +\epsilon_H}{2(1-\epsilon_H)}\;,
\end{align}
where time dependency of the slow-roll parameters is neglected. This difference is mainly due to neglecting the term $\frac{\delta_{H}\dddot{\phi}}{H\ddot{\phi}}$ in the definition of $\frac{z''}{z}$. This yields a difference in the derivative terms of the order of $0.02\sim 0.04$, which in turn is responsible for a difference in $n_{s}$ of the order of $4\sim 8\%$. Using $\nu_{\text{\tiny{SL}}}$ instead of the full term tends to drive the resulting $n_s$ downwards.

The tendencies of the two aforementioned errors are opposite, and so they might sometimes cancel each other. This makes it possible to get an accurate result using the standard SL expression for a specific potential, but studying a collection of such potentials reveals the incomplete nature of this cancellation.

Table \ref{ns of different methods} shows the different results using different methods of deriving the scalar index. We use three different analytical methods: (1) Eq.~\eqref{eq:SLR} - The SL original method, extracting a term for the scalar index as a function of the potential and its derivatives, (2) Eq. \eqref{eq:TrueNs} - The SL original method, but not relating slow-roll quantities to potential and derivatives, and (3) Using the same methods as the SL analysis, with the definition for $\nu$ as in Eqs.~(\ref{TrueNu},\ref{AccurateZoZ-2},\ref{eq:TrueNs}). From this analysis, it seems the origin of the most significant error is the inaccurate relations between slow-roll parameters and their potential and derivatives counterparts. Second in significance is the definition of $\nu$ with the full $\frac{z''}{z}\tau^{2}$ expression, along with the proper derivation of $\frac{\partial \nu}{\partial \log(k)}$.  The evaluation of $-\tau a H (1-\epsilon_{H})=1$ is off by $\sim 0.04\%$ and the difference between $\psi(\tfrac{3}{2})$ and $\psi(\nu)$  yields a correction of the order of $\sim 0.01\%$.

There might be additional factors that stem from the temporal dependence of $\nu$ in the MS equation. However, these mostly affect the running of the spectral index, and are harder to estimate accurately.

Taking these approximations into account, lowers the discrepancy to the order of $0.5\%$, in a consistent manner.
Another possible explanation is that the time-dependence in \eqref{AccurateZoZ}, modifies the corresponding $\omega^{2}_{k}(\tau)=\left(k^{2}-\frac{\tilde{C}}{\tau^{2}}\right)$ to $\omega^{2}_{k}(\tau)=\left(k^{2}-\frac{f(\tau)}{\tau^{2}}\right)$.
This could lead to modified solutions for the MS equation. An example of this phenomenon is given in \cite{Martin:2000ei}, where the Hankel functions were replaced by the Whittaker functions (albeit these models are observationally excluded). It is worth mentioning that this avenue was studied analytically by Dodelson \& Stewart \cite{Stewart:2001cd,Dodelson:2001sh}. They derived an expression for the scalar index in cases where the slow-roll hierarchy breaks down. However, this analysis was not checked numerically. Additional derivation attempts aiming at yielding better precision analytical expression for the scalar index $n_s$ were made in \cite{Gong:2001he,Schwarz:2001vv}. Specifically \cite{Schwarz:2001vv} supplies an analysis of the predicted level of accuracy as a function of the horizon flow functions $\epsilon_1\equiv \epsilon_H$ and $\epsilon_2\equiv 2(\epsilon_H +\delta_h)$, in figure \ref{Different_regions}. The different approximation schemes were put to the numerical test in the context of our models. Figure \ref{different_ns_SEG} shows that all methods of approximation yield results varying in accuracy and precision levels, it also shows however that the SEG approximation is the best candidate to improve on since on average they yield errors of less than $1\%$.
~ Studying results where relative errors in $n_s$ are over $1\%$, for each expression and locating it on the $\epsilon_1-|\epsilon_2|$ diagram in figure \ref{different_ns_METHODS2} reveals that the analysis offered in \cite{Schwarz:2001vv} is not completely applicable to our models. Figure \ref{e1e260} shows that for the models studied, even though the conditions outlined in \cite{Schwarz:2001vv} are met, and $\epsilon_1 \epsilon_2 \times \Delta N <10^{-2\sim 3}$ for $\Delta N=60$, the relative error between numerical result and SEG-CH expression can be above $1\%$.  
\begin{table}
		\begin{center}
		\begin{tabular}{|c|c||c|c|c|c|}
			\hline
				Ex. no. & & Num. value &Eq.~\eqref{eq:SLR} &Eq. \eqref{eq:TrueNs}& Eq.~(\ref{TrueNu},\ref{AccurateZoZ-2},\ref{eq:TrueNs})\\
			\hline		
			1 &$n_{s}$ & $0.9698$ & $0.9833$ & $1.05$ & $0.9650$\\
			  &rel. error& $0$/NA & $1.38\%$& $7.99\%$& $-0.49\%$\\
			\hline		
			2 &$n_{s}$ & $0.9495$ & $0.9643$ & $1.027$ & $0.9474$\\
			  &rel. error& $0$/NA & $1.54\%$& $7.8\%$& $-0.21\%$\\
			\hline
			3&$n_{s}$ &$ 0.9661 $&$ 0.9803 $&$ 1.031 $&$ 0.9695$\\
			&rel. error& $0$/NA & $1.4\%$& $6.6\%$& $0.35\%$\\
			\hline
		\end{tabular}
	\caption[{\bf $n_s$ derived in different methods}]{Shown are different results for different methods of calculating the scalar index $n_s$. These were calculated for the 3 example potentials mentioned in Table \ref{Examples}. The first is the numerical result. Next is the standard Stewart \& Lyth expression Eq.~\eqref{eq:SLR}. Another result is given by using \eqref{eq:TrueNs}, using $\nu_{SL}$ but without substituting potential and derivative expressions for slow-roll parameters. Finally, we use Eqs.~(\ref{AccurateZoZ-2},\ref{TrueNu},\ref{eq:TrueNs}), to accurately assess the scalar index.\label{ns of different methods}}
	\end{center}
\end{table}
\subsubsection{So what is it good for?}
Absolutely nothing. This whole exercise in relating PPS directly to the inflationary potential and derivative was supposed to yield two advantages:
\begin{itemize}
	\item[1.] To be able to compute the PPS directly from the inflationary potential in a fast way. This is partially achieved with \cite{Dvorkin:2009ne}. However in the age of High-Performance-Computing (HPC), clusters and Graphic-Processor-Unit (GPU) based computing, the advantage of this approach is marginal at best.
	\item[2.] To be able to solve the inverse problem: Given a PPS as recovered from observations to be able to directly calculate the inflationary potential. This notion has been annihilated by the above analysis since it becomes evident that a direct connection through $\varepsilon_H\simeq \varepsilon_V$ etc. is not valid, and even if possible, deconvolving the inflationary potential via the approach in \cite{Dvorkin:2009ne} would be extremely costly in time and resources.  
\end{itemize}


\graphicspath{{Chapter5/}}

\chapter{Results}

\subsection{Models with $r=0.001$}
\begin{figure*}[!h]
\includegraphics[width=1\textwidth]{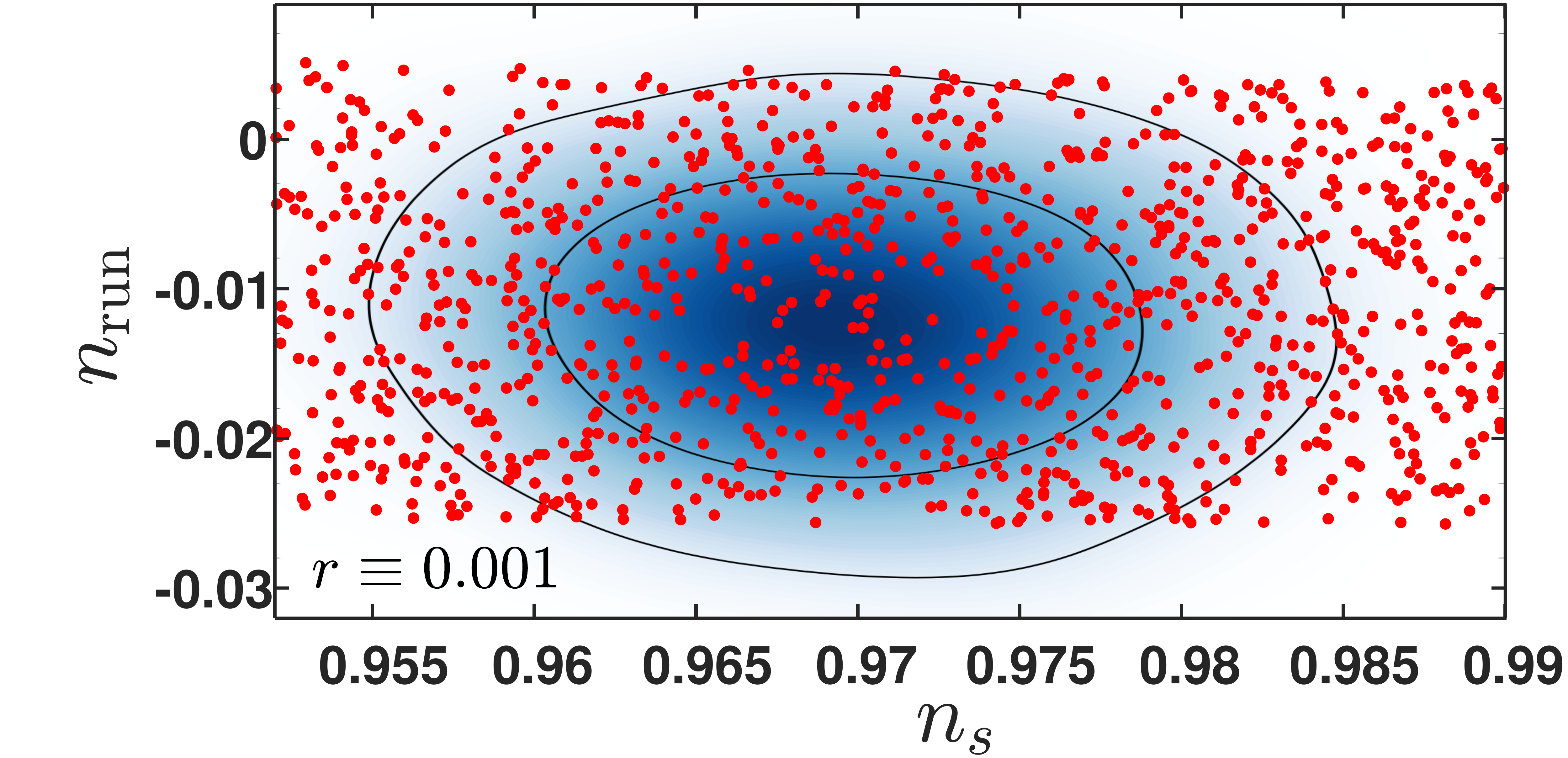}
\caption[{\bf Covering $n_s-\alpha_s$ with $r=0.001$ models.}]{Shown are the results of evaluating $n_s$ and $n_{\text{\tiny{run}}}$ for about 1100 models for which $r_0=0.001$.  The contour curves are the $68\%$ and $95\%$ confidence estimators, obtained from a CosmoMC  $\Lambda$CDM + index running model run \cite{Lewis:2002ah} using the Planck \& Bicep joint data analysis \cite{Ade:2015tva}.  The pivot scale used in the analysis is $k_{pivot}=0.05\; h\; Mpc^{-1}$, which is the same scale as in \cite{Ade:2015tva}.  \label{figure:FirstResult}}
\end{figure*}

In this section, we present the results of evaluating cosmological parameters for many small field models. In Fig.~\ref{figure:FirstResult} we show an example for which we calculate $n_s$ and $\alpha_s$ for about 1100 models with a fixed scalar to tensor ratio $r_0=0.001$. The results are shown on a $n_{s}-\alpha_s$ joint probability graph with the $68\%,95\%$ contours that are the probability estimators as yielded by a CosmoMC \cite{Lewis:2002ah} $\Lambda$CDM +index running model run, with the most recent Bicep \& Planck data (including WMAP 9-year mission)  \cite{Ade:2015tva}.

The reason for initially choosing the value of $r_0=0.001$ (and not a higher value, for example, $r=0.01$) was the following. We discovered that as we increased the values of $r$, the inflaton potentials needed to be more complicated and additional parameters were required. Also, we encountered several technical difficulties which we were able to resolve for the lower values of $r$.  Solving these difficulties and constructing a reliable framework for numerical calculations of the CMB observables was an essential step towards building models with higher values of $r$.

We allow the values of $n_s$ to vary quite substantially, rather than restrict them to the narrow range that is allowed by the data. Our idea is that when $r$ and $n_{\text{\tiny{run}}}$  are free to vary, the constraints on $n_s$ are relaxed in a significant way. The reason is that there is some degeneracy among the parameters. This is validated in the preliminary analysis that we present here. In addition, despite the fact that some models have yielded an almost flat (and some even a blue) $n_s$ and therefore are in conflict with the data, we found their analysis useful because insight regarding the departure of precisely calculated results from what the analytic SL term \eqref{eq:SLR} predicts was gained, as was discussed in chapter \ref{Chap:Fresh_ns}.
\subsubsection{Evaluating cosmological parameters for fixed $r_0$}
The $n_{s}-\alpha_s$ plane was covered with models which yield a fixed value of $r_0=0.001$. The cosmological parameters of some 3500 potentials were calculated. Fig.~\ref{figure:FirstResult} shows cosmological parameters of $\sim$ 1100 models. A significant number of the models yield values of $n_{s}$ and $\alpha_s$ within the $68\%$ and $95\%$ likelihood region. The most probable value for $\frac{V''}{V}=-0.0052 \pm 0.0034$. This is within the $68\%$ CL Planck results, with or without including high-$l$ polarization data. The third coefficient values are given by $\frac{V'''V'}{V^2}=0.0138 \pm 0.0065 $, which is in better agreement with the result without high-$l$ data. However the 2015 Planck analysis \cite{Ade:2015lrj} sets $\epsilon_4\equiv 0$ which might bias the results slightly. In the 2013 analysis \cite{Planck:2013jfk} this was not done, and our results agree with their analyses, including our values for $\frac{V^{(4)}V'}{V^2}$. Additional factors that contribute to the difference in analyses are the approximate connection between Hubble flow functions $\epsilon_i$ and the potential derivative quantities $\epsilon_V, \eta_V,\xi^2_V$. An interesting feature of these models is the departure of precisely calculated results from what the analytic SL expression \eqref{eq:SLR} predicts, as discussed in chapter \ref{Chap:Fresh_ns}. It might be possible to cover the $n_{s}-\alpha_s$ allowed region with models with a higher scalar-to-tensor ratio. However, the treatment of models which yield higher $r$ is  more complex since, by increasing $r$, one is forced to consider a larger $\Delta\phi$ range CMB region. The CMB region (see Fig.~\ref{fig:1OverSqrt(2eps)}) is roughly 3 times larger in $\phi$ for models with $r_0=0.01$, thus it will typically result in a running of running of the power spectrum. 

\begin{figure*}[!h]
\includegraphics[width=1\textwidth]{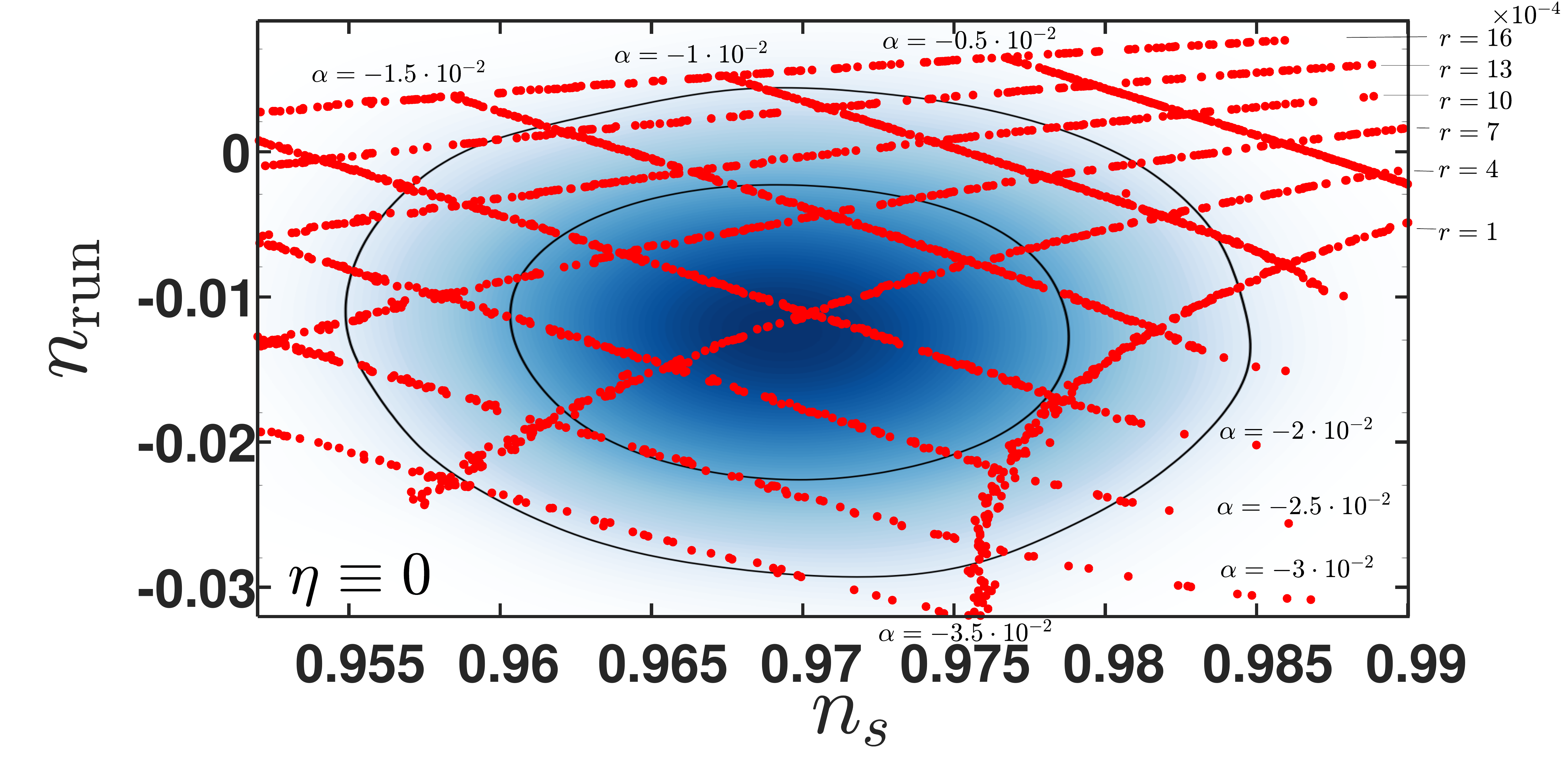}
\caption[{\bf Mapping $n_s-\alpha_s$ phase-space with $\eta_0=0$ }]{Covering the $n_s-n_{\text{\tiny{run}}}$ plane with constant $r$  and constant $\alpha_s$ characteristics, for $\eta_0 =0$}\label{figure:SecondResult}
\end{figure*}

\subsubsection{Evaluating cosmological parameters for fixed $\eta_0$}
The effects of varying $r_0$ on the resulting power spectrum were studied. In order to do this $\eta_0$ was set to $0$ for simplicity, and the $n_{s}-n_{\text{\tiny{run}}}$ plane was covered with models of varying $r_{0}$ and $\alpha_{0}$. Fig.~ \ref{figure:SecondResult} shows the results of this study.\bigskip

Notice that the effect of varying both $r_0$ and $\alpha_0$ on the changes in the value of $n_{s}$ is more pronounced than expected. Usually when $\eta_0\equiv 0$ one expects $n_{s}-1$ to first order to be $\propto -\frac{3r_{0}}{8}$ and thus  $\Delta n_s/\Delta r_0\simeq 10^{-4}\sim 10^{-5}$. At second order, we expect $n_{s}-1$ to be $\propto \frac{\alpha_{0}}{15}$ and thus $\Delta n_s/\Delta\alpha_0\simeq 10^{-3}$, whereas in this case, the change in $n_{s}$ is of the order of $10^{-2}$. A possible explanation to this phenomenon is a discrepancy between the analytic predictions made using \eqref{eq:SLR} and the precise calculations (see chapter \ref{Chap:Fresh_ns}).
\section{6th degree models with $r=0.01$}
We apply the methods discussed in Chapter \ref{methods} to the  degree six polynomial inflationary potentials that yield $r=0.01$. We calculate the most likely coefficients and extract the resulting most likely polynomial inflationary potential. The PPS resulting from this inflationary potential is then calculated in order to confirm that the most likely coefficients reconstruct the most likely observables.
\subsection{Results for degree six polynomials that yield $r=0.01$}
\begin{figure}[!ht]
\includegraphics[width=1\textwidth]{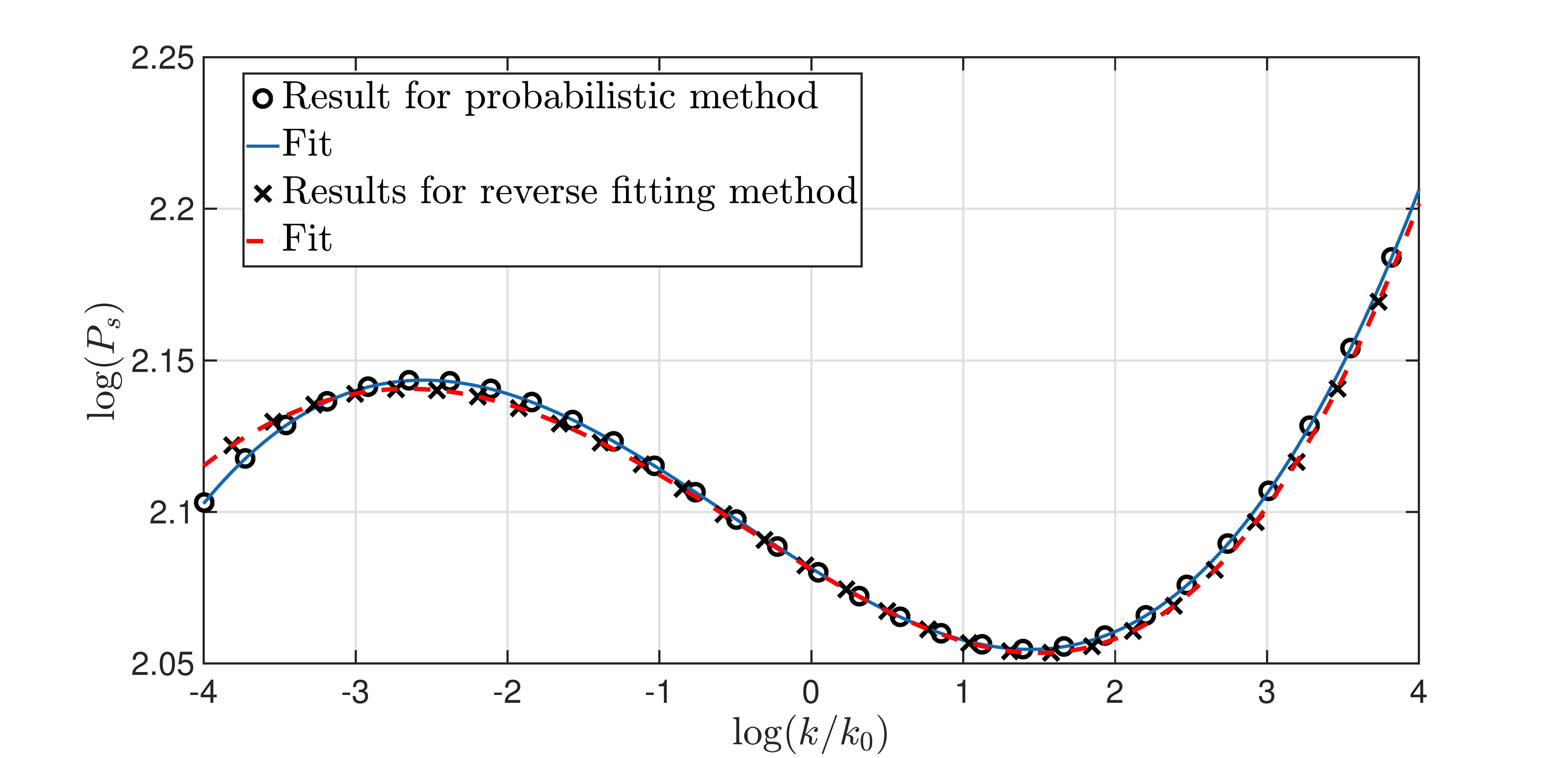}
\caption[{\bf PPS reconstruction for most likely $r=0.01$ model.}]{Reconstruction of the PPS from the most likely potential with $r=0.01$, as calculated by the multinomial (reverse fitting) method (X's and red dash), as well as the probabilistic method (circles and blue line). The CMB observables are well within the 68\% confidence levels of the MCMC analysis for both. However, the probabilistic method seems to yield more precise results.\label{Reconstruct_001} }
\end{figure}
In Fig. \ref{main_001} we showed a cover for the joint likelihood map of $n_s - \alpha_s$, of about $2000$ potentials with $r=0.01$. The cover is approximately uniform, thus we were able to assign likelihoods to every potential we study, as previously discussed.

By a process of marginalization, as discussed in Section \ref{methods}, we extract the most likely coefficients, which yield the likeliest observables. This process is represented graphically in Fig. \ref{Prob_001}. The results are shown in Table \ref{RES_001} and the PPS reconstruction is shown in Fig \ref{Reconstruct_001}. The advantage of this method is that it also determines the deviation from the average value. This can be used as an indicator for the level of tuning that is required to construct the most likely small field model. A discussion of tuning in field theoretic models can be found in \cite{Barbieri:1987fn}, as well as in \cite{Ellis:1986yg} and \cite{Fowlie:2014xha}. In most cases, the tuning level can be viewed as simply $\tfrac{\Delta x_i}{x_i}$, which in this case are given by $(0.375,0.27,5.5)$ for $a_2,a_3,a_4$ respectively. The width of the Gaussian fits for $\{a_2,a_3,a_4\}$ are $\{0.015,0.041,0.112\}$ respectively. These widths represent the effective area in parameter space that yields observables within the $68\%$ CL. Which is another measure of the tuning required in the degree 6 polynomial models.

~ Recalculating the CMB observables that this most likely model yields, we find $n_s=0.9687,\alpha_s=0.0089,\beta_s=0.0176$. These values are very close to the most likely values found from the previously discussed MCMC analysis of the BICEP2+Planck data. The resulting scalar index fits the most likely value in Table \ref{Table_cosmomc} exactly, while $\alpha_s$ and $\beta_s$ deviate from these values by no more than $12.5\%$  . We found that this is a relic of the binning method. Adding more models to the simulated data and refining the binning process results in even better proximity to the desired values.

Using the method of multinomial evaluation (\ref{multi}), we found the multinomial coefficients for each of the model degrees of freedom. For instance, for $a_2$ we have:
\begin{align}
	 \begin{array}{ccc}
	B&=&\left(\begin{array}{ccc}
	 -20.97 &-0.936& -37.716\\
	&19.19&30.402\\
	&&-407.53
	\end{array}\right)\\
	A&=&\left(40.918,0.955,79.253\right)\\
	p_0&=&- 19.938
	\end{array} .
\end{align}
Since $n_s \sim 1$, and $\alpha_s$ and $\beta_s$ are of the order of $10^{-2}$, the above result suggests that $a_2$ is primarily dominated by $n_s$. Similarly, we have found that $a_3$ is dominated by a linear combination of $\alpha_s$, and $\beta_s$, and $a_4$ is primarily dominated by $\beta_s$.
This method yields the most likely CMB observables with comparable accuracy to the previous method upon recalculation: $n_s=0.9684,\alpha_s=0.0077,\beta_s=0.020$.
\subsection{Most likely potentials}
\begin{table}[!ht]
\begin{center}
\begin{tabular}{|c||c|c||}
\hline
Observable&\multicolumn{2}{|c|}{Recalculated}\\
\hline
&Probabilistic Method& Multi-fit\\
\hline\hline
$n_s$&$0.9687$&$0.9684$\\
\hline
$\alpha_s$&$0.0089$&$0.0076$\\
\hline
$\beta_s$&$0.0176$&$0.020$ \\
\hline
\end{tabular}
\caption[{\bf Comparison of PPS from different methods}]{A comparison of recalculated power spectra observables from the results of the two extraction methods.\label{RECALC_001}}
\end{center}
\end{table}
Since $n_s$ is better constrained, we opt for the analysis that yields a more precise value of $n_s$. The leading degree 6 polynomial which yields $r=0.01$ at the proper pivot scale, is thus given by:
\begin{align}
	V=V_0\left(1 - 0.035\phi + 0.04\phi^2 -0.15\phi^3 +0.02\phi^4 +0.76\phi^5 -0.78 \phi^6\right).
\end{align}
Upon initial investigation, it seems these models produce a relatively flat tensor power spectrum. This might motivate future research of the tensor power spectrum predictions and constraints.
\subsection{Observable dependence}
\begin{table}[!ht]
\begin{center}
\begin{tabular}{|c||c|c|c||c}
\hline
&\multicolumn{2}{|c|}{Gaussian extraction}&Multinomial fit\\
\hline
$r=0.01$&$\mu$ (average)& $\sigma$ (standard deviation)& value\\
\hline\hline
$a_2$&$0.0402$ &$0.0156$&$0.01866$\\
\hline
$a_3$&$-0.152$ &$0.0414$&$-0.0235$\\
\hline
$a_4$&$0.0215$ &$0.1123$&$-0.3452$\\
\hline
\end{tabular}
\caption[{\bf Most likely coefficients}]{The most likely coefficients extracted by the process of likelihood assignment and marginalization, as well as by using the multinomial method. \label{RES_001}}
\end{center}
\end{table}
An interesting finding is an inter-dependence of the three observables $n_s,\alpha_s,\beta_s$. For models that yield $r=0.01$, there is a quadratic relation between the observables, such that $\beta_s=\beta_s(n_s,\alpha_s)$. It should be stressed that this is a phenomenon associated with the models and not with the observational data. This is supported by the small $n_s,\alpha_s,\beta_s$ paired-covariance found in \cite{Cabass:2016ldu}, implying weak dependence among observables in the data itself.
\begin{figure}[!h]
\includegraphics[width=1\textwidth]{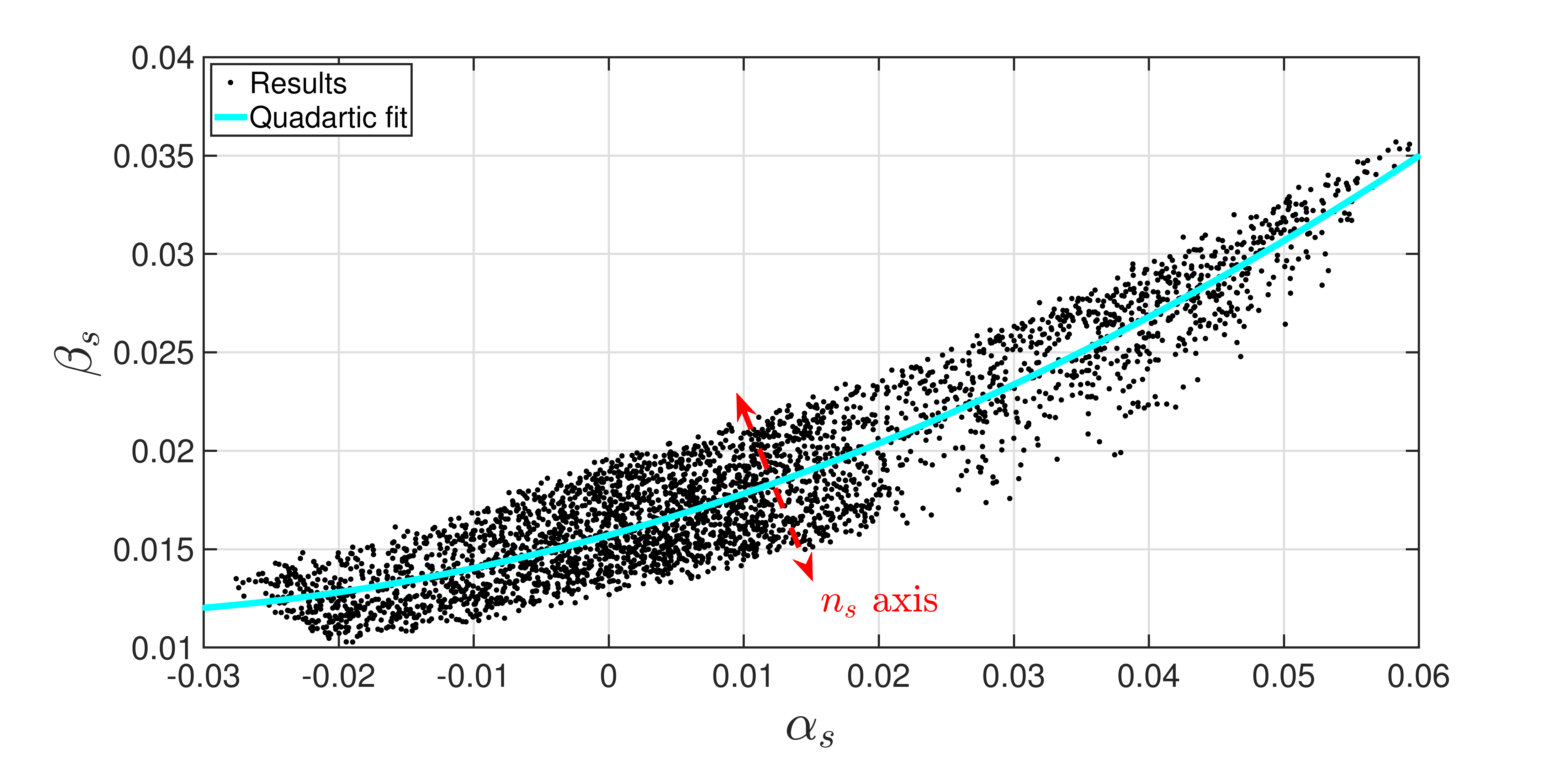}
\caption[{\bf Dependence of $\beta$ on other observables.}]{Dependence of $\beta_s$ on the other observables, exposes an approximate quadratic relation between $\alpha_s$ and $\beta_s$. The width of the resulting band indicates the deviation from a quadratic relation, which is correlated to $n_s$.
\label{Dependency2} }
\end{figure}

 One might think that the previous findings in \cite[Figure~23]{Ade:2015xua} indicate that $n_s$ and $\alpha_s$ are dependent. However, the graph shows a dependence between $\alpha_s$ and $n_{s,0.002}$ which is the scalar index evaluated at $k=0.002\; h\;Mpc^{-1}$. Taking some initial $n_s$ evaluated at $k_{0}$, it follows that $n_s$ evaluated at some other scale depends on the index running $\alpha_s$. Indeed, when one examines the color coding in \cite[Figure~23]{Ade:2015xua}, which represents $n_s$ at the pivot scale, it is clear that $n_s$ and $\alpha_s$ are independent.
\section{6th degree models with $r=0.03$}
\subsection{Methods}
We analyse the most recently available observational data \cite{Aghanim:2018eyx} by using CosmoMC \cite{Lewis:2002ah}, extracting the likelihood curves of the scalar index, its running and the running of running, $n_s,\alpha_s,\beta_s$ respectively. We then simulate a large number of inflationary models with polynomial potentials and calculate the primordial power spectrum (PPS) observables $n_s$, $\alpha_s$, $\beta_s$, that they predict. Each simulated potential is assigned a likelihood by the combined likelihood of the observables that it yields, as discussed in detail in \cite{Wolfson:2018lel}. We restrict the models that we consider to those predicting a power spectrum that can be fitted well by a degree 3 polynomial. This corresponds to the scalar index $n_s$, the index running $\alpha_s$, and the running of running $\beta_s$. We do that by fitting the PPS and evaluating the fitting error
\begin{align}
	\Delta^2=\frac{1}{n}\sum_{k=1}^n\left[\log(P_S(k))-f(k)\right]^2,
\end{align}
where $f(k)$ is the fitting curve. The threshold for considering a specific model in our analysis is $\Delta^2<10^{-6}$.

An additional complication arises due to the higher amount of tuning that is required for these models. The coefficient $a_1$ is fixed as $a_1=-\sqrt{r/8}$ \cite{BenDayan:2009kv}, so when the value of $r$ is higher, then $dV/d\phi$ at the CMB point has a larger magnitude. This has the effect of decreasing the number of e-folds generated per field excursion interval. If $r$ is increased to $Cr$, then $\frac{dN}{d\phi}$ is decreased by a factor  $\sim \frac{1}{\sqrt{C}}$
close to the CMB point. Since the first 8 or so e-folds of inflation are fairly constrained by observations, the amount of freedom in constructing the potential is reduced. It follows that either greater tuning is required to construct valid potentials, or one should consider a higher degree polynomial as suggested in \cite{Hotchkiss:2011gz}. Ultimately the choice is a matter of practical convenience. We opted for using degree 6 polynomials.

We employ two methods of retrieving the most likely polynomial potential. First,  we extract by marginalisation the most likely coefficients $\{a_p\}$. The other method amounts to performing a multinomial fitting of the coefficients as a function of the observables and then inserting the most likely observables to recover the corresponding coefficients. This method is explained in detail in \cite{Wolfson:2018lel}.

The `most likely model', is the model with a potential that generates the most likely CMB observables  $n_S\simeq 0.9687,\alpha_s\simeq 0.008,\beta_s\simeq 0.02$, as produced by the MCMC analysis of the most recent data available to date.
\begin{figure}[!h]
\includegraphics[width=0.8\textwidth]{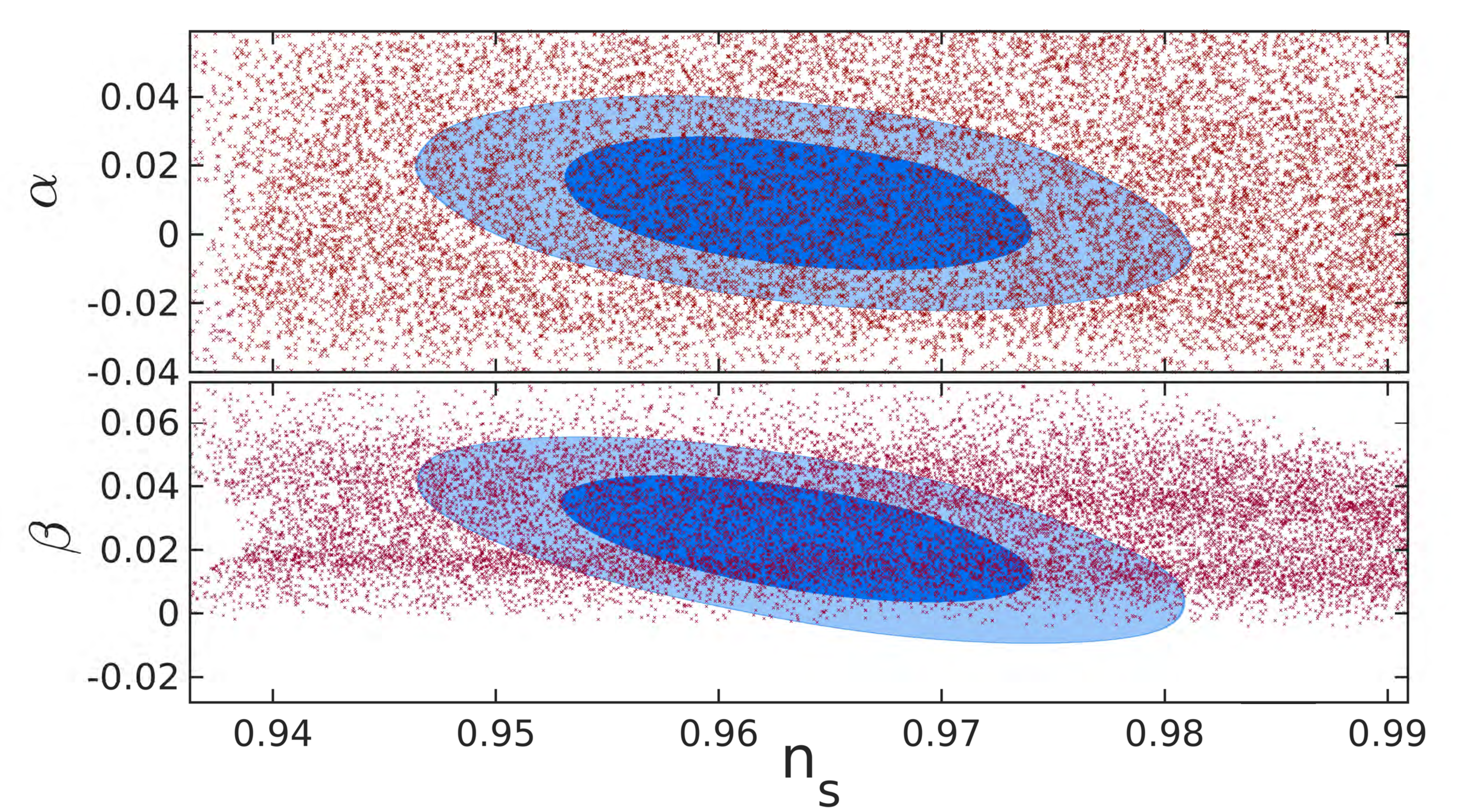}
\caption[{\bf Covering $n_s-\alpha_s$ with $r=0.003$ models.}]{Covering the observable phase space with small field models that predict $r\simeq 0.03$. The roughly uniform cover of the $95\%$ CL areas ensures an accurate likelihood transfer from MCMC results to models.\label{Main_result}}
\end{figure}

\subsection{Results}
\begin{figure}[!h]
\includegraphics[width=1\textwidth]{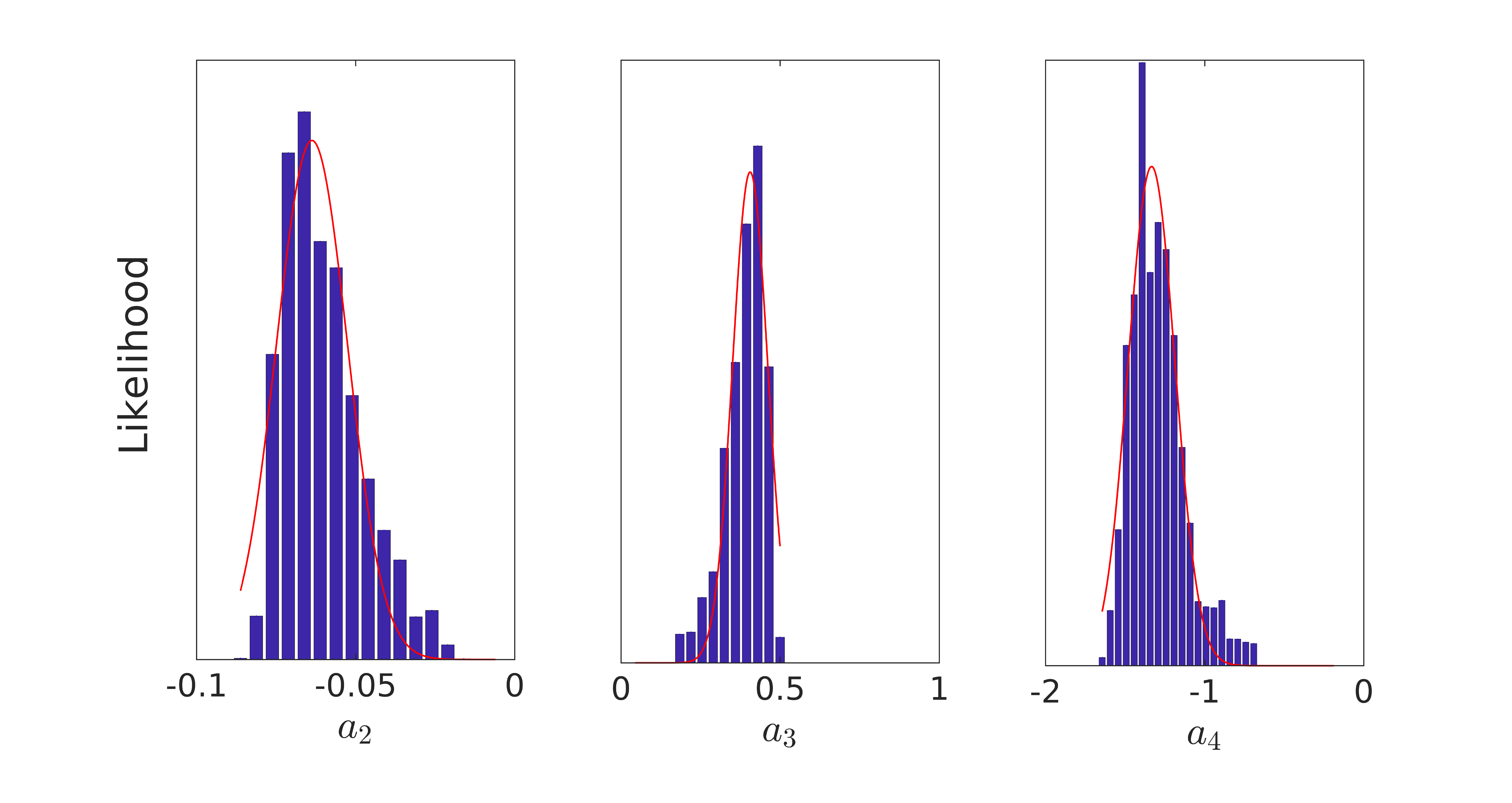}
\caption[{\bf Likelihood analysis for $r=0.03$ models.}]{Likelihood analysis for model coefficients $a_2,a_3,a_4$, using Gaussian fits to recover mean values. The skewness of the results reflect properties of the model, not of the underlying MCMC analysis. The width of the Gaussian fits are given by $(0.016,0.08,0.21)$ and are an indication of the tuning level required for these models.\label{Likelihood_results}}
\end{figure}

We produced many models that predict $r\simeq0.03$  and, additionally, predict PPS observables within the likely values. A roughly uniform cover of both the $(n_s,\alpha_s)$ and $(n_s,\beta_s)$ allowed values is shown in Fig.~\ref{Main_result}. This enables us to assign a likelihood to each simulated model, as discussed in detail in \cite{Wolfson:2016vyx,Wolfson:2018lel} and in the methods section. Consequently, it is possible to perform a likelihood analysis of models.  The results of this likelihood analysis is an approximately Gaussian distribution of the free coefficients as is shown in Fig. \ref{Likelihood_results}. Since the peaks of Gaussian fit (red line in Fig.~\ref{Likelihood_results}) do not entirely coincide with the peaks of the distribution and the distribution tails are not symmetric, we conclude that the distribution has a significant skewness.
The required tuning level is also evident from Fig.~\ref{Likelihood_results} and is given by $(\Delta a_2,\Delta a_3, \Delta a_4)=(0.016,0.08,0.21)$. Using Gaussian analysis and taking into account the skewness, we recover the most likely coefficients, which yield the following  degree six polynomial small field potential:
\begin{align}
	V=V_0\left[1-\sqrt{\frac{0.03}{8}}\phi -0.069\phi^2 + 0.431\phi^3 -1.413\phi^4 +2.455\phi^5 -1.487\phi^6\right]. \label{Most_likely_gauss}
\end{align}

Due to the skewness of the distribution, the values obtained by the Gaussian fit deviate by a significant amount from the most likely values of the observables. For instance, $n_s$ as determined by the potential in \eqref{Most_likely_gauss} is $\sim 0.98$ which is about $2\%$ away from the most likely value. For this reason we use this method of analysis to evaluate the required  tuning levels, whereas the most likely model is extracted by the multinomial fitting method.

Using the popular Stewart-Lyth (SL) theoretical values for $n_s$ and $\alpha_s$ \cite{Stewart:1993bc,Lyth:1998xn} as derived directly from the inflationary potential around the pivot scale, one finds  values that deviate by a significant amount from the Planck values. The SL values correspond to a very blue power spectrum and large running,
\begin{align}
	\left.n_s\right|_{SL}\simeq 1.55 \\
	\nonumber \left.\alpha_s\right|_{SL}\simeq -0.216 .
\end{align}
This discrepancy, that was discussed in \cite{Wolfson:2016vyx}, is related to the magnitude of $r$ for our class of models. When $r$ is smaller than $\sim 10^{-4}$ in such models, the original model building procedure that relied on the SL values, which is outlined in \cite{BenDayan:2009kv}, is valid and produces approximately the correct values of the observables. However, when values of $r$ are larger, one cannot trust the analytic SL estimates.

\begin{figure}[!h]
\includegraphics[width=1\textwidth]{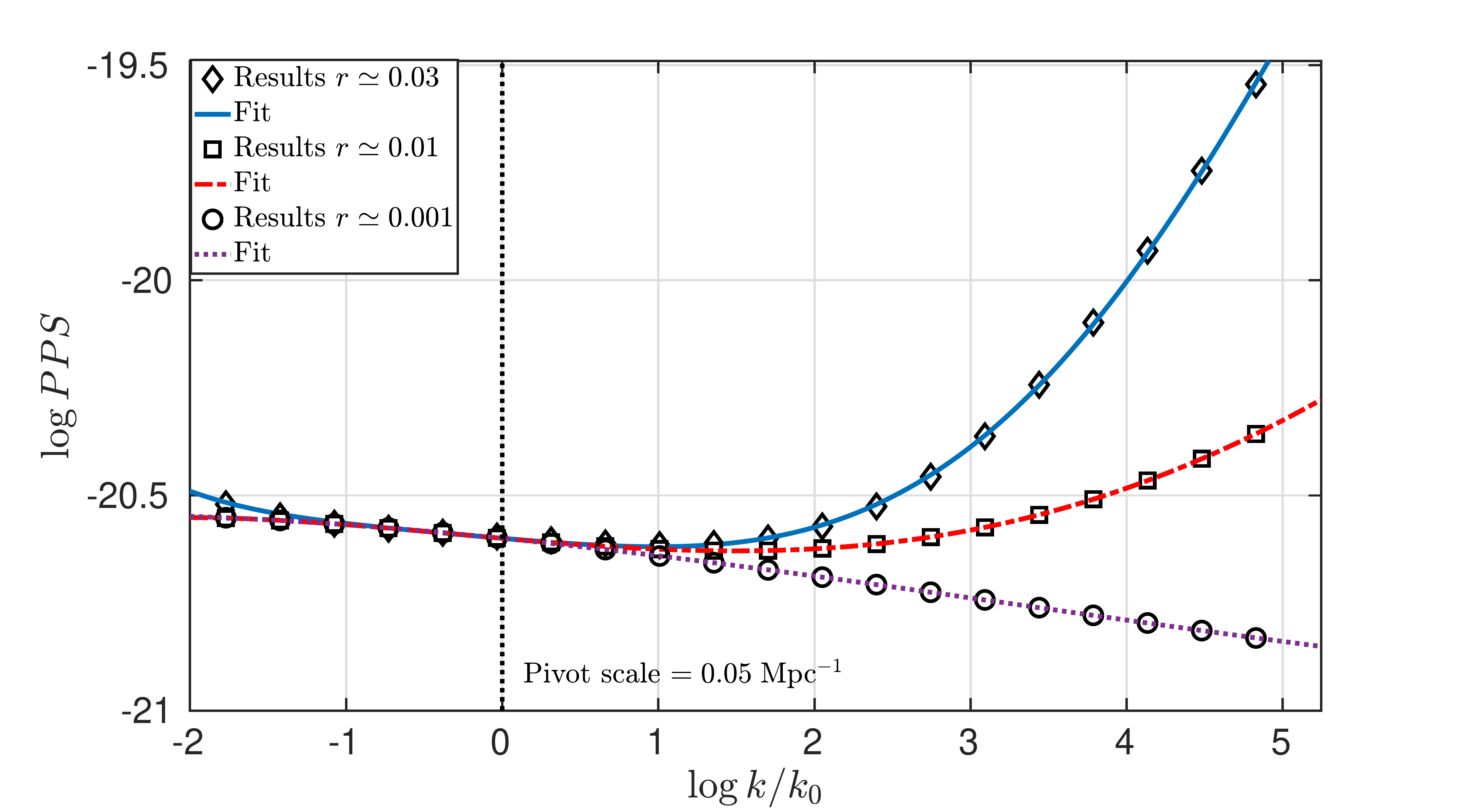}
\caption[{\bf PPS results of most likely models with $r=0.001,0.01,0.03$.}]{Primordial power spectra of three  models. A model with a degree six polynomial potential that predicts $r\simeq 0.03$ (diamonds and blue line). A model with a degree six polynomial potential that predicts $r\simeq0.01$ (squares and red dash-dot)  and a model with a degree five polynomial potential that predicts $r=0.001$ (circles and purple dots). The pivot scale for all three is $k_0=0.05 \mathrm{Mpc^{-1}}$ and the results are overlayed at that scale for ease of comparison.\label{fig:PPS}}
\end{figure}

In Fig.~\ref{fig:PPS} the power spectra  generated by three inflationary models are shown. (1) A model with degree 5 polynomial potential that predicts $r\simeq 0.001$; (2) A model with degree 6 polynomial potential that predicts $r\simeq 0.01$; and finally (3) A model with degree 6 polynomial potential that predicts $r\simeq 0.03$.

Representing each coefficient $a_p$ as a function of the observables $(n_s,\alpha_s,\beta_s)$ and evaluating them at $(0.9687,0.008,0.02)$, leads to the following potential,
\begin{align}
	V=V_0\left[1-\sqrt{\frac{0.03}{8}}\phi -0.067\phi^2 +0.413\phi^3 -1.419\phi^4 +2.512\phi^5 -1.523\phi^6\right].
\end{align}
The values $n_s,\alpha_s,\beta_s$ that this model predicts deviate from the most likely values by $(0.06\%,10\%,19\%)$. However, the deviations are within the $95\%$ CL of all recent MCMC analyses.



\chapter{Conclusion}\label{Conclusions}
We have found that small field models of the degree 6 polynomial form can produce a significant GW signal at the level of $r=0.03$ while conforming to known CMB observables. We have further shown that analytic methods of evaluating the PPS in these models fail and a full numerical treatment is necessary. While the idea of these models is relatively new, it was first suggested almost ten years ago \cite{BenDayan:2009kv}. However, our work was the first to put these models to the numerical test. Due to the failings of analytical methods of PPS evaluation, the model building technique originally suggested in \cite{BenDayan:2009kv} was shown to be insufficient thus our work is the first to show the validity of these models. These models are gaining prominence among considered inflationary models, due to the swampland conjecture \cite{Garg:2018reu,Kehagias:2018uem,Ben-Dayan:2018mhe}, that seem to favour small field models, over the large field variety.\bigskip \\
In explicitly showing analytic approaches of evaluating the PPS to be insufficient in certain cases, we question the validity of such approximations in other cases as well. The discussion in section \ref{Chap:Fresh_ns}, also shows beyond a doubt that the second order Lyth-Riotto term for the index running given in Eq.~\eqref{eq:Lyth-Riotto-alpha} cannot, by definition, be correct. This means that all models previously analysed by using this expression are now in need of re-examination. This is further suggestive of a need for numerical reassessment of such codexes as \cite{Martin:2013tda}, that heavily rely on such analytic tools.\bigskip \\

Our initial hypothesis was that a degree 5 polynomial model with $r=0.03$ should be able to cover the full PPS observable phase space, as allowed by current CMB experiments. Our numerical analysis has disproved this hypothesis. Instead, we have shown that for full coverage of the observables parameter space with models which yield $r=0.03$, either degree 6 polynomial models while allowing a running of running is necessary, or one must consider a degree 7 polynomial at least, such as in work done in \cite{Hotchkiss:2011gz}. Be that as it may, full coverage is, in fact, possible by small field models with $r=0.03$, which was the end goal of this research.\bigskip \\

Whether one favours large field models or small field models, we have created the numerical tools that now seem to be indispensable in studying inflationary potentials in the age of precision cosmology. The limit of analytic tools have been known for a long time, we are by no means the first to point out this discrepancy and the possible theoretical pitfalls (cf. \cite{Wang:1997cw}). However, we are the first to show the meaning of this disparity in actual results and their repercussions in terms of precision and model building. Our only hope is that future researchers find it necessary to evaluate their results numerically, at the high standard of precision we have now achieved.\bigskip \\
\noindent \begin{minipage}{0.45\linewidth}~\end{minipage}*    *    *\noindent \begin{minipage}{0.45\linewidth}~\end{minipage}\bigskip \\
If I may be so bold as to suggest my personal conclusions, one should aspire to make connections with people of different disciplines, in physics, and otherwise. This serves the purposes of mutual enrichment and comradery. Sometimes one needs the layperson for a set of fresh eyes on a problem. Sometimes an adversarial point of view is needed, to test new and somewhat `crazy' ideas. But above all:\\
\begin{large}
\begin{center}
\textbf{PHYSICISTS DO NOT GIVE UP!}
\end{center}
\end{large}

\ifpdf
    \graphicspath{{Chapter6/Figs/Raster/}{Chapter6/Figs/PDF/}{Chapter6/Figs/}}
\else
    \graphicspath{{Chapter6/Figs/Vector/}{Chapter6/Figs/}}
\fi



\begin{spacing}{0.9}


\def\nphysb{Nucl.~Phys.~B}   		
\def\prb{Phys.~Rev.~B}       		
\def\prl{Phys.~Rev.~Lett.}   		
\def\rmp{Rev.~Mod.~Phys.}    		
\def\rpp{Rep.~Prog.~Phys.}   		
\def\pnas{Proc.~Natl.~Acad.~Sci.}   	
\def\ijmpb{Int.~J.~Mod.~Phys.~B} 	
\def\anp{Ann.~Phys.}			
\def\jetp{J.~Exp.~Theor.~Phys.}	
\def\jpa{J.~Phys.~A}			
\def\bams{Bull. Amer. Math. Soc.}	
\def\plb{Phys. Lett. B}		
\def\jcap{JCAP}			
\def\phrep{Phys. Rept.}		
\def\apj{ApJ}				
\def\prd{Phys. Rev. D}			
\def\mnras{MNRAS}			
\def\arxiv{ArXiv e-prints}		
\def\pr{Physical Review}		

\bibliographystyle{unsrt} 

\cleardoublepage
\bibliography{References/listArx} 



\end{spacing}


\begin{appendices} 

\chapter{Derivation of Mukhanov-Sasaki equation} 

\subsubsection*{Deriving the equation of motion for $\phi$ in conformal time}
In this section we will derive the equation of motion for the scalar field, and the perturbation thereof, in conformal-Newtonian gauge.\\
Using the conformal time metric $ds^2=a^{2}\left(d\tau^2-d\vec{x}^2\right)$ we define:
\begin{align}
	g_{\mu\nu}=a^2\left(\tau\right)\eta_{\mu\nu}
\end{align}
Where $\eta_{\mu\nu}$ is Minkowski's metric.\\\\
We now take the action to be:
\begin{align}
	S=\int d^4 x\sqrt{-g}\left[\tfrac{1}{2}g^{\mu\nu}\partial_{\mu}\phi\partial_{\nu}\phi - V\left(\phi\right)\right]
\end{align} 
Varying in respect to $\phi$ we get:
\begin{align}
	\delta S=\int d^4 x \left(\tfrac{1}{2}g^{\mu\nu}\right(\partial_{\mu}(\delta \phi)\partial_{\nu}\phi+\partial_{\nu}(\delta \phi)\partial_{\mu}\phi\left)-\frac{dV}{d\phi}\delta\phi\right)\sqrt{-g}\\ 
	\nonumber +\delta\sqrt{-g}\left[\tfrac{1}{2}g^{\mu\nu}\partial_{\mu}\phi\partial_{\nu}\phi - V\left(\phi\right)\right]
\end{align}
Since we take the metric and the scalar field to be independent coordinates, $\frac{dg_{\mu\nu}}{d\phi}=0$ and so we consider only the first term. Applying the symmetry $(\nu\leftrightarrow \mu)$ of the metric we get: 
\begin{align}
	\delta S =\int d^4 x \left(g^{\mu\nu}\partial_{\nu}\phi\partial_{\mu}\delta\phi-\frac{dV}{d\phi}\delta\phi\right)\sqrt{-g}	
\end{align}
Integrating by parts we are left with:
\begin{align}
	\int d^4 x \left(-\partial_{\mu}\left(g^{\mu\nu}\sqrt{-g}\partial_{\nu}\phi\right)\delta \phi-\sqrt{-g}\frac{dV}{d\phi}\delta\phi\right)+\left. g^{\mu\nu}\sqrt{-g}\partial_{\nu}\phi\delta \phi\right|_{p1}^{p2}
\end{align}
And the second term is just some constant, depending only on the boundary conditions, and not on any variational contribution, thus we are allowed to disregard it.\\
Thus we are left with:
\begin{align}
\partial_{\mu}\left(g^{\mu\nu}\sqrt{-g}\partial_{\nu}\phi\right)+\sqrt{-g}\frac{dV}{d\phi}=\\ 
\nonumber \sqrt{-g}\partial_{\mu}\left(g^{\mu\nu}\partial_{\nu}\phi\right) + g^{\mu\nu}\partial_{\nu}\phi\partial_{\mu}\left(\sqrt{-g}\right)+\sqrt{-g}\frac{dV}{d\phi}=0\\
\Rightarrow \partial_{\mu}\left(g^{\mu\nu}\partial_{\nu}\phi\right) + g^{\mu\nu}\partial_{\nu}\phi\frac{\partial_{\mu}\left(\sqrt{-g}\right)}{\sqrt{-g}}+\frac{dV}{d\phi}=0\label{Variation calculus for the action in phi}
\end{align}
which, up to now is a completely general result.\\\\
Plugging in the conformal metric we get:
\begin{flalign}
	\sqrt{-g}=a^4 ;\hspace*{20pt} \partial_{\mu}=\left(\frac{d}{d\tau},\vec{\nabla}\right);\hspace*{20pt}\partial^{\mu}=a^{-2}\left(\frac{d}{d\tau},-\vec{\nabla}\right)
\end{flalign}
And we are left with:
\begin{flalign}
	a^{-2}\left(\phi''-\nabla^{2}\phi\right)-2a^{-3}a'\phi'+a^{-2}\phi'\frac{4a'}{a}+\frac{dV}{d\phi}\\
	\Rightarrow \left(\phi''-\nabla^{2}\phi\right)+2H\phi'+a^{2}\frac{dV}{d\phi}=0
\end{flalign}
which reverts back to Eq.(\ref{equation of motion for phi in conformal time}) if we consider a homogeneous, time dependant $\phi=\phi(\tau)$\\
However, this equation gives us additional information, as we will see in the continued derivation, since it contains the Laplacian term.\\\\
Now, looking back at Eq.(\ref{Variation calculus for the action in phi}), which, as stated is so far completely general, we want to apply perturbation theory approach and vary the coordinate slightly:
\begin{flalign}
	\phi=\phi^{0}(\tau)+\delta\phi\left(\tau,\vec{x}\right)
\end{flalign}
Doing this, is not enough, since there is an interplay between energy density (as affected by $\phi$ and functions thereof) and the metric. So we will also use a perturbed metric, and we will use the Conformal-Newtonian gauge in which:
\begin{flalign}
	g_{\mu\nu}=a(\tau)^{2}\left(\eta_{\mu\nu}+h_{\mu\nu}\right)
\end{flalign}
Where as a rule the perturbation is very small compared to the zero order metric.\\\\
We will derive the equations using $h_{\mu\nu}$ to give the most general result, and in the last possible moment, we will replace the abstract perturbation with a specific explicit one.\\\\
So, considering the perturbed quantities stated above we have:
\begin{flalign}
	\partial_{\mu}\left[\left(g_{0}^{\mu\nu}+a^{-2}h^{\mu\nu}\right)\partial_{\nu}\left(\phi^{0}+\delta\phi\right)\right]\\
	\nonumber + \left(g_{0}^{\mu\nu}+a^{-2}h^{\mu\nu}\right)\partial_{\nu}\left(\phi^{0}+\delta\phi\right)\frac{\partial_{\mu}\sqrt{-g}}{\sqrt{-g}}\\
	\nonumber + \left.\frac{dV}{d\phi}\right|_{\phi=\phi^{0}}+\left.\frac{d^{2}V}{d\phi^{2}}\right|_{\phi=\phi^{0}}\delta\phi=0
\end{flalign}
Keeping only first order in perturbation (either in $h^{\mu\nu}$ or in $\delta\phi$) we arrive at:
\begin{flalign}	\partial_{\mu}\left(g_{0}^{\mu\nu}\partial_{\nu}\phi^{0}\right)+\partial_{\mu}\left(a^{-2}h^{\mu\nu}\partial_{\nu}\phi^{0}\right)+
\partial_{\mu}\left(g_{0}^{\mu\nu}\partial_{\nu}\delta\phi\right)+\\
\nonumber g_{0}^{\mu\nu}\partial_{\nu}\phi^{0}\left.\frac{\partial_{\mu}\sqrt{-g}}{\sqrt{-g}}\right|_{h=0}+a^{-2}h^{\mu\nu}\partial_{\nu}\phi^{0}\left.\frac{\partial_{\mu}\sqrt{-g}}{\sqrt{-g}}\right|_{h=0}+\\
\nonumber g_{0}^{\mu\nu}\partial_{\nu}\delta\phi\left.\frac{\partial_{\mu}\sqrt{-g}}{\sqrt{-g}}\right|_{h=0}+ g_{0}^{\mu\nu}\partial_{\nu}\phi^{0}\delta\left(\frac{\partial_{\mu}\sqrt{-g}}{\sqrt{-g}}\right)+\\
\nonumber \left.\frac{dV}{d\phi}\right|_{\phi=\phi^{0}}+\left.\frac{d^{2}V}{d\phi^{2}}\right|_{\phi=\phi^{0}}\delta\phi=0
\end{flalign}
We can simplify this a tiny bit by demanding the zero order term to vanish independently, to get:
\begin{flalign}
\partial_{\mu}\left(a^{-2}h^{\mu\nu}\partial_{\nu}\phi^{0}\right)+
\partial_{\mu}\left(g_{0}^{\mu\nu}\partial_{\nu}\delta\phi\right)+
 a^{-2}h^{\mu\nu}\partial_{\nu}\phi^{0}\left.\frac{\partial_{\mu}\sqrt{-g}}{\sqrt{-g}}\right|_{h=0}+\\
\nonumber g_{0}^{\mu\nu}\partial_{\nu}\delta\phi\left.\frac{\partial_{\mu}\sqrt{-g}}{\sqrt{-g}}\right|_{h=0}+ g_{0}^{\mu\nu}\partial_{\nu}\phi^{0}\delta\left(\frac{\partial_{\mu}\sqrt{-g}}{\sqrt{-g}}\right)+ \left.\frac{d^{2}V}{d\phi^{2}}\right|_{\phi=\phi^{0}}\delta\phi=0
\end{flalign}
Where the subscript $\left.\cdot\right|_{h=0}$ means setting the metric perturbation to zero, and the term $\delta\left(\frac{\partial_{\mu}\sqrt{-g}}{\sqrt{-g}}\right)$ stands for "take only the first order perturbation of this quantity"\\\\
So far this calculation is completely general, but now we can't escape choosing a gauge for the perturbation term. In \cite{Mukhanov1992203} it has been shown that the "correct" gauge in this case, i.e. a gauge that does not introduce any relic of the perturbed gauge into the equation of motion is the Conformal-Newtonian one, in which:
\begin{flalign}
	h_{\mu\nu}=2\left(\begin{array}{cccc}
	\Phi&&&\\
	&\Psi&&\\
	&&\Psi&\\
	&&&\Psi
	\end{array}\right)
\end{flalign}
It can readily be shown that the first order inverse of the metric $g_{\mu\nu}=a^{2}\left(\eta_{\mu\nu}+h_{\mu\nu}\right)$, is given by:
\begin{flalign}
	g^{\mu\nu}=a^{-2}\left(\begin{array}{cccc}
	1-2\Phi&&&\\
	&-1-2\Psi&&\\
	&&-1-2\Psi&\\
	&&&-1-2\Psi
	\end{array}\right)
\end{flalign}
Thus the inverse can be decomposed into:
\begin{flalign}
	g^{\mu\nu}=a^{-2}\left(\eta^{\mu\nu}+h^{\mu\nu}\right)
\end{flalign}
with 
\begin{flalign}
	h^{\mu\nu}=-2\left(\begin{array}{cccc}
	\Phi&&&\\
	&\Psi&&\\
	&&\Psi&\\
	&&&\Psi
	\end{array}\right)
\end{flalign}
(differently put: $h^{\mu\nu}=-h_{\mu\nu}$ )\\\\
With this explicit form we can now approach the term for $\sqrt{-g}$:
\begin{flalign}
	g_{\mu\nu}=a^{2}\left(\begin{array}{cccc}
	1+2\Phi&&&\\
	&-1+2\Psi&&\\
	&&-1+2\Psi&\\
	&&&-1+2\Psi\\
	\end{array}\right)
\end{flalign}
Thus:
\begin{flalign}
	\sqrt{-g}=\sqrt{a^{8}\left(1+2\Phi\right)\left(1-2\Psi\right)^{3}}=a^{4}\sqrt{\left(1+2\Phi\right)\left(1-2\Psi\right)^{3}}=\\
	\nonumber \overset{\texttt{1st order}}{\simeq}a^{4}\left(1+\Phi -3\Psi\right)
\end{flalign}
And so the variational approach result yields:
\begin{flalign}
	0=-4H\Phi \phi'^{(0)} -2\left(\Phi\phi''^{(0)}\right) +2H\delta\phi'\\
	\nonumber +\left(\delta\phi''-\nabla^{2}\delta\phi\right)-\left(\phi'^{(0)}\left(\Phi'+3\Psi'\right)\right)+a^{2}\frac{d^{2}V}{d\phi^{2}}\delta\phi
\end{flalign}
And replacing the first two terms, by using the equation for the unperturbed metric and scalar field we get:
\begin{flalign}	0=2H\delta\phi'+\left(\delta\phi''-\nabla^{2}\delta\phi\right)-\left(\phi'^{(0)}\left(\Phi'+3\Psi'\right)\right)+a^{2}\frac{d^{2}V}{d\phi^{2}}\delta\phi+2a^{2}\Phi\frac{dV}{d\phi}\label{Equation of motion for perturbed phi}
\end{flalign}
\subsubsection*{Deriving the equation of motion for the perturbed metric}
The aim of this section is to derive the equation of motion for the perturbed metric, as a function of the scalar field $\phi$. While in the previous passage we stated the action, and via variational calculus, recovered the dynamics, varying the action of the Hilbert-Einstein action while aesthetically pleasing, amounts to additional unnecessary work.\\
The approach we will follow next, is a perturbative one, but of the Einstein field equations, rather then the action.
\begin{flalign}
	R^{\mu}\,_{\nu}-\tfrac{1}{2}g^{\mu}\,_{\nu}\mathbf{R}\equiv G^{\mu}\,_{\nu}=8\pi G T^{\mu}\,_{\nu}
\end{flalign}
It can be shown, that this equation can be expanded like so:
\begin{flalign}
	\sum_{k=0}^{\infty}\frac{\delta^{k}}{k!}\cdot\,^{(k)}G^{\mu}\,_{\nu}=\sum_{k}\frac{\delta^{k}}{k!}8\pi G \cdot\,^{(k)}T^{\mu}\,_{\nu},
\end{flalign}
where the superscript $k$ for the tensors $G^{\mu}\,_{\nu}$ and $T^{\mu}\,_{\nu}$ state the perturbative degree, such that $\,^{(0)}T^{\mu}\,_{\nu}$ denotes the zero order perturbation, $\,^{(1)}T^{\mu}\,_{\nu}$ the first, and so on.
One can look at the first order perturbation like so:
\begin{flalign}
		\delta G^{\mu}\,_{\nu}=8\pi G \delta T^{\mu}\,_{\nu},
\end{flalign}
where here $\delta(\cdot )$ denotes the first order correction.\\\\
Some fancy machinery is now called for, as we need to construct the first order perturbation of the Ricci tensor and scalar, as well as derive the first order perturbation to the Stress-Energy tensor $T^{\mu}\,_{\nu}$
\subsubsection*{Stress-Energy tensor first order perturbation}
Consider Eq.(\ref{Stress-Energy tensor for scalar field}), we now want to introduce the perturbation in scalar field as well as metric, saving only first order elements, and discarding the zero order equation we are left with:
\begin{flalign}
\delta T^{\mu}\,_{\nu}=g^{\mu\alpha}\partial_{\alpha}\phi^{(0)}\partial_{\nu}\delta\phi +g^{\mu\alpha}\partial_{\alpha}\delta\phi\partial_{\nu}\phi^{(0)}\\ \nonumber +\delta(g^{\mu\alpha})\partial_{\alpha}\phi^{(0)}\partial_{\nu}\phi^{(0)}
-\left[\tfrac{1}{2}g^{\alpha\beta}\partial_{\alpha}\phi^{(0)}\partial_{\beta}\delta\phi+\tfrac{1}{2}g^{\alpha\beta}\partial_{\beta}\phi^{(0)}\partial_{\alpha}\delta\phi \right.  
\\ \nonumber \left.+\tfrac{1}{2}\delta\left(g^{\alpha\beta}\right)\partial_{\alpha}\phi^{(0)}\partial_{\beta}\phi^{(0)}-\frac{dV}{d\phi}\delta\phi\right]\delta^{\mu}_{\nu}
\end{flalign}
Taking into account our gauge and assumptions on $\phi$, the term in the square brackets can be readily evaluated as:
\begin{flalign}
	\left[\frac{1}{a^{2}}\phi'^{(0)}\delta\phi'-\frac{\Phi\phi'^{(0)2}}{a^{2}}-\frac{dV}{d\phi}\delta\phi\right]
\end{flalign} 
Transforming the perturbed term to:
\begin{flalign}
	\delta T^{\mu}\,_{\nu}=g^{\mu\alpha}\partial_{\alpha}\phi^{(0)}\partial_{\nu}\delta\phi +g^{\mu\alpha}\partial_{\alpha}\delta\phi\partial_{\nu}\phi^{(0)}+\delta(g^{\mu\alpha})\partial_{\alpha}\phi^{(0)}\partial_{\nu}\phi^{(0)}
\\ \nonumber -\left[\frac{1}{a^{2}}\phi'^{(0)}\delta\phi'-\frac{\Phi\phi'^{(0)2}}{a^{2}}-\frac{dV}{d\phi}\delta\phi\right]\delta^{\mu}\,_{\nu}
\end{flalign}
Decoding this yields:
\begin{flalign}
	\left\{\begin{array}{cc}
		\delta T^{0}\,_{0}=&\frac{1}{a^{2}}\left[\phi'^{0}\delta\phi'-\Phi\phi'^{(0)2}+a^{2}\frac{dV}{d\phi}\delta\phi\right]\\
		\delta T^{0}\,_{i}=&\frac{\phi'^{(0)}\delta\phi_{;i}}{a^{2}}\\
		\delta T^{i}\,_{j}=&-\frac{1}{a^{2}}\left[\phi'^{(0)}\delta\phi'-\Phi\phi'^{(0)2}-a^{2}\frac{dV}{d\phi}\delta\phi\right]\delta^{i}\,_{j}
	\end{array}\right.
\end{flalign}
Doing the same kind of procedure to find the perturbed metric, we are left with the following Christoffel symbols:
\begin{align}
	\begin{array}{lcr}
	\Gamma^{0}\,_{00}=H+\Phi' &\hspace*{20pt}& \Gamma^{0}\,_{0i}=\Gamma^{i}\,_{00}=\tfrac{d\Phi	}{dx^{i}}\\
	\\
	\Gamma^{0}\,_{ij}=\delta_{ij}\left(H-2H\left(\Phi+\Psi\right)-\Psi'\right)&&\Gamma^{i}\,_{0j}=\Gamma^{i}\,_{j0}=H-\Psi'\\
	\\
	\Gamma^{i}\,_{ij}=-\tfrac{d\Psi}{dx^{j}}&&\Gamma^{i}\,_{jj}=\tfrac{d\Psi}{dx^{i}}-2\tfrac{d\Psi}{dx^{i}}\delta_{ij}
	\end{array}
\end{align}
which in turn yields the follwoing Ricci tensor:
\begin{align}
	\begin{array}{ccc}
		R_{00}=&\hspace*{10pt}&-3H' +\nabla^{2}\Phi+3\left(\Psi'+\Phi'\right)+3\Psi''\\
		\\
		R_{0i}=&& 2\partial_{i}\Psi'+2H\partial_{i}\Phi\\
		\\
		R_{ij}=&& \delta_{ij}\left[H'+2H^{2}-\Psi''+\nabla^{2}\Psi -2\left(H'+2H^{2}\right)\left(\Psi+\Phi\right)-H\Phi'-5H\Psi'\right]\\
		\\
		&&+\partial_{i}\partial_{j}\left(\Psi-\Phi\right)
	\end{array}
\end{align}
It follows that the Ricci scalar $\mathbf{R}$, first order is given by:
\begin{align}
	\mathbf{R}=\frac{1}{a^{2}}\left[-6\left(H'+H^{2}\right)+2\nabla^{2}\Phi-4\nabla^{2}\Psi+12\left(H'+2H^{2}\right)\Phi+6\Psi''+6H\left(\Phi'+3\Psi'\right)\right]
\end{align}
Leading (at last) to the Einstein tensor, where we have already noted the perturbed part alone:
\begin{align}
	\begin{array}{lcl}
		\delta G^{0}\,_{0}&=&\frac{2}{a^{2}}\left[-3H\left(H\Phi+\Psi'\right)+\nabla^{2}\Psi\right]\\
		\\
		\delta G^{0}\,_{i}&=&\frac{2}{a^{2}}\partial_{i}\left[H\Phi+\Psi'\right]\\
		\\
		\delta G^{i}\,_{j}&=&\frac{-2}{a^{2}}\left[\left(2H'+H^{2}\right)\Phi+H\Phi'+\Psi''+2H\Psi'+\tfrac{1}{2}\nabla^{2}\left(\Phi-\Psi\right)\right]\delta^{i}\,_{j}\\\\
		&&+\tfrac{1}{a^{2}}\partial_i\partial_j\left(\Phi-\Psi\right)
	\end{array}
\end{align}
Equating the off diagonal spatial-spatial terms we get:
\begin{align}
	\nabla^{2}\left(\Phi-\Psi\right)=0
\end{align}
Meaning the function $\left(\Phi-\Psi\right)$ is a harmonic function defined over $\mathbb{R}^{1+3}$, thus, by virtue of Liouville's theorem, it is constant over all $\mathbb{R}^{1+3}$. Demanding the perturbations to vanish at infinity sets the function $\left(\Phi-\Psi\right)$ to zero over all space(time), thus we are left with $\Phi=\Psi$, which leaves us with a single degree of freedom.\\\\
Rewriting the results, taking into account our new understanding we are left with:
\begin{align}
	\begin{array}{lcl}
		\delta G^{0}\,_{0}&=&\frac{2}{a^{2}}\left[-3H\left(H\Phi+\Phi'\right)+\nabla^{2}\Phi\right]\\
		\\
		\delta G^{0}\,_{i}&=&\frac{2}{a^{2}}\partial_{i}\left[H\Phi+\Phi'\right]\\
		\\
		\delta G^{i}\,_{j}&=&\frac{-2}{a^{2}}\left[\left(2H'+H^{2}\right)\Phi+3H\Phi'+\Phi''\right]\delta^{i}\,_{j}
	\end{array}
\end{align}
Using the perturbed Einstein equation:
\begin{align*}
	\delta G^{\mu}\,_{\nu}=M_{pl}^{-2}\delta T^{\mu}\,_{\nu}
\end{align*}
We are left with these equations:
\begin{align}
	\begin{array}{lcl}
	2\left[-3H\left(H\Phi+\Phi'\right)+\nabla^{2}\Phi\right]&=&M_{pl}^{-2}\left[\phi'\delta\phi'-\Phi\phi'^{2}+a^{2}\frac{dV}{d\phi}\delta\phi\right]\\
		\\
		2\partial_{i}\left[H\Phi+\Phi'\right]&=&M_{pl}^{-2}\partial_{i}\left(\phi'\delta\phi\right)\\
		\\
2\left[\left(2H'+H^{2}\right)\Phi+3H\Phi'+\Phi''\right]&=&M_{pl}^{-2}\left[\phi'\delta\phi'-\Phi\phi'^{2}-a^{2}\frac{dV}{d\phi}\delta\phi\right]
	\end{array}
\end{align}
Or, using the connection $\phi'^{2}=2M_{pl}^{2}\left(H^{2}-H'\right)$:
\begin{align}
	\begin{array}{lcl}
	\nabla^{2}\Phi -3H\Phi' -\left(H'+2H^{2}\right)\Phi&=&\tfrac{M_{pl}^{-2}}{2}\left[\phi'\delta\phi'+a^{2}\frac{dV}{d\phi}\delta\phi\right]\\
		\\
		2\partial_{i}\left[H\Phi+\Phi'\right]&=&M_{pl}^{-2}\partial_{i}\left(\phi'\delta\phi\right)\\
		\\
\Phi''+3H\Phi'+\left(H'+2H^{2}\right)\Phi&=&\tfrac{M_{pl}^{-2}}{2}\left[\phi'\delta\phi'-a^{2}\frac{dV}{d\phi}\delta\phi\right]
	\end{array}
\end{align}
Using the second equation to identify $\delta\phi$, we take the third equation, subtract the first from it, and finally using Eq.(\ref{equation of motion for phi in conformal time}), we eliminate the term for $a^{2}\tfrac{dV}{d\phi}$ to arrive at this expression:
\begin{align}\label{Mukhanov-Sasaki - raw}
	\Phi''-\nabla^{2}\Phi+2\left(H-\tfrac{2\phi''}{\phi'}\right)\Phi'+2\left(H\-\tfrac{2\phi''}{\phi'}H\right)\Phi=0
\end{align}
Or in an equivalent form:
\begin{align}\label{Mukhanov-Sasaki - alternative}
	\Phi''-\nabla^{2}\Phi+2\left(\tfrac{a}{\phi'}\right)'\left(\tfrac{a}{\phi'}\right)^{-1}\Phi'+2\phi'\left(\tfrac{H}{\phi'}\right)'\Phi=0
\end{align}
Switching to a new variable $u\equiv\left(\tfrac{a}{\phi'}\right)\Phi$, with $\theta\equiv \frac{H}{a\phi'}$ this equation now takes the form:
\begin{align}\label{Mukhanov-Sasaki equation}
	u''-\nabla^{2}u-\left(\frac{\theta''}{\theta}\right)u=0
\end{align}
This is the so-called Mukhanov-Sasaki equation.\\\\
The careful reader might easily recognize a simple equation for a time-dependant harmonic oscillator.\\\\
Moving to Fourier space, the reciprocal of $\theta$ is used as the pump filed thus taking the Fourier representation of this equation, equating each mode to zero separately yields:
\begin{align}\label{Mukhanov-Sasaki fourier decomp.2}
	u_{k}''+\left(k^{2}-\frac{Z''}{Z}\right)u_{k}=0
\end{align}  
Where here $Z$ is defined such that $Z=\frac{a\phi'}{H}$


\chapter{The INSANE code - usage examples}\label{INSANE_CODE_APPENDIX}
This appendix contains an example of \code{params.in} file that is used to run the INSANE code.
There are several options to consider. Since the code is fast, it makes it possible to run the code on a list of inputs, given in a predefined
format, and produce an output in a predefined format as well. An empty template of \code{params.in} is supplied to this end. The user however will need to construct a parser that parses the incoming list and produces the desired output.
\section*{An example \code{params.in} file:}
\begin{Verbatim}
"""
Created on Wed Mar 20 08:49:25 2019

@author: ira
"""

# note that in the python file we have:
# import numpy as np
# import sympy as sym


# Healpy integration yes/no
Healpy=False

# pyCamb integration yes/no
pycamb=False

pycambPATH='../../CAMB/CAMB-1.0.3/'

# path is absolute or relative?
absolute=False

H0=np.sqrt(V0/float(3))
phidot0 = 0

# number of efolds in the inflation (>50)
efoldsago=60

# If this is an actual physical inflation to consider or something else
# (like studying feature-response etc.)
physical=True

# If this is NOT physical we need to specify wave number to start and end with
EndK=1600
StartK=1



# number of efolds to analyse (i.e. number of efolds in CMB)
efolds=np.log(2500)

# set True if you want to take models of a list.
from_list=False

# The specific model
V0=1
l1= -0.035355339059327
l2= 0.018658438192659
l3= -0.023535429113224
l4= -0.345249995103359
l5= 1.216702780626785
l6= -0.974291597599925
l7=0
phi0=-0.192599999999995


# is the model supplied symbolic or polynomial?
symbolic=True

# if True use the following syntex for example:
phi=sym.Symbol('phi')
V=1+l1*phi+l2*phi**2 +l3*phi**3 +l4*phi**4 +l5*phi**5 +l6*phi**6 +l7*phi**7

# precision for time integration in background solution
Tprecision=0.01

# level of feedback in background solution: 0=silent,1=verbose,2=graphic
feedback=0

#######################################################
#
# MS params
#
#######################################################


# time precision for MS solver
MStimePrec=0.01

# Number of k-modes to solve
MSkPrec=40

# Whether to draw k's such that they are evenly spaced on a log-log scale
Klog=True

# MS solver feedback level
MSfeedback=0

# Take starting time at CMB point=exactly N efolds ago ('exact'), or when phi=0 ('aprox')
CMBpoint='aprox'


# Analysis parameters
pivotScale=0.05

# Maximal fitting degree - the analyzer will try to fit with increasingly complex polynomials
# until either fitting error is small or exhausted the maximal degree. 
# None means deg is 20 by default
deg=None

# Analysis feedback level
AnFeedback=2


\end{Verbatim}
\section*{A usage example without \code{params.in}:}
\begin{Verbatim}
# -*- coding: utf-8 -*-

"""
Created on Sun Jun 28 13:18:31 2015

@author: Spawn
"""
from __future__ import division
from __future__ import print_function
import platform
from MsSolver15 import MsSolver
from CAMBenv import CAMBenv
import healpy as hp
from BackgroundGeometry import BackgroundSolver
import numpy as np
from scipy import io
import time as TIME
import sympy as sym


def parseCoeffs(string):
        """
        returns a tuple of the coefficients l1...l5,phi0 that have been stored in
        txt format as "l1;l2;l3;l4;l5;phi0"

        Parameters
        -----------
        string the string to parse

        Returns
        -----------
        a 6 element list [l1,l2,l3,l4,l5,phi0] of floats after parsing

        """
        STR=string
        print(STR)
        if isinstance(STR, str):
            STR=str(STR)
            STR=STR.split(";")
            return([float(STR[0]),float(STR[1]),float(STR[2]),float(STR[3]),float(STR[4]),float(STR[5])])
        else:
            e=Exception()
            e.message="not a proper string"
            raise e
        return([])


startTIME=TIME.time()
plat=platform.system()
mode='aprox'
efolds=10
efoldsago=60
EndK=1600
StartK=1
deg=3

"""""""""""""""""""""""""""""""""""""""
Best candidate with r=0.01
"""""""""""""""""""""""""""""""""""""""
l1= -0.035355339059327
l2= 0.018658438192659
l3= -0.023535429113224
l4= -0.345249995103359
l5= 1.216702780626785
l6= -0.974291597599925
l7=0
phi0=-0.192599999999995

V0=1
"""""""""""""""""""""""""""""""""""""""
"""""""""""""""""""""""""""""""""""""""

phi=sym.Symbol('phi')
V=1+l1*phi+l2*phi**2 +l3*phi**3 +l4*phi**4 +l5*phi**5 +l6*phi**6 +l7*phi**7

"""
Define initial conditions
"""
phidot0 = 0
H0=1/np.sqrt(3)

solver=BackgroundSolver(l1,l2,l3,l4,l5,H0,phi0,1,phidot0,\
	Tprecision=0.05,poly=False,Pot=V,mode='verbose')
solver.Solve()
print(solver.message)
if ((solver.message=="Inflation ended - |eta|~1") or \
	(solver.message=="Inflation ended - epsilon~1")):
    print("im in")
    if (solver.a[len(solver.a)-1]>50):
        try:
            Solver=MsSolver(solver.a,solver.H,solver.phi,solver.phidot\
		,solver.t,solver.epsilon,solver.eta,solver.xisq,solver.vd4v,\
		solver.vd5v,solver.v,solver.eh,solver.deltah,solver.delpp,\
		Tprecision=0.01,Kprecision=40,efoldsAgo=efoldsago,\
		efoldsNum=efolds,log=True,mode=mode)
            Solver.prepareToWork(mode='silent')
            Solver.buildUKT(mode='silent')
            if (Solver.message=="Not a slow roll evolution"):
                result="{0};{1};{2};{3};{4};{5};{6};{7};{8}; -- {9}"\
		.format(V0,l1,l2,l3,l4,l5,phi0,phidot0,H0,Solver.message)

            else:
                try:
                    Solver.analysis(mode='graphic',deg=8,pivotScale=0.05)
                    result="{0};{1};{2};{3};{4};{5};
				{6};{7};{8};{9};{10};{11};
				{12};{13};{14};{15};{16};{17};{18};{19};{20}"\
			.format(l1,l2,l3,l4,l5,Solver.coeffs[0],Solver.coeffs[1]\
			,Solver.coeffs[2],Solver.coeffs[3],Solver.coeffs[4],\
			Solver.coeffs[5],Solver.phiCMB,phi0,phidot0,H0,efolds,\
			efoldsago,Solver.Slope,Solver.ns,Solver.nrun,Solver.error)
                except Exception as e:
                    print(e.message)
                    result=e.message
        except Exception as e:
            result="{0};{1};{2};{3};{4};{5};{6};{7};{8};{9};{10};error message:'{11}'"\
			.format(V0,l1,l2,l3,l4,l5,phi0,phidot0,H0,efolds,efoldsago,e.message)

    else:
        print("not enough efolds for physical inflation")
        result="{0};{1};{2};{3};{4};{5};{6};{7};{8};
		 -- not enough efolds for physical inflation"\
	.format(V0,l1,l2,l3,l4,l5,phi0,phidot0,H0)
else:
        print(solver.message)
        result="{0};{1};{2};{3};{4};{5};{6};{7};{8}; -- {9}"\
	.format(V0,l1,l2,l3,l4,l5,phi0,phidot0,H0,solver.message)
v0=Solver.coeffs[0]
print('V0={0}'.format(Solver.coeffs[0]))
print('normalized a1={0}'.format(v0*Solver.coeffs[1]))
print('normalized a2={0}'.format(v0*Solver.coeffs[2]))
print('normalized a3={0}'.format(v0*Solver.coeffs[3]))
print('normalized a4={0}'.format(v0*Solver.coeffs[4]))
print('normalized a5={0}'.format(v0*Solver.coeffs[5]))
print('ns={0}'.format(Solver.ns))
print('nrun={0}'.format(Solver.nrun))
print('nrunrun={0}'.format(Solver.nrunrun))
print('error={0}'.format(Solver.error))

elapsed=TIME.time()-startTIME

try:
    cs=CAMBenv('../../CAMB/CAMB-1.0.3/','rel')
except Exception as e:
    cs=CAMBenv('../../CAMB/CAMB-1.0.3','rel')

cs.Insert_PS(Solver.lk0,Solver.lps)

cs.PRINT()


m=hp.synfast(cs.Cls,320)
hp.mollview(m)

\end{Verbatim}

\end{appendices}

\printthesisindex 

\end{document}